\documentclass[linenumbers,trackchanges]{aastex7}

\newcommand{\pf}{Phys. Fluids}
\newcommand{\jpp}{J. Plasma Phys.}
\newcommand{\prsla}{Proc. R. Soc. Lond.}

\usepackage{lineno}
\linenumbers

\usepackage{mathrsfs}
\usepackage{hyperref} % remove ugly link borders
\begin{document}

\title{Turbulence and its Potential Impact on Solar Chromospheric and Coronal Heating}

\author[orcid=0000-0002-4642-6192,sname='Zank']{G. P. Zank}
\affiliation{Center for Space Plasma and Aeronomic Research (CSPAR), The University of Alabama in Huntsville, Huntsville, AL 35805, USA}
\affiliation{Department of Space Science, The University of Alabama in Huntsville, Huntsville, AL 35899, USA}
\email[show]{garyp.zank@gmail.com}  

\author[orcid=0000-0001-5278-8029,sname='Li']{Xiaocan Li} 
%\altaffiliation{Las Campanas Observatory}
\affiliation{Los Alamos National Laboratory, Los Alamos, NM 87545, USA}
\email{xiaocanli@lanl.gov}

\author{K. Khanal}
\affiliation{Center for Space Plasma and Aeronomic Research (CSPAR), The University of Alabama in Huntsville, Huntsville, AL 35805, USA}
\affiliation{Department of Space Science, The University of Alabama in Huntsville, Huntsville, AL 35899, USA}
\email{kk0099@uah.edu}

\author{Alphonse C. Sterling}
\affiliation{NASA/Marshall Space Flight Center, Huntsville, AL 35805, USA}
\email{alphonse.sterling@nasa.gov}

\author[sname=Nakanotani]{M. Nakanotani}
\affiliation{Center for Space Plasma and Aeronomic Research (CSPAR), The University of Alabama in Huntsville, Huntsville, AL 35805, USA}
\affiliation{Department of Space Science, The University of Alabama in Huntsville, Huntsville, AL 35899, USA}
\email{mn0052@uah.edu}  

\author[sname=Zhao]{L.-L. Zhao}
\affiliation{Center for Space Plasma and Aeronomic Research (CSPAR), The University of Alabama in Huntsville, Huntsville, AL 35805, USA}
\affiliation{Department of Space Science, The University of Alabama in Huntsville, Huntsville, AL 35899, USA}\email{lz0009@uah.edu}

%\author{N. Subashchandar}
%\affiliation{Center for Space Plasma and Aeronomic Research (CSPAR), The University of Alabama in Huntsville, Huntsville, AL 35805, USA}
%\affiliation{Department of Space Science, The University of Alabama in Huntsville, Huntsville, AL 35899, USA}
%\email{nss0010@uah.edu}

\author{L. Adhikari}
\affiliation{Center for Space Plasma and Aeronomic Research (CSPAR), The University of Alabama in Huntsville, Huntsville, AL 35805, USA}
\affiliation{Department of Space Science, The University of Alabama in Huntsville, Huntsville, AL 35899, USA}
\email{la0004@uah.edu}

\author{M. Yalim}
\affiliation{Center for Space Plasma and Aeronomic Research (CSPAR), The University of Alabama in Huntsville, Huntsville, AL 35805, USA}
\affiliation{Department of Mathematical Sciences, The University of Alabama in Huntsville, Huntsville, AL 35899, USA}\email{msy0002@uah.edu}

\author[orcid=0000-0002-4454-147X,sname='Zank']{P.S.  Athiray}
\affiliation{Center for Space Plasma and Aeronomic Research (CSPAR), The University of Alabama in Huntsville, Huntsville, AL 35805, USA}
\affiliation{Department of Space Science, The University of Alabama in Huntsville, Huntsville, AL 35899, USA}
\email{sp0196@uah.edu}

\author{Fan Guo}
\affiliation{Los Alamos National Laboratory, Los Alamos, NM 87545, USA}
%\affiliation{}
\email{guofan@lanl.gov}

\author{R.L. Moore}
\affiliation{Center for Space Plasma and Aeronomic Research (CSPAR), The University of Alabama in Huntsville, Huntsville, AL 35805, USA}
\email{ronald.l.moore@nasa.gov}

\begin{abstract}
	
Low-frequency turbulence in the solar chromosphere remains poorly understood. We address 1) the sources of low-frequency turbulence that potentially heat the chromosphere, and 2) how turbulence is transported and dissipated throughout the chromosphere and lower corona. We use particle-in-cell simulations to investigate mixed polarity magnetic fields corresponding to emergent magnetic carpet field in coronal holes or quiet Sun regions for strong (imbalanced) and weak (balanced) guide magnetic fields. The initial mixed polarity magnetic field transitions rapidly to a turbulent state dominated by advected small-scale nonlinear structures, with a minority slab turbulence population and the emergent   field is largely annihilated. Turbulence is anisotropic for imbalanced magnetic field and more isotropic for balanced cases. We develop a transport model for turbulence advected and dissipated throughout the chromosphere by randomly distributed energy-containing scale dynamical flows described by log-normal statistics. We compute the expectations for the total energy per unit volume $\langle y \rangle (h)$ J m${}^{-3}$, the 
Els\"asser specific energy $\langle \langle {Z^{\infty}}^2 \rangle \rangle (h)$ m${}^2$ s${}^{-2}$, the heating rate  $\langle \dot{\cal H} \rangle (h)$ J m${}^{-3}$ s${}^{-1}$, and the correlation length $\langle \lambda \rangle (h)$ km  
as functions of height $h$ above the photosphere. 
Turbulent energy is injected into the low corona by a random ``patchwork'' of sites across the transition region surface. The expected energy injection rates $\langle \dot{S} \rangle$ J m${}^{-2}$ s${}^{-1}$ for the chromosphere and at the base of the corona exceed the estimated energy requirements needed to heat both the chromosphere and corona. Similarly, we show that spicules can be heated gradually with increasing height by entrained magnetic carpet and photospheric turbulence.
\end{abstract}



\section{Introduction} \label{sec:intro}

The solar chromospheric and coronal heating problems are inextricably linked \citep{Withbroe_Noyes_1977, Withbroe_1988}. One can estimate that a rate of energy dissipation of $\sim 4 \times 10^6$ erg cm${}^{-2}$ s${}^{-1}$ or $4 \times 10^3$ J m${}^{-2}$ s${}^{-1}$ is required to maintain the observed chromospheric temperature \citep{Withbroe_Noyes_1977}. By contrast, an energy deposition rate of $3 \times 10^5$ erg cm${}^{-2}$ s${}^{-1}$  or $3 \times 10^2$ J m${}^{-2}$ s${}^{-1}$ is necessary to sustain the observed coronal heating beyond $\sim 1.03 R_{\odot}$ \citep{Withbroe_1988}, $R_{\odot}$ being the radius of the Sun. These estimates have since been modestly revised upward  to $1.4 \times 10^7$ erg cm${}^{-2}$ s${}^{-1}$ or $1.4 \times 10^4$ J m${}^{-2}$ s${}^{-1}$  by \cite{Anderson_Athay_1989a}.  Any mechanism(s) that heats the low solar atmosphere must therefore account for both the chromospheric and  coronal heating rates, recognizing that the chromosphere is strongly mediated by radiative cooling and the corona is strongly mediated by both radiative cooling and collisional thermal conduction \citep[e.g.,][]{Moore_Fung_1972}. A few models have attempted to incorporate both chromospheric and coronal heating using simple phenomenological heating descriptions \citep[e.g.,][]{Hammer_1982a, Hammer_1982b, Withbroe_1988} of the form $\exp [-(r - R_{\odot})/H]$ ($r$ being heliocentric radius, $H$  a scale height over which heating is assumed to occur), or weak acoustic shock heating \citep{Hammer_1982b}, or a more elaborate Alfv\'en wave heating model \citep{Suzuki_Inutsuka_2006}. However, the question of how the chromosphere is heated, like that for coronal heating, is far from settled and whether the heating mechanisms for both regions are related or unrelated is an open question. 

In reviewing coronal heating mechanisms, both \cite{DeMoortel_Browning_2015} and \cite{Klimchuk_2015} argue that magnetic reconnection is in some way responsible for the heating, \cite{Klimchuk_2015} going so far as to assert that ``all coronal heating is impulsive.'' To some extent, these considerations largely motivated Parker's \citep{Parker_1972, Parker_1983, Parker_1988} nanoflare model. Here, photospheric motions twist or braid magnetic field lines, thereby allowing for reconnection and the creation of current sheets that inject suprathermal charged particles (small-scale Alfv\'en jets) with energies $\sim 1/2 mV_A^2$ (or temperatures up to $10^8$ - $10^9$ K for strong solar magnetic fields) into the thermal plasma. However, the energy released by and the heating rate from nanoflares is constrained by the Parker angle $\theta_P$ (the (acute) angle between the reconnecting magnetic field lines in the nanoflares), with $\theta_P \sim 10^{\circ}$ - $20^{\circ}$ for either active or quiet regions \citep{Parker_1983, Klimchuk_2015}. 

By contrast, the dissipation of low-frequency MHD turbulence as a coronal heating mechanism has attracted considerable interest, especially over the past decade with the Parker Solar Probe (PSP) mission. Perhaps the first turbulence transport formalism-based model for coronal heating was introduced by \cite{Matthaeus_etal_1999_coronalheating}, in which some fraction of outwardly propagating Alfv\'en waves are reflected by large-scale density gradients in the corona, resulting in counter-propagating waves that interact non-linearly to produce zero-frequency modes. The cascading of the turbulent fluctuations to higher wave numbers results in the eventual dissipation of magnetic (and kinetic) energy at small scales that heats the coronal plasma. PSP observes Alfv\'en waves with high normalized cross-helicity $|\sigma_c| \sim 1$ in both super-Alfv\'enic \citep[e.g.,][]{Zhao_etal_2020a} and sub-Alfv\'enic \citep{Zank_etal_2022, Zank_etal_2024} flows in the young solar wind. The typical dominance of outwardly propagating Alfv\'en waves means that a balanced flux of counter-propagating Alfv\'en waves is absent, i.e., $|\sigma_c| \sim 1$, which weakens or eliminates the non-linearity required to cascade magnetic energy from large to small scales and thus the ability to heat the coronal plasma.  To address this criticism, an alternative MHD turbulence model for coronal heating was introduced by \cite{Zank_etal_2018, Zank_etal_2021} who argue that the small plasma beta environment of the low solar atmosphere implies that turbulent fluctuations are primarily quasi-2D non-propagating structures such as magnetic islands, and Alfv\'enic fluctuations form a minority component \citep{zank_matthaeus_1993_incompress, Zank_Matthaeus_1992b, Zank_etal_2017a}. The non-propagating 2D turbulent fluctuations undergo a corresponding cascade to small scales at which dissipation associated with multiple small-scale current sheets occurs to heat the plasma. PSP observations \citep{Zank_etal_2024} have identified a dominant magnetic island component in MHD-scale turbulent fluctuations in the sub-Alfv\'enic solar wind \citep{Kasper_etal_2021, Zank_etal_2022, Zhao_etal_2022, Bandyopadhyay_etal_2022, Alberti_etal_2022, Zhang_etal_2022, Liu_etal_2023, Jiao_etal_2024} together with a minority and predominantly outward propagating Alfv\'enic component \citep{Zank_etal_2022}. Both the Alfv\'en wave or slab turbulence models \citep{oughton_etal_2001_reducedmhdmodel, dmitruk_etal_2001, Dmitruk_etal_2002, Suzuki_Inutsuka_2005, Cranmer_etal_2007, Cranmer_etal_2013, Wang_etal_2009, Chandran_Hollweg_2009, Verdini_etal_2010, Matsumoto_Shibata_2010, Chandran_etal_2011, usmanov_etal_2011_swmodturbtransheat, Lionello_etal_2014, Usmanov_etal_2014, Woolsey_Cranmer_2014, Shoda_etal_2018, Chandran_Perez_2019, Chandran_2021}  and the turbulence-based 2D nonlinear structures models \citep{Zank_etal_2018, Zank_etal_2021, Adhikari_etal_2020a, Adhikari_etal_2022c, Telloni_etal_2022a, Telloni_etal_2022b, Telloni_etal_2023a, Telloni_etal_2024, Adhikari_etal_2024a, Adhikari_etal_2024b} for coronal heating have since been  been considerably expanded.

For both turbulence models \citep{Matthaeus_etal_1999_coronalheating, Zank_etal_2018}, the origin of the turbulent fluctuations (Alfv\'enic and/or 2D/magnetic flux ropes) remains uncertain and speculative. Rapid transverse motions in the photosphere may initiate Alfv\'enic modes (or even fast magnetosonic modes that closely resemble Alfv\'en waves in the low plasma beta regime and are sometimes referred to as compressional Alfv\'en waves), which can then propagate into the sub- and super-Alfv\'enic corona along open magnetic field lines. Conversely, \cite{Zank_etal_2018, Zank_etal_2021} ascribe the origin of quasi-2D magnetic flux ropes and non-propagating structures  to reconnection of mixed polarity magnetic carpet loops just above the photosphere  in the chromosphere and above. The problem then is to advect the turbulent nonlinear structures into the low corona from the chromosphere. 

The ``traditional dichotomy'' \citep{DeMoortel_Browning_2015} between waves/turbulence and reconnection is likely inappropriate, and indeed turbulence and reconnection can more fruitfully be viewed as different sides of the same coin. The transfer and eventual dissipation  of magnetic energy in the dominant quasi-2D MHD model of turbulence \citep{Matthaeus_Lamkin_1986, zank_matthaeus_1993_incompress, Zank_Matthaeus_1992b, hunana_zank_2010_apj, Zank_etal_2017a, Zank_etal_2020} is determined by small- and multi-scale turbulent reconnection of nonlinear structures and the generation of small-scale current sheets \citep{Servidio_etal_2009, Servidio_etal_2010, Servidio_etal_2011}. Parker Solar Probe (PSP) \citep{Bale_etal_2023, Raouafi_etal_2023} and Solar Dynamics Observatory \citep{Uritsky_etal_2023} have provided suggestive observational evidence for interchange reconnection low in the solar corona, which appears to be consistent with the model advanced by \cite{Zank_etal_2018, Zank_etal_2021}. Zank et al.\ suggest persistent reconnection between mixed polarity small scale loops that form the magnetic carpet low in the solar atmosphere (50\% of mixed polarity magnetic carpet loops have heights less than $\sim 1$ Mm) as the source of quasi-2D turbulence.  Interchange reconnection in the chromosphere between the far more numerous mixed polarity loops and the occasional open magnetic field can also be expected to occur,  observations of which provide evidence for turbulent reconnection shaping energy transfer in the low solar atmosphere.  Some numerical experiments have begun to explore the relation between reconnection, turbulence, and the generation of quasi-2D fluctuations \citep{Rappazzo_Parker_2013} and/or Alfv\'en modes \citep{Yang_etal_2025} in the young solar wind. \cite{Rappazzo_Parker_2013} introduce 2D magnetic field perturbations at the base of a line-tied loop with a strong homogeneous mean magnetic field. The initially large-scale orthogonal magnetic field fluctuations decay but preserve the quasi-2D characteristics as energy is cascaded and dissipated via the generation of current sheets [see also \cite{Yalim_etal_2023}]. Conversely, \cite{Yang_etal_2025} utilize 3D MHD simulations of bursty interchange reconnection, finding that Alfv\'en waves are spontaneously excited in the reconnecting current sheet region to propagate bi-directionally along unreconnected magnetic field lines (see the cartoon Figure 1 in \cite{Zank_etal_2020b} that illustrates the same effect or mechanism for the formation of switchbacks). \cite{Bale_etal_2023} use PSP magnetic field observations to show quite persuasively that such interchange reconnection occurs low in the corona to produce bursts of switchbacks propagating outward. A potential scenario for the generation of Alfv\'en waves \citep{Yang_etal_2025} and switchbacks \citep{Zank_etal_2020b}  via interchange reconnection at low heights is via the reconnection of mixed polarity magnetic carpet loops that reach up to and just above the transition region with open magnetic field.  An important related study by \cite{Pontin_etal_2024} suggests that flux cancellation on small scales at a multitude of sites in the low solar atmosphere may be important in contributing to the heating of the chromosphere and corona. In examining the energy release rates due to flux cancellation at various sites, using both analytic and numerical MHD simulations, \cite{Pontin_etal_2024} find fluxes of the order of $10^6$ - $10^7$ erg cm${}^{-2}$ s${}^{-1}$, consistent with the constraints identified by \cite{Withbroe_Noyes_1977, Anderson_Athay_1989a}. 
The question is then whether reconnection occurs between the far more abundant mixed polarity loops in the chromosphere and what kind of turbulence is produced as a consequence.

The magnetic carpet describes the small-scale photospheric magnetic field of the quiet Sun and covers virtually the entire surface ($\sim 99$\%) \citep{Title_Schrijver_1998}. The magnetic carpet possesses randomly oriented positive and negative flux features and is constantly evolving with a flux replacement time scale of about 1 - 2 hours \citep{Hagenaar_etal_2008}. The magnetic carpet emerges from ephemeral regions \citep{Schrijver_etal_1997}, which possess magnetic field strengths of $\sim 100$ G, and has a weak to very little correspondence with either sunspot cycle or latitude. \cite{Martinez_Gonzalez_etal_2010} describe the observed continuous emergence of magnetic flux from the photosphere into the chromosphere to form low-lying loops \citep{Martinez_Gonzalez_etal_2007}. Although small-scale, and with relatively weak magnetic fields, \cite{Martinez_Gonzalez_etal_2010} estimate (conservatively) a magnetic injection rate of at least $1.04 \times 10^6$ -- $2.2 \times 10^7$ erg cm${}^{-2}$ s${}^{-1}$, which is comparable to the required energy deposition rate for the chromosphere quoted above. Finally, IRIS observations of the solar transition region identified ``a plethora of short, low-lying loops'' that are highly dynamic \citep{Hansteen_etal_2014}. MHD simulations \citep{Isobe_etal_2008} indicate that small-scale loops can reach chromospheric heights where they experience reconnection. IRIS observed numerous small-scale jets lasting 20 -- 80 seconds with widths $\leq 200$ km and speeds of 80 - 250 km/s emanating from the network of the transition region and chromosphere \citep{Tian_etal_2014}. 
These small-scale features are evidence for magnetic reconnection occurring frequently in the low solar atmosphere. 

The existence of 
\begin{enumerate}
\item highly dynamical, constantly emerging mixed polarity small-scale, low-lying loops across 99\% of the Sun's surface,  
\item with a magnetic energy injection rate of $1.4 \times 10^6$ -- $2.2 \times 10^7$ erg cm${}^{-2}$ s${}^{-1}$, 
\item in a low plasma beta environment, 
\item with observed small-scale jets originating in the chromosphere, almost certainly due to small-scale reconnection,
\end{enumerate}
confronts us with the question of what kind of magnetized turbulence is generated in the chromosphere? Furthermore, once magnetized turbulence is generated locally, how is it then transported throughout the chromosphere and possibly into the low corona in the absence of a large-scale bulk flow since the chromosphere has, until relatively recently \citep[e.g.,][]{Vernazza_etal_1981, Maltby_etal_1986, Fontenla_etal_1993}, been treated typically as a static atmosphere? More recent models \citep[e.g.,][]{Carlsson_etal_2019, Wedemeyer_etal_2004, Mathur_etal_2022} now recognize that the chromosphere is highly dynamical comprised of numerous forms of temporal flows such as shock waves and spicules, for example, despite the absence of a steady but slow bulk wind-like flow. 
\cite{Yang_etal_2025} use 3D MHD simulations to show that bursty interchange magnetic reconnection excites Alfv\'en waves propagating bi-directionally along open magnetic field, very similarly to the switchback generation mechanism proposed by \cite{Zank_etal_2020}. While certainly plausible, and Alfv\'en waves can easily transport magnetic energy in the chromosphere and beyond, most of the magnetic energy is injected into the chromosphere in the form of small-scale loops of mixed polarity with little open field across the solar surface. 

%% The "ht!" tells LaTeX to put the figure "here" first, at the "top" next
%% and to override the normal way of calculating a float position.
%% The asterisk after "figure" tells the compiler to span multiple columns
%% if a two column style is selected.
%\begin{figure*}[ht!]
%\plotone{Figure cartoon_draft_GPZ.jpg}
%\caption{To be added.
%\label{fig:1}}
%\end{figure*}

\begin{figure}
%%\epsscale{.80}
\begin{center}
\includegraphics[width= 0.5\textwidth]{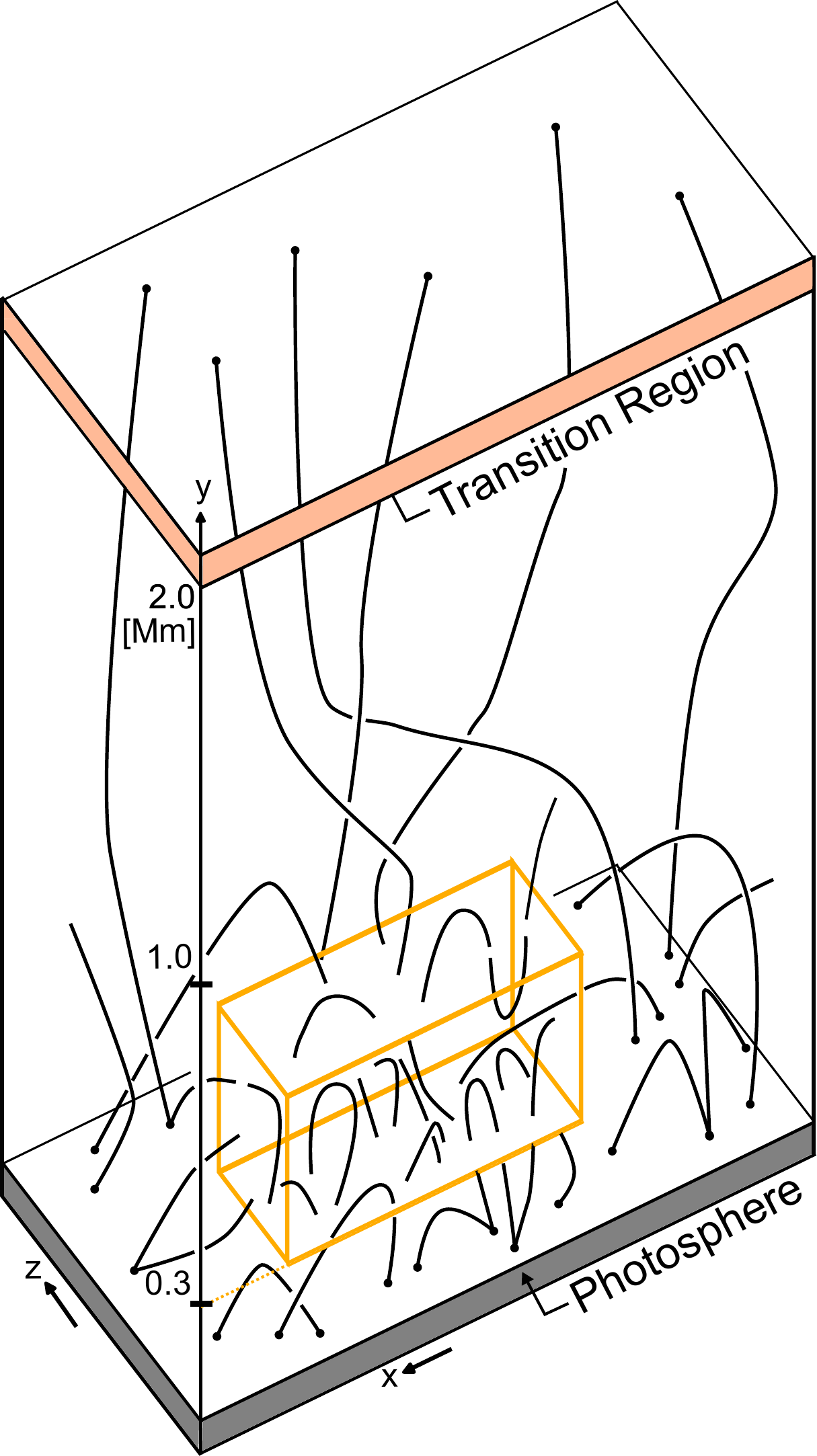}
\end{center}
\caption{\small A cartoon illustrating the computational box (the orange rectangular cuboid) embedded in the chromosphere. The photosphere lies in the $(x,z)$ plane below the cuboid through which emergent flux passes, including mixed polarity magnetic carpet loops of all characteristic scales, large-scale non-magnetic carpet loops, and some open magnetic field. Nominally, the box is located at a height of between 0.3 and 1.0 Mm, above which at approximately 2Mm is the very narrow transition layer. The imbalance in the mixed polarity magnetic flux within the computational box is modeled by assuming initially a ratio between a guide magnetic field $B_g$ oriented in the $y$-direction orthogonal to the $(x,z)$-plane, and the magnetic field strength in the initial reconnection plane $B_0$, also in the $(x,z)$-plane. See text for details.  } \label{fig:1}
\end{figure}

Here we examine the nature of magnetized turbulence generated by reconnection between magnetic loops of mixed polarity in a low-beta plasma. To investigate this requires that we make a decision about how best to simulate the problem numerically. On the one hand, a 3D magnetohydrodynamic simulation has the advantage of allowing us to utilize reasonable physical parameters corresponding to the chromosphere. The drawback, which is critical to the problem under consideration here, is that MHD cannot capture accurately the physics of reconnection, both in its initiation and its subsequent evolution to smaller and smaller spatial and temporal scales. By contrast, a 3D Particle-in-Cell (PIC) simulation containing both ion and electrons as separate particle species has the ability to fully capture all aspects of reconnection initiation and its turbulent evolution across inertial range and kinetic scales. The drawback is that PIC simulations typically use artificial parameters  to ensure code stability, and these parameterizations frequently bear little resemblance to the actual physical parameters of the regions being simulated. Such PIC parameters typically include a large Alfv\'en speed (although non-relativistic), and an idealized electron plasma to ion gyrofrequency ratio, and plasma beta values. PIC codes have adopted such artificial or unphysical parameters because simulations show the reconnection rate, which controls how fast the magnetic energy is dissipated, is insensitive to these parameters. For example, \cite{Li_etal_2019} argue that their PIC simulations can potentially explain solar flare acceleration primarily because the low ion and electron plasma beta values used in the code are close to those in the solar corona despite other parameters being quite different from those found in the corona. Our focus with these simulations is to understand how turbulence is initiated in the chromosphere through the emergence of mixed polarity magnetic field and identify the nature of the evolving turbulence in the chromosphere as turbulent reconnection and related dissipative processes proceeds across scales that include the inertial range (i.e., greater than the ion inertial length/ion cyclotron time scales) and into kinetic scales (i.e., bounded by the ion and electron inertial/cyclotron scales). Since it is important to accurately simulate the physics of these processes, only a 3D PIC code can address our goals 
unlike a 3D MHD simulation. We will show that reconnection of mixed polarity field generates numerous multi-scale magnetic islands that form 3D flux ropes together with numerous thin current sheets. The turbulence simulation is followed until the turbulence is fully developed. 
One important ingredient that is absent in our simulations is the presence of neutral gas. Simulations of reconnection in a partially ionized strongly magnetized region of the chromosphere have been presented by \cite{Ni_etal_2018}, for example, finding that reconnection in a non-equilibrated (i.e., between ionization and recombination) ion-neutral multi-fluid plasma is faster than occurs in an equilibrated plasma. A recent investigation of MHD turbulence in the partially ionized local interstellar medium by \cite{Nakanotani_Zank_2025a} shows that quasi-2D turbulence, primarily magnetic islands, are not damped by the neutral gas, unlike slab turbulence, making the presence of neutral gas important in the evolution of magnetic turbulence.  We do not include the physics of plasma-neutral gas coupling, ionization and recombination in our PIC simulations as this would add significantly to the computational demands already facing the model. However, we plan to explore these and related aspects more closely based on both 3D MHD and hybrid codes. 

While the generation of magnetized turbulence is critical to the heating of the chromosphere and corona, equally critical to their heating is the feasibility of transporting turbulence throughout the region and up to the low corona. Most of the magnetic carpet loops ($\geq 50$\%) lie below $\sim 1$ Mm \citep{Cranmer_vanBallegooijen_2010}, meaning that the generation of turbulence via the magnetic carpet \citep{Zank_etal_2018} will be restricted to the lower chromosphere. The question of how the chromosphere can be heated more-or-less uniformly at all heights up to and including the low chromosphere then arises since the chromosphere does not possess a mean upward flow analogous to the solar wind that appears to originate just above the transition region. As discussed recently by several authors, especially \cite{Carlsson_etal_2019}, the current view is that the chromosphere is highly dynamical, essentially on energy-containing scales, despite corresponding to a static atmosphere on larger scales. Although a mean large-scale flow from the photosphere to the transition region and just above (the low corona) does not exist, numerous localized flows are present, such as e.g.,  flows associated with the emergence of magnetic carpet loops, spicules, and acoustic shocks. These larger-scale but essentially turbulent flows (perhaps constituting the kinetic energy part of the  energy-containing range in the chromosphere) can carry the turbulence generated by the magnetic carpet (and also entrain photospheric turbulence) throughout the chromosphere and even up to and through the transition region and into the low corona. Hereafter, we discuss reconnection-driven turbulence, which includes two subsections that consider  different polarity states initially. Section \ref{sec:Discussion} utilizes the results drawn from the simulations to develop a simple transport formalism for magnetohydrodynamic (MHD) turbulence in the chromosphere and up to the base of the corona. This allows us to compute the evolution of the Els\"asser turbulent energy density, the Els\"asser specific energy, the correlation length, and the heating rates due to a turbulent flow field in the chromosphere. From these results, we compute the expected values of the  Els\"asser turbulent energy density, the  specific energy, the correlation length, and the heating rate numerically in the chromosphere and analytically at the base of the solar corona. We then discuss the implications of these results for heating both the chromosphere and coronal plasma, based on the well-known energy constraints identified by \cite{Athay_1966, Withbroe_Noyes_1977, Anderson_Athay_1989a}. 

Finally, we explore briefly the possibility that the heating observed in type II spicules \citep{dePontieu_etal_2007, dePontieu_etal_2011, Klimchuck_2012} can be explained via the transport and dissipation of magnetized turbulence. For completeness, some comments about the possible role of prominences/filaments and/or mini-filaments in both generating and transporting chromospheric turbulence are provided in the final subsection prior to the Conclusions.

\section{Simulation approach}
For the simulations, we utilize the 3D VPIC Particle-in-Cell (PIC) code \citep{Bowers_etal_2008, Li_etal_2019}, and remind the reader about the discussion at the end of \S\ref{sec:intro} regarding the choice of physical parameters in PIC codes. The VPIC code solves Maxwell's equations and the relativistic Vlasov equation. We assume a guide magnetic field $B_g  \hat{\bf y}$ in the $\hat{\bf y}$-direction so that we can regard the $\hat{\bf x}$-$\hat{\bf z}$-plane as the reconnection plane. We consider a very low plasma beta simulation with the electron and proton plasma beta values being $\beta_e = \beta_p = 0.02$, where $\beta_{e,i} = 8 \pi nkT_{e,i}/ B_0^2$, an ion to electron mass ratio of $m_i/m_e = 25$, $d_i = c/\omega_{pi} = c/\sqrt{4\pi n_i e^2/m_i}$ is the ion inertial length, $\omega_{pi}$ being the ion plasma frequency, $n_{i,e}$ the ion/electron number density, $k$ Boltzmann's constant, and $T_{i,e}$ the ion/electron temperature. The Alfv\'en speed is taken to be $V_A = B_0/\sqrt{4\pi n_0m_i} = 0.2c$, $c$ the speed of light, and the electron thermal speed is $v_{th,e} /c = 0.1$ and the ratio of the electron plasma frequency to the electron gyrofrequency is $\omega_{pe}/\Omega_{ce} = 1$. We note that the Alfv\'en speed is far larger than is physically appropriate to the solar chromosphere, which is highly variable ranging from $O(1)$ to $O(2 \times 10^3)$ km/s, as is the ratio $\omega_{pe}/\Omega_{pe}$, which ranges from $O(1)$ to $O(100)$, from just above the photosphere to just below the transition layer. The plasma beta too is highly variable but is small with increasing height until the transition region. Despite these parameter differences, VPIC provides a physically accurate description of the reconnection process.

The simulations are initiated from a force-free current sheet with ${\bf B} = B_0 \tanh (z/\lambda) \hat{\bf x} + B_0 \sqrt{\mbox{sech}^2 (z/\lambda) + b_g^2}\hat{\bf y}$ where $B_0$ is the strength of the reconnecting magnetic field, $\lambda = d_i$ is the half-thickness of the current sheet, and $b_g \equiv B_g/B_0$. We set $b_g = 1$ and 0.2 in the simulations below. Initially, the proton and electron distributions are assumed to be Maxwellian with uniform density $n_0$ and temperature $T_i = T_e = T_0$ and $kT_0 = 0.01m_ec^2$. The electrons have a bulk velocity drift of $u_e$ to satisfy Ampere's law. The box size of the 3D simulation is $150d_i \times 75d_i \times 62.5d_i$ with a grid size of $n_x \times n_y \times n_z = 3072 \times 1536 \times 1280$ and 150 particles/species are used per cell. Periodic boundary conditions are applied along the $x-$ and $y-$directions for both the electromagnetic fields and  particles, and a perfectly conducting boundary is applied along the $z-$direction for the electromagnetic fields together with a reflecting boundary for the particles. The simulation is initiated with the introduction of a long wavelength perturbation in $B_z$, which induces reconnection. 

Illustrated in Figure \ref{fig:1} is a cartoon of the computational setup in the context of the magnetic carpet. As illustrated, the lower surface at $y = 0$ is the photosphere and the transition region is located at $\sim 2$ Mm. The magnetic carpet emerges from the photosphere, as illustrated by the solid dots with multiple magnetic field lines emerging from or entering a dot (hence the term ``carpet''), with the loops being of mixed polarity.  Most of the magnetic field is in the form of loops emerging from and returning to the photosphere with some forming field lines that extend up to and beyond the nominal computational box shown in Figure \ref{fig:1} located near $\sim 0.5$ - 1 Mm above the photosphere. Model simulations by \cite{Cranmer_vanBallegooijen_2010} suggest that roughly 50\% of the magnetic carpet loop heights are within about 1 Mm of the photosphere with little variation in loop height with respect to the flux imbalance ratio. The overall results of \cite{Cranmer_vanBallegooijen_2010}  at the 75th percentile are consistent with the observations presented by \cite{Wiegelmann_Solanki_2004}. Within the computational box, we have outwardly and inwardly directed flux that may be imbalanced and the strength of the imbalance is modeled according to the following approach.

To understand the effects of mixed polarity magnetic fields, we examine two cases. 
\begin{enumerate}
\item If adjacent loops have the same polarity, the angle between the magnetic directions is small, meaning that this corresponds to a strong guide field case. To model the case of a box having a modestly large imbalanced magnetic flux, we assume that the ratio of the guide field to the reconnecting magnetic field  $b_g = B_g/B_0 = 1$. 
\item Conversely, if adjacent loops initially have opposite polarity, the angle between the magnetic field directions is large. In this case, the initial magnetic guide field will be weak. For the case of almost balanced magnetic flux within the box, we assume a ratio $b_g = B_g/B_0 = 0.2$.
\end{enumerate}
This is not unlike the flux imbalance fraction used by \cite{Cranmer_vanBallegooijen_2010}, which is defined by the net magnetic flux density $B_{net} = |B_+ + B_-|$ ($\pm$ referring to outward and inward oriented magnetic field respectively) divided by the absolute flux density $B_{abs} =  B_+ + |B_-|$. The flux imbalance is typically small in quiet-Sun regions ($\leq 0.3$) and larger values ($\geq 0.7$) are found more typically in coronal holes where open flux can dominate \citep{Wiegelmann_Solanki_2004, Hagenaar_etal_2008, Abramenko_etal_2009, Cranmer_vanBallegooijen_2010, Cranmer_etal_2013}. 

\begin{figure}
\includegraphics[width=\textwidth]{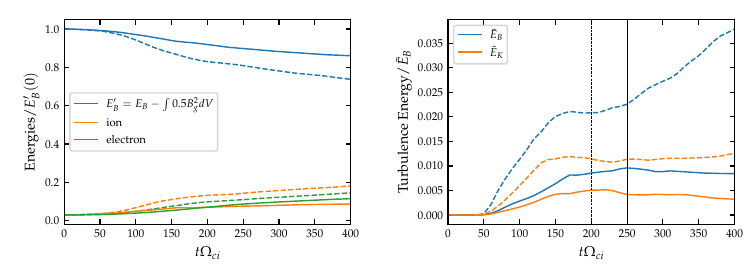}
\caption{\small {\bf Left}: Plots of the evolving magnetic (blues curve), ion (orange), and electron (green) energies as a function of the normalized time $t \Omega_{ci}$, all of which are normalized to the initial magnetic energy, excluding the magnetic guide field energy, $E_B^{\prime} (t = 0)$. The solid lines are for $b_g=1.0$, and the dashed lines are for $b_g=0.2$. {\bf Right}:  Plots of the normalized turbulent magnetic (blue) and kinetic (orange) energy densities $\tilde{E}_B$ and $\tilde{E}_K$ respectively as a function of the normalized time. The vertical black lines identify the time after which the turbulence is fully developed and on which the subsequent analysis is based. } \label{fig:2}
\end{figure}

\section{Reconnection-driven turbulence} \label{sec:simulations} 

In the following subsections, we consider two cases, the strong and the weak guide field, which can be regarded as appropriate to open coronal holes and the quiet Sun respectively. For both cases, we perturb the initial force-free current sheet using a long wavelength perturbation in the reconnection plane along the $\hat{z}$-direction. With the initiation of reconnection, the current sheet is destabilized thanks to the tearing mode instability. As a consequence, the current sheet begins to break up into multiple magnetic flux ropes of different scales, which interact and merge, thereby producing secondary flux ropes or magnetic islands that are produced continuously in the 3D reconnection layer (the $(x,z)$-plane). The result is a highly turbulent reconnection layer that is very inhomogeneous. 

The highly inhomogeneous character of the magnetic field and plasma makes it very difficult to define a global mean magnetic field $\bar{\bf B}$ in the simulation domain. The mean magnetic field $\bar{\bf B}$ is obtained by averaging the magnetic fields along the guide field direction $\hat{\bf y}$. A mean flow field $\bar{\bf U}$ is similarly defined except that it is weighted by $\sqrt{\rho}$, which allows us to calculated the kinetic energy spectrum.  The mean field is needed to identify the turbulent magnetic and kinetic fluctuations, identified as $\tilde{\bf B} = {\bf B} - \bar{\bf B}$ and  $\tilde{\bf U} = {\bf U} - \bar{\bf U}$ respectively, where $\langle {\bf B} \rangle = \bar{\bf B}$,  and $\langle \tilde{\bf B} \rangle = 0$, and similarly for ${\bf U}$. 

\subsection{Imbalanced magnetic field flux case: Open coronal hole surrogate} \label{subsec:open}

In this section, we take the guide field ratio to be $B_g/B_0 = 1$, which serves as a surrogate for a region that contains open magnetic field in the sense that the emergence of the magnetic carpet is interspersed by a population of open magnetic field lines. This would correspond to higher values of the flux imbalance fraction. 

The fluctuating magnetic and kinetic turbulence energy densities can be expressed as $\tilde{E}_B = \tilde{B}^2 /(2 \mu_0 \rho_0)$ and $\tilde{E}_K = \rho \tilde{U}^2 / 2 \equiv w^2/2$, where $\mu_0$ is the permeability of free space and $\rho_0$ is the mean mass density of the fluid. Plotted in Figure \ref{fig:2} (left panel) is the evolution of the energy in the simulation box of volume $V$, the blue curve showing the magnetic energy density in the non-guide field, i.e., $E_B^{\prime} = E_B - \int B_g^2/2 dV$ normalized to the initial magnetic energy $E_B^{\prime} (t = 0)$. The energy evolution in ions (orange curve) and electrons (green) is seen to be significantly less. The decrease in magnetic energy is reflected in a corresponding almost equal increase in ion and electron energy. Figure \ref{fig:2} (right) shows the fluctuating magnetic and kinetic energies normalized to the energy density of the mean magnetic field $\bar{E}_B = \bar{B}^2 /(2\mu_0 \rho_0)$, as a function of normalized time. Furthermore, the right plot shows that the initial rapid evolution in turbulent energy gives way to an approximate plateau after about $250/\Omega_{ci}$. Thereafter,  we treat the turbulence as being in a statistically steady-state, during which time it is dominated by the magnetic energy density. 
Recall that the ion and electron plasma beta values are 0.02. Both magnetic and kinetic energy exhibit a slow decline thereafter. The analysis presented below is for times at and later than  $250/\Omega_{ci}$. 

\begin{figure}
%%\epsscale{.80}
\begin{center}
\includegraphics[width=\textwidth]{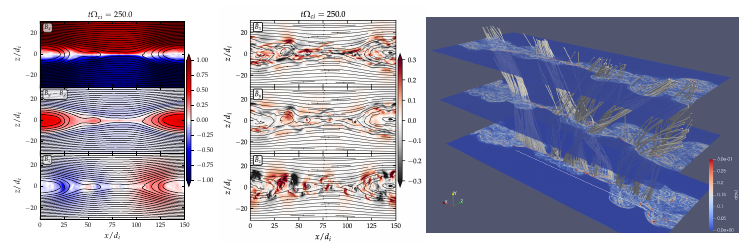}
\end{center}
\caption{\small {\bf Left}: The three components of the mean magnetic field $\bar{\bf B} = (\bar{B}_x, \bar{B}_y, \bar{B}_z)$ in the $(x,z)$-plane at time $t \Omega_{ci} = 250$. Note that the guide field $B_g$ has been subtracted from $\bar{B}_y$. The magnetic field lines are represented as streamlines of the 2D vector fields $(\bar{B}_x, \bar{B}_z)$, and the color coding refers to the strength of the magnetic field component with red in the $y$-positive and blue in the $y$-negative direction. {\bf Middle}: The fluctuating magnetic field components $(\tilde{B}_x, \tilde{B}_y, \tilde{B}_z)$ in the $y=0$ plane. The black lines are the streamlines of the in-plane magnetic field. Multiple magnetic islands are present in each of the slices. The color denotes magnetic field component intensity and orientation. Component $\tilde{B}_x$  admits sharp transitions across thin current sheets. {\bf Right}: 3D depiction of the fluctuating current density and magnetic field lines originating from the base of the simulation that correspond to flux ropes. The fluctuating current density possesses considerable small-scale structure that resemble magnetic islands in each of the planar cuts. In 3D, the combination of the magnetic field lines in the nominal ``$y$-direction''  and small scale magnetic islands yields complex braided  magnetic flux ropes. A movie showing the dynamical evolution of fluctuating current density and magnetic field lines is included with the online version of the paper.} \label{fig:3}
\end{figure}

Figure \ref{fig:3} (left) shows the three components of the mean magnetic field at time $t \Omega_{ci} = 250$ in the $(x,z)$-plane, illustrating the inhomogeneity of the mean field. Evidently, the mean magnetic field direction is not generally aligned with the $y$-direction except possibly for certain localized $(x,y,z)$ volumes. The top panel shows $\bar{B}_x$, the middle panel $\bar{B}_y - B_g$, and the bottom $\bar{B}_z$, all plotted in the $x$-$z$-plane normalized by the ion inertial length $d_i$. By plotting the in-plane magnetic field lines as streamlines, the significant large-scale variation imposed by magnetic islands and current sheets is revealed. The color coding refers to the strength of the magnetic field component with red in the $y$-positive and blue in the $y$-negative direction. Prior to $t \Omega_{ci} = 250$, the reconnection layer (not shown) is filled with magnetic flux ropes that kink, merge, interact, and produce secondary flux ropes continuously within the layer. By $t \Omega_{ci} = 250$, there is essentially only a single mean-field magnetic flux rope remaining in the system and a single mean current sheet. However, as we discuss below, the fluctuating magnetic field is highly turbulent. 

We note that the early phase $t \Omega_{ci} < 250$ is of interest in the context of the emergence of mixed polarity magnetic flux from the photosphere into the chromosphere. With a flux emergence time of $\sim 1 - 2$ hours \citep{Hagenaar_etal_2008} within ephemeral regions (magnetic field strengths of $\sim 100$ G), the evolution to a state of fully developed turbulence is very rapid once a perturbation is initiated ($t \sim 2\pi 250 /\Omega_{ci} (B = 100 \: \mbox{G}) = 1.6 \times 10^{-3}$s in the simulation. We stress that this number for the time taken for the onset of fully developed turbulence cannot be regarded as realistic in any way given the limitations of PIC codes -- see the discussion at the end of the Introduction). In our simulation, the perturbation is essentially Alfv\'enic in that we perturb  $B_z$. The simulation suggests that the magnetic carpet emerging into the chromosphere should be highly variable on short time scales, and observations suggest that low-temperature loops do exhibit intense variability.

We plot the same 2D slices in terms of the fluctuating magnetic field components $\tilde{B}_x$, $\tilde{B}_y$, and $\tilde{B}_z$ in the middle panels of Figure \ref{fig:3}. The black lines identify the streamlines of the in-plane magnetic field (which is why some magnetic field lines appear to have a start and end point). Numerous magnetic islands can be seen in each of the three slices. Recall that the mean magnetic field does not possess a simple $y$-oriented configuration and hence the magnetic islands, although possessing some structure in the slices, have their primary orientation ordered by the inhomogeneous mean field. Several sharp transitions in the $z$-direction across thin current sheets can be seen in the $B_x$ component of the magnetic field (e.g., at $(100,0)$), which are difficult to eliminate. Accordingly, we need to be careful in the subsequent spectral analysis with spectra that are functions of the wavenumber $k_z$. 

Consider now a 3D rendering of the fluctuating current density $\tilde{\bf J} = 1/\mu_0 \: \nabla \times \tilde{\bf B}$, illustrated in Figure \ref{fig:3} (right), showing three superimposed planar cuts through the $y$-axis taken at different heights. Some mean magnetic field lines oriented roughly along the $y$-axis are plotted over the fluctuating current density, originating from the lowest of the three planes. This is a good approximation in the center of the reconnection layer where the anti-parallel field is mostly dissipated. The mean magnetic field lines form bundles, some of which acquire a pronounced braided structure, and are mean magnetic field flux ropes. The current density is very complex and highly inhomogeneous with height, and it has multiple complex structures separated by numerous very thin current sheets. 

\begin{figure}
%%\epsscale{.80}
\begin{center}
\includegraphics[width=\textwidth]{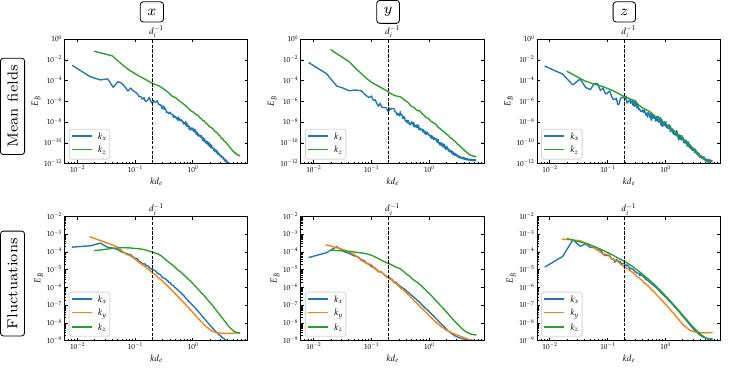}
\end{center}
\caption{\small {\bf Top}: The PSDs for the mean magnetic field components $\bar{B}_x, \bar{B}_y, \bar{B}_z$ as a function of wavenumbers $k_x, k_z$. Owing to the presence of very small-scale current sheets, both the $\bar{B}_x (k_z)$ and $\bar{B}_y (k_z)$ PSDs are ill-defined. The PSD for $\bar{B}_z (k_z)$ exists and is well-defined. 
{\bf Bottom}: The PSDs for the fluctuating magnetic field components $\tilde{B}_x, \tilde{B}_y, \tilde{B}_z$ as a function of wavenumbers $k_x, k_y, k_z$. Similar to the mean field, both the $\tilde{B}_x (k_z)$ and $\tilde{B}_y (k_z)$ PSDs are ill-defined, and the PSD for $\bar{B}_z (k_z)$ exists and is well-defined.} \label{fig:4}
\end{figure}

The mean field is variable and is not oriented simply in the $y$-direction because the guide field is modestly strong, $B_g = B_0$ (see Figure \ref{fig:3}). Hence, the parallel direction is not very well-defined if, as we do, we assume that the $y$-direction is parallel to the guide field. In the center of the current sheet, however, the anti-parallel component is largely dissipated, and we can therefore treat the guide field as the parallel direction in that region. We can therefore expect that spectra that are functions of the wavenumber $k_y$ will reflect the presence of both some perpendicular wavenumbers and parallel wavenumbers.  Nonetheless, ideally the wavenumbers $k_x$ and $k_z$ should correspond roughly to perpendicular wavenumbers, i.e.,  $k_{\perp} = \sqrt{k_x^2 + k_z^2}$, and $k_y$ to roughly the parallel wavenumber $k_{\parallel}$. It transpires that the mean $\bar{B}_z$ and fluctuating $\tilde{B}_z$ components are well-defined for $k_x$, $k_y$, and $k_z$ and the remaining mean $\bar{B}_x, \bar{B}_y$ and fluctuating magnetic field components $\tilde{B}_x, \tilde{B}_y$, are well-defined in $k_x$ and $k_y$ only. The mean and fluctuating  components are plotted as PSDs in the top and bottom panels of Figure \ref{fig:4} respectively, including the ill-defined cases of $\bar{B}_x (k_z), \bar{B}_y (k_z)$ and  $\tilde{B}_x (k_z), \tilde{B}_y (k_z)$. Similar issues arise in the $k_z$ spectra for kinetic energy and Els\"asser variables, and we do not show similar plots again.

\begin{figure}
\includegraphics[width=\textwidth]{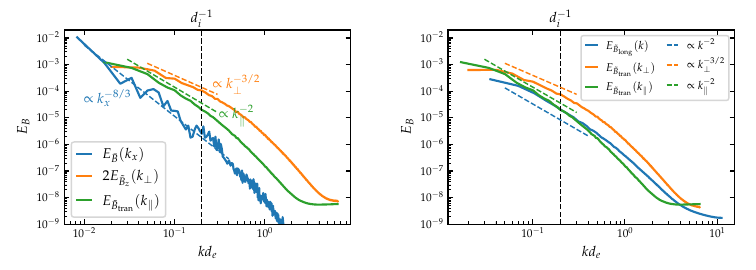}
\caption{\small {\bf Left}: Power spectral densities (PSDs) $E_{\bar B} (k_{x})$  (blue), $2 \times  E_{{\tilde B}_z} (k_{\perp})$ (orange), and $E_{{\tilde B}_{\text{tran}}} (k_{\parallel})$ (green) for the mean and fluctuating transverse magnetic fields as a function of wavenumber $k_x$, perpendicular wavenumber $k_{\perp} = \sqrt{k_x^2 + k_z^2}$ and parallel wavenumber $k_{\parallel} = k_y$. Here the subscript ``tran'' refers to transverse fluctuations only such that $\tilde{B}_{\text{tran}} = \sqrt{\tilde{B}_x^2 + \tilde{B}_z^2}$. {\bf Right}: As in the left panel except that we plot an estimated spectrum for the longitudinal or compressive fluctuating component $E_{{\tilde B}_{\text{long}}} (k)$ (blue) and $E_{{\tilde B}_{\text{tran}}} (k_{\perp})$ (orange). See text for details. The wavenumbers are normalized to the electron inertial length. } 
 \label{fig:5}
\end{figure}

Based on Figure \ref{fig:4}, we can plot the PSDs of $\bar{B}_z$ and $\tilde{B}_z$ as a function of $k_{\perp} = \sqrt{k_x^2 + k_z^2}$ since we can compute both $\bar{B}_z (k_x, k_z)$ and  $\tilde{B}_z (k_x, k_z)$. The mean field PSD $E_{\bar B} (k_x)$ is plotted in Figure \ref{fig:5} (left), as shown by the blue curve with an approximate power law distribution $k_x^{-8/3}$ in the inertial range. The fluctuating $\tilde{B}_z$ PSD, $E_{\tilde{B}_z }(k_{\perp})$, represents small-scale 2D or flux-rope turbulent magnetic field fluctuations and is plotted in Figure \ref{fig:5} (left) in orange (actually showing $2 \times E_{\tilde{B}_z }(k_{\perp})$). $E_{\tilde{B}_z }(k_{\perp})$ is $\propto k_{\perp}^{-3/2}$. Also plotted in Figure \ref{fig:5} (left) as a green curve is the PSD of the fluctuating transverse slab fluctuations, $E_{{\tilde B}_{\text{tran}}} (k_{\parallel})$,  where $\tilde{B}_{\text{tran}}^2 (k_{\parallel}) \equiv \tilde{B}_{\perp}^2 (k_{\parallel}) = \tilde{B}_x^2 (k_{\parallel}) + \tilde{B}_z^2 (k_{\parallel})$ and $k_{\parallel} \simeq k_y$. This can be interpreted as the  slab turbulence spectrum and is an approximate power law distribution  $\propto k_{\parallel}^{-2}$ in the inertial range. 

The right panel of Figure \ref{fig:5} plots again $E_{{\tilde B}_{\text{tran}}} (k_{\parallel})$ together with a green curve. The orange curve is approximate in that we plot the PSD of $\tilde{B}_{\perp}^2 (k_{\perp})$ using an approximated $\tilde{B}_x (k_{\perp})$. We have $\tilde{B}_z (k_{\perp})$ from the bottom rightmost plot of Figure \ref{fig:4} and $\tilde{B}_x (k_x)$ from the bottom leftmost plot, but evidently $\tilde{B}_x (k_z)$ is problematic for the reasons alluded to above. One might expect that $\tilde{B}_z$ and $\tilde{B}_x$ are basically symmetric transverse components and that the small-scale magnetic island/flux rope modes should be functions of $k_z$ and $k_x$ only. Since the PSDs of $\tilde{B}_z (k_x)$ and $\tilde{B}_z (k_z)$ are almost identical in the left panel of Figure \ref{fig:4}, we assume that this is true for $\tilde{B}_x$ so that the PSDs of $\tilde{B}_x (k_x) \sim \tilde{B}_x (k_z)$. Subject to these assumptions, we plot the PSD of $\tilde{B}_{\perp}^2 (k_{\perp})$ as a function of  $k_{\perp}$ i.e., the quasi-2D component PSD $E_{\tilde{B}_{\perp}} (k_{\perp})$, which is an approximate power law $\propto k_{\perp}^{-3/2}$in the inertial range.

The PSDs for the longitudinal or compressive fluctuating component $\tilde{B}_y$ shown in the middle panel of Figure \ref{fig:4} are almost identical for $k_x$ and $k_y \sim k_{\parallel}$ but as before, the PSD for the $k_z$ component is ill-defined. Assuming that the PSDs satisfy $\tilde{B}_y (k_x) \simeq \tilde{B}_y (k_z)$, we can construct the PSD for the longitudinal component wavenumber in $k = \sqrt{k_x^2 + k_y^2 + k_z^2}$. These compressive spectra are plotted in  Figure \ref{fig:5} (right) as a blue curve and possess an approximate power law spectrum $k^{-2}$.

\begin{figure}
\includegraphics[width=\textwidth]{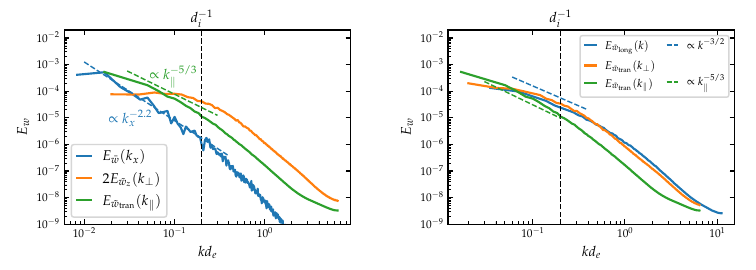}
\caption{\small  {\bf Left}: PSDs $E_{\bar w} (k_{x})$  (blue), $2 \times  E_{{\tilde w}_z} (k_{\perp})$ (orange), and $E_{{\tilde w}_{\text{tran}}} (k_{\parallel})$ (green) for the mean and fluctuating magnetic fields as a function of wavenumber $k_x$, perpendicular wavenumber $k_{\perp} = \sqrt{k_x^2 + k_z^2}$ and parallel wavenumber $k_{\parallel} = k_y$. Here the subscript ``tran'' refers to transverse fluctuations only such that $\tilde{w}_{\text{tran}} = \sqrt{\tilde{w}_x^2 + \tilde{w}_z^2}$. {\bf Right}: As in the left panel except that we plot an estimated spectrum for the longitudinal or compressive fluctuating component $E_{{\tilde w}_{\text{long}}} (k)$ (blue) and $E_{{\tilde w}_{\text{tran}}} (k_{\perp})$ (orange). See text for details.  The wavenumbers are normalized to the electron inertial length. \label{fig:6}}
\end{figure}

Consider now the kinetic energy PSDs where the spectrum is calculated using ${\bf w} \equiv \sqrt{\rho} {\bf U}$ and $\rho$ and ${\bf U}$ are the plasma density and velocity. Following the same procedure as used for the magnetic field, we show the PSDs for the mean kinetic energy $E_{\bar w} (k_x)$ (blue curve) and the fluctuating kinetic energy $2 E_{{\tilde w}_z} (k_{\perp})$ (orange) and $ E_{{\tilde w}_{\text{tran}}} (k_{\parallel}) = E_{{\tilde w}_{\perp}} (k_{\parallel})$ in Figure \ref{fig:6} (left). By using a similar approximation, {\em viz}., that  the PSDs of $\tilde{w}_x (k_x)$ and $\tilde{w}_x (k_z)$ are approximately equal, we plot the quasi-2D spectrum $E_{{\tilde w}_{\perp}} (k_{\perp})$ in Figure \ref{fig:6} (right), together with the longitudinal kinetic energy PSD $E_{{\tilde w}_{\text{long}}} (k)$. Both $E_{{\tilde w}_{\perp}} (k_{\perp})$ and $E_{{\tilde w}_{\text{long}}} (k)$ are an approximate power law $\propto k^{-3/2}$ in the inertial range, but the inertial range is less well-defined than for the magnetic energy spectra. In each case, the spectra are flatter than their magnetic field counterparts, with $E_{\bar w} (k_x) \propto k_x^{-2.2}$ and the fluctuating slab spectrum $E_{{\tilde w}_{\perp}} (k_{\parallel}) \propto k_{\parallel}^{-5/3}$, and the quasi-2D spectra $E_{{\tilde w}_z} (k_{\perp})$ and $E_{{\tilde w}_{\perp}} (k_{\perp})$ are even flatter. 

\begin{figure}
\begin{center}
\includegraphics[width= 0.65\textwidth]{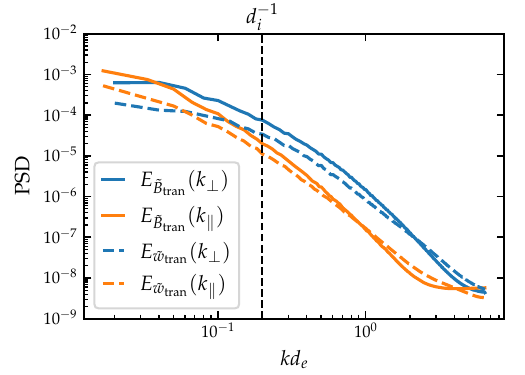}
\end{center}
\caption{\small A comparison of the  magnetic and kinetic energy spectra, showing the 
power spectral densities $E_{{\tilde B}_{\text{tran}}} (k_{\perp})$ (solid blue curve), $E_{{\tilde B}_{\text{tran}}} (k_{\parallel})$, (solid orange) and $E_{{\tilde w}_{\text{tran}}} (k_{\perp})$ (dashed blue line), $E_{{\tilde w}_{\text{tran}}} (k_{\parallel})$ (dashed orange) for the transverse fluctuations as a function of the perpendicular $k_{\perp}$ and parallel $k_{\parallel}$ wavenumbers, normalized to the electron inertial length. 
} \label{fig:7}
\end{figure}

Finally, Figure \ref{fig:7} compares the fluctuating magnetic and kinetic energy PSDs for the quasi-2D $E_{{\tilde B}_{\perp}} (k_{\perp})$, $E_{{\tilde w}_{\perp}} (k_{\perp})$ and slab $E_{{\tilde B}_{\perp}} (k_{\parallel})$, $E_{{\tilde w}_{\perp}} (k_{\parallel})$ components. Evidently, the kinetic energy spectra are flatter than their magnetic energy counterparts, the magnetic energy density for both quasi-2D and slab fluctuations is larger than the corresponding kinetic energies, and finally the quasi-2D magnetic fluctuations form the energetically dominant component. The integrated power $\langle {\tilde B}_{\perp}^2 (k_{\perp}) \rangle=6\times 10^{-5}$ T${}^2$ and $\langle {\tilde B}_{\perp}^2 (k_{\parallel}) \rangle = 4.4\times 10^{-5}$ T${}^2$, giving a ratio of 1.36. That the kinetic fluctuation spectra are steeper than the magnetic spectra suggests that velocity fluctuations are damped more efficiently at smaller scales. 

\begin{figure}
\includegraphics[width=\textwidth]{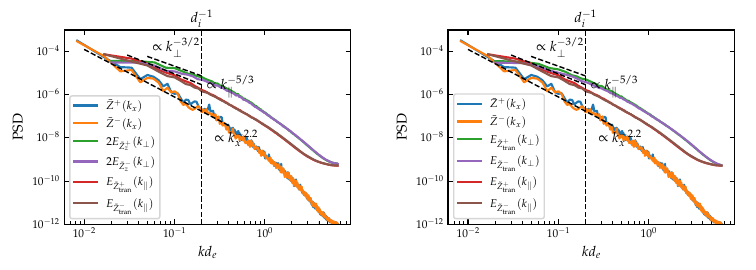}
\caption{\small  {\bf Left}: PSDs of the mean and fluctuating Els\"asser energies $E_{\bar z}^{\pm} (k_{x})$  (blue and orange curves respectively), $2 \times  E_{{\tilde z}_z^{\pm}} (k_{\perp})$ (green and purple), and $E_{{\tilde z}_{\text{tran}}^{\pm}} (k_{\parallel})$ (red and burgundy) as a function of wavenumber $k_x$, perpendicular wavenumber $k_{\perp} = \sqrt{k_x^2 + k_z^2}$ and parallel wavenumber $k_{\parallel} = k_y$.  {\bf Right}: As in the left panel except that we plot an estimated spectrum for $E_{{\tilde z}_{\text{tran}}^{\pm} } (k_{\perp})$ (green and purple). See text for details.  The wavenumbers are normalized to the electron inertial length.} 
 \label{fig:8}
\end{figure}

The magnetic and velocity components can be combined through the Els\"asser variables ${\bf Z}^{\pm} = {\bf U} \pm {\bf B}/\sqrt{\mu_0 \rho}$. Similar to the magnetic and kinetic energy spectra, we can compute the PSDs of the mean and fluctuating Els\"asser energies. Figure \ref{fig:8} (left) the PSDs of $\bar{Z}^{\pm} (k_x)$, $\tilde{Z}_z^{\pm} (k_{\perp})$, and $\tilde{Z}_{\perp}^{\pm} (k_{\parallel})$. The right panel of Figure \ref{fig:8} uses the prior assumptions for the fluctuating PSDs of $\tilde{z}_x^{\pm} (k_x)$ and   $\tilde{z}_x^{\pm} (k_z)$ to obtain $E_{{\tilde z}_{\perp}^{\pm}} (k_{\perp})$. Of note is that the quasi-2D PSDs $E_{{\tilde z}_{\perp}^{+}} (k_{\perp}) = E_{{\tilde z}_{\perp}^{-}} (k_{\perp}) \propto k_{\perp}^{-3/2}$.  Below these spectra are the slab PSDs, $E_{{\tilde z}_{\perp}^{+}} (k_{\parallel}) = E_{{\tilde z}_{\perp}^{-}} (k_{\parallel})$, both possessing spectra $\propto k_{\parallel}^{-5/3}$. The ratio of the fluctuating quasi-2D to slab Els\"asser energy can be estimated as $\langle  \tilde{Z}_{\perp}^{+ 2} (k_{\perp})  + \tilde{Z}_{\perp}^{- 2} (k_{\perp}) \rangle / \langle \tilde{Z}_{\perp}^{+ 2} (k_{\parallel}) +  \tilde{Z}_{\perp}^{- 2} (k_{\parallel}) \rangle\approx 1.25$. 
It is possible that the ratio will change in larger simulations with more scale separation in the inertial range, but this is difficult to test using PIC simulations.
The cross helicity for both quasi-2D and slab fluctuations is $\simeq 0$.

\subsection{Almost balanced magnetic flux case: Quiet sun surrogate} \label{subsec:quiet}

Here we take the guide field ratio to be $B_g/B_0 = 0.2$, which, as discussed above,  serves as a surrogate for the quiet Sun.  In the quiet Sun, the emerging magnetic carpet occurs in a region of almost equally mixed magnetic field polarity, hence the assumption $B_g/B_0 = 0.2$. In this case, it is even more difficult to define a mean field than it was when $B_g/B_0 = 1.0$.

The setup of the simulation is very similar to that described in section \ref{subsec:open} and is not repeated here. The energy evolution of the simulation is shown as dashed colored lines in Figure \ref{fig:2} (left and right). Compared to the $B_g/B_0 = 1$ case, the magnetic energy decreases more rapidly, and more magnetic energy is converted into plasma kinetic energy (Figure \ref{fig:2}, left). As illustrated in Figure \ref{fig:2}, right, the energy in turbulent magnetic fluctuations increases significantly a little after about $t \Omega_{ci} = 250$, leading us to choose the time $t \Omega_{ci} = 200$ for the analysis. After $t \Omega_{ci} = 250$,  the bidirectional outflows interact strongly, leading to an increase in the turbulent magnetic energy $E_B$ (blue dashed line).

\begin{figure}
%%\epsscale{.80}
\begin{center}
\includegraphics[width=\textwidth]{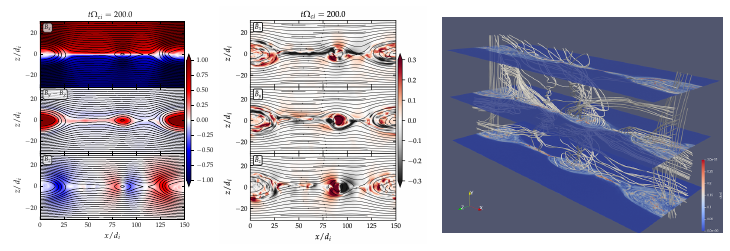}
\end{center}
\caption{\small  Similar to Figure \ref{fig:3} but for the case $B_g/B_0 = 0.2$. A movie showing the dynamical evolution of fluctuating current density and magnetic field lines is included with the online version of the paper.
} \label{fig:9}
\end{figure}

Since the guide field is weak, it is now not really possible to define a global mean magnetic field. Like the $B_g/B_0 = 1$ case, the turbulent reconnection layer is highly inhomogeneous and very thin current sheets are present. The fluctuating magnetic field and current density possess considerable structure, as illustrated in Figure \ref{fig:9}, which closely resembles the strong guide field case. Figure \ref{fig:9}  (left), evaluated at time $t \Omega_i = 200$, shows the mean field components $\bar{\bf B} = (\bar{B}_x, \bar{B}_y, \bar{B}_z)$ in the $(x,z)$-plane through a representative cut across the $y$-axis. The large-scale variation imposed by magnetic islands and current sheets is again apparent. The middle plot of Figure \ref{fig:9} shows fluctuating magnetic field components $(\tilde{B}_x, \tilde{B}_y, \tilde{B}_z)$ in the $y=0$ plane. Several sharp transitions in the $z$-direction across thin current sheets can be seen in different components, which are difficult to eliminate. The fluctuating fields appear to be less turbulent than in the $B_g/B_0 = 1$ case, but the 3D plot in the right panel of Figure \ref{fig:9} shows a clearly turbulent reconnection layer with highly complex structure and chaotic magnetic field lines, making it even more difficult to define a mean field direction.

Similar to the $B_g/B_0 = 1$ case, we can compute the PSDs of the mean and fluctuating magnetic and kinetic energies as functions of $k_x$, $k_y$, and $k_z$. Owing to the presence of very small-scale current sheets, the PSDs of  $\bar{B}_{x,y} (k_z)$ are ill-defined whereas the PSD for $\bar{B}_z$ has $\langle \bar{B}_z^2 \rangle (k_x) \simeq \langle \bar{B}_z^2 \rangle (k_z)$. For the fluctuating magnetic field components, the PSDs for $\langle \tilde{B}_x^2 \rangle (k_x)$ and $\langle \tilde{B}_x^2 \rangle (k_y)$ are basically comparable (of the same form and slightly different amplitudes) and  $\langle \tilde{B}_x^2 \rangle (k_z)$ is again ill-defined.  Similarly,  $\langle \tilde{B}_y^2 \rangle (k_x) \simeq  \langle \tilde{B}_y^2 \rangle (k_y)$ while $\langle \tilde{B}_z^2 \rangle (k_x) \simeq \langle \tilde{B}_z^2 \rangle (k_y) \simeq \langle \tilde{B}_z^2 \rangle (k_z)$. The fluctuating kinetic energy PSDs exhibit behavior similar to that of the fluctuating magnetic energy PSDs.

\begin{figure}
%%\epsscale{.80}
\begin{center}
\includegraphics[width=\textwidth]{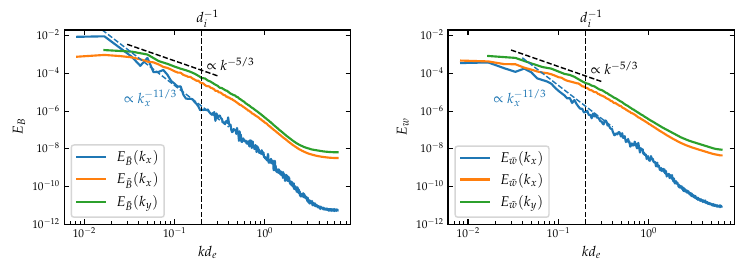}
\end{center}
\caption{\small {\bf Left}: Power spectral densities  $E_{\bar{ B}} (k_x)$, $E_{\tilde{ B}} (k_x)$, and $E_{\tilde{ B}} (k_y)$ for the mean and fluctuating magnetic field. {\bf Right}: The corresponding PSDs for the mean and fluctuating kinetic energies $\bar{w}$ and $\tilde{w}$.  } 
\label{fig:10}
\end{figure}

Unlike the strong guide field case, the parallel and perpendicular directions cannot be defined properly, so we plot  $E_{\bar{B}} (k_x)$, $E_{\tilde{B}} (k_x)$, and $E_{\bar{B}} (k_y)$ in Figure \ref{fig:10} (left) after removing the $k = 0$ mode and combining the components. The mean field PSD $E_{\bar{B}} (k_x) \propto k_x^{-11/3}$. By contrast, the PSDs in $k_x$ and $k_y$ for the fluctuating magnetic field are almost identical and both $E_{\tilde{B}} (k_x) \simeq E_{\tilde{B}} (k_y) \propto k_{x,y}^{-5/3}$. The spectra for the fluctuating magnetic field strongly suggest that there is little to distinguish between any nominal directions and that the turbulent magnetic field is essentially isotropic. In the absence of a well-defined mean magnetic field, the fluctuating magnetic field is composed primarily of non-propagating structures such as small-scale magnetic flux ropes, as illustrated in Figure \ref{fig:9}.

The right panel of Figure \ref{fig:10} shows the kinetic energy  with the mean flow field possessing a spectrum $\propto k_x^{-11/3}$, and, like the magnetic field case, the fluctuating kinetic energy PSDs $E_{\tilde{w}} (k_x) \simeq E_{\tilde{w}} (k_y)  \propto k_{x,y}^{-5/3}$ possess a Kolmogorov spectrum. Once again, the comparable amplitudes of the $k_x$ and $k_y$ spectra are suggestive of isotropic kinetic energy turbulence. 
 
\begin{figure}
%%\epsscale{.80}
\begin{center}
\includegraphics[width= 0.65\textwidth]{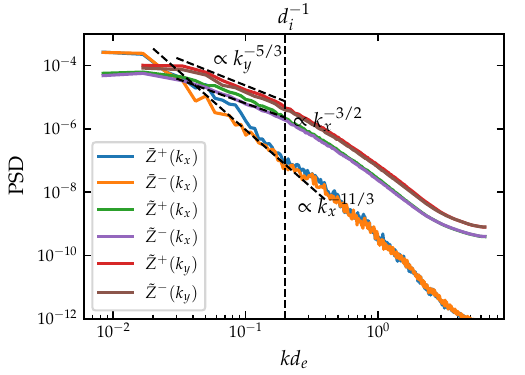}
\end{center}
\caption{\small PSDs $E_{\bar{ z}^{\pm}} (k_x)$, $E_{\tilde{ z}^{\pm}} (k_x)$, and $E_{\tilde{ z}^{\pm}} (k_y)$ for the mean and fluctuating Els\"asser variables as a function of normalized $k_x$ and $k_y$. The $k = 0$ component is removed.  } \label{fig:11}
\end{figure}

Figure \ref{fig:11} shows the PSDs of the magnitude of the Els\"asser variables, $E_{\bar{z}^{\pm}} (k_x)$ for the mean Els\"asser field and  $E_{\tilde{z}^{\pm}} (k_x)$ and $E_{\tilde{z}^{\pm}} (k_y)$ for the fluctuating field exhibit power law spectra in their respective wavenumbers. Specifically, $E_{\bar{ z}^+} (k_x) \simeq E_{\bar{ z}^-} (k_x) \propto k_x^{-11/3}$ after removing the $k_x = 0$ contribution. However, the PSDs of $| \tilde{z}^{\pm}|$ exhibit different power law slopes in $k_x$ and $k_y$, with $E_{\tilde{ z}^+} (k_x) \propto k_x^{-5/3}$,  $E_{\tilde{ z}^-} (k_x) \propto k_x^{-3/2}$, 
and $E_{\tilde{ z}^+} (k_y) \sim E_{\tilde{ z}^-} (k_y) \propto k_y^{-5/3}$. Some steepening occurs after the inverse ion inertial length scale $d_i^{-1}$. 

\section{Discussion} \label{sec:Discussion}

Two sets of simulations were considered, reflecting essentially different levels of mixed polarity magnetic field environments as expressed through the ratio of an initial guide magnetic field strength to a reconnection plane magnetic field strength $B_g /B_0$. Physically, one can think of a ratio with larger values corresponding to a primarily singly oriented set of magnetic field lines whereas a smaller ratio would describe an environment with more mixed magnetic field polarity. 
A coronal hole with a concentration of open magnetic field lines into which the mixed polarity magnetic carpet emerges would have a relatively large ratio $B_g /B_0$, thereby motivating our simulation using $B_g /B_0 = 1$. This could equally apply to the base of large coronal loops that might also exhibit a concentration of directed magnetic field. By contrast, quiet Sun regions might not have a concentration of large-scale open (or closed) magnetic field, i.e., neither an open field region nor a large loop region. In this case, the absence of a significant  large-scale guide magnetic field means that the magnetic carpet emerges into a relatively unstructured low lying part of the solar atmosphere. To investigate quiet-Sun-like regions, we adopt a ratio of $B_g /B_0 = 0.2$ that reflects a region dominated by mixed magnetic field polarity, associated with the emerging magnetic carpet. 

As illustrated in Figure \ref{fig:1}, the simulation takes place in a box located somewhere in the chromosphere, nominally taken to be about 1Mm or less above the photosphere, since \cite{Cranmer_vanBallegooijen_2010} show that about 50\% of all  magnetic carpet loops lie within this range of heights. To best simulate reconnection in the chromosphere, we adopted a PIC-code approach rather than an MHD simulation since this more accurately captures the initiation and physics of reconnection. The drawback of PIC simulations is that it is difficult to access realistic physical parameters in the simulations, but the simulations appear to be relatively insensitive to these unphysical parameters \citep{Li_etal_2019}. 

\subsection{Chromospheric turbulence}

For the $B_g /B_0 = 1$ case, a reasonably defined out-of-reconnection-plane axis can be found, allowing for the perpendicular and parallel wavenumbers $k_{\perp}$ and $k_{\parallel}$ to be approximated. The simulations show that the low-frequency turbulence characteristics are different in directions perpendicular and parallel to the mean magnetic field directions. Specifically, the magnetic energy density spectra for the transverse magnetic field components are anisotropic. The fluctuating quasi-2D PSD $\langle \tilde{B}_{\perp}^2 \rangle (k_{\perp}) = E_{\tilde{B}_{\perp}} (k_{\perp})$ has a spectrum $\propto k_{\perp}^{-3/2}$ and  a larger amplitude than the slab spectrum $\langle \tilde{B}_{\perp}^2 \rangle (k_{\parallel})$, which exhibits a low-frequency spectrum $\propto k_{\parallel}^{-2}$. 
As illustrated in Figures \ref{fig:5} and \ref{fig:7},
quasi-2D turbulence is dominant and slab turbulence constitutes a minority component in the $B_g/B_0 = 1$ case.  Figure \ref{fig:6} also shows that the energy in turbulent transverse quasi-2D magnetic fluctuations is significantly larger than the corresponding kinetic energy in incompressible transverse quasi-2D velocity fluctuations. The same is true for the transverse slab magnetic and velocity fluctuations. Associated with the spectral anisotropy is a variance anisotropy in the ratio quasi-2D : slab of $\sim 1.25$. Thus, the simulations for the $B_g/B_0 = 1$ case show that the turbulence is anisotropic and dominated by the quasi-2D component. 

As discussed, this result holds within a chromosphere possessing some open field or closed field large loops extending well into the corona where a relatively well defined guide magnetic field exists. The essential result to emerge is that these regions will possess magnetic turbulence that is generated by 1) repeated reconnection between the predominant mixed magnetic carpet loops, and 2)  interchange reconnection between some magnetic carpet loops with open magnetic field. 
As illustrated in Figures \ref{fig:3}, \ref{fig:5}, \ref{fig:6} and \ref{fig:7}, although a large-scale mean magnetic field can be identified, the fluctuating magnetic field and kinetic energy are dominant, indicating that the initial mixed polarity magnetic field has largely been annihilated to be replaced by fully developed turbulence.
The  reconnection process between mixed polarity magnetic carpet loops generates quasi-2D turbulence whereas the latter process 2) can initiate inwardly/down and outwardly/up propagating Alfv\'enic fluctuations that form the slab turbulence component. 

For the quiet Sun case, $B_g/B_0 = 0.2$, the simulations reveal that significant and extensive small-scale flux ropes are present, interacting and evolving throughout the interaction region. As with the $B_g/B_0 = 1$ case, the simulation becomes rapidly dominated by magnetic and velocity fluctuations and the mean fields are largely annihilated. For this case, directions parallel and perpendicular to a mean field are extremely difficult to define and while one can introduce wavenumbers $k_x$ and $k_y$,  unlike the former $B_g/B_0 = 1$ example, these wavenumbers no longer correspond to directions transverse to or aligned to a mean field direction. Our simulations show that $\langle \tilde{B}_x^2 \rangle (k_x) \simeq \langle \tilde{B}_x^2 \rangle (k_y)$, both of which exhibit Kolmogorov-like spectra,  Figure \ref{fig:10}. We find that there is little to distinguish between nominal directions and that the turbulent field is essentially isotropic. Figures \ref{fig:10} and \ref{fig:11} indicate that in the absence of a well-defined mean magnetic field, the fluctuating magnetic field is composed primarily of non-propagating structures that are randomly oriented. In this respect, the distinction between the quiet Sun and open Sun regions lies not in the character of the fluctuations so much as in the geometry of those structures, the former being distributed isotropically and the latter being quasi-2D. The fluctuating kinetic energy spectra (Figure \ref{fig:10}) resemble the magnetic field spectra, but the Els\"asser energy spectra (Figure \ref{fig:11}) exhibit some differences in the forward and backward (${\bf z}^{\pm}$) components. For our purposes, the central result from the $B_g/B_0 = 0.2$ case is that the turbulence is dominated by randomly oriented magnetic structures such as small-scale magnetic flux ropes. 

Small-scale magnetic flux ropes or quasi-2D structures  form an advected, nonlinearly interacting population of turbulent fluctuations \citep{Matthaeus_etal_84, Zank_Matthaeus_1992b, zank_matthaeus_1993_incompress, Zank_etal_2012, Zank_etal_2017a}. Slab turbulence is comprised of forward and backward propagating Alfv\'en waves. Since small-scale magnetic flux ropes are advected fluctuations \citep{Zhao_etal_2020, Zhao_etal_2021a, Zhao_etal_2025a}, the presence of a flow through the transition region will ensure their transport from the chromosphere to the low corona without complication.
By contrast, Alfv\'en waves produced by turbulent interchange reconnection of open and  the closed magnetic carpet mixed polarity loops can produce forward and backward Alfv\'enic fluctuations (and/or fast magnetosonic modes, their being closely related to Alfv\'en waves in the low plasma beta regime) as described in \cite{Zank_etal_2020b} (see their Figure 1 cartoon and also the simulations of \cite{Yang_etal_2025}). From this process, we would expect that the  normalized cross helicity for slab turbulence $\sigma_A$ would be $\sigma_A \simeq 0$, i.e., balanced slab turbulence. The situation is complicated by outwardly propagating Alfv\'en waves that will interact with Alfv\'en waves reflected by the transition region since 
Alfv\'enic fluctuations incident on the transition region  experience strong reflection (estimates suggest that $> 95$\% of the incident flux will be reflected -- see \cite{Ferraro_Plumpton_1958, Asgari-Targhi_Ballegooijen_2012, Zank_etal_2021, Cranmer_Molnar_2023, Nakanotani_etal_2026}). 
Since the reflected Alfv\'en flux is counter-propagating with respect to the upward flux, this effectively initiates a nonlinear cascade that generates 2D zero-frequency non-propagating modes \citep{Shebalin_etal_1983} in the chromosphere. 
\cite{Zank_etal_2021} suggested  that the efficient reflection of Alfv\'en waves at the transition region presents an additional further possibility for the origin of 2D turbulence. Specifically, the majority of Alfv\'enic fluctuations will be trapped in the chromosphere, unable to easily transmit through the transition region. 
The timescale for this process can be estimated as $\tau_A^{-1} \sim k V_A (1 - \sigma_A^2)^{1/2}$ \footnote{This is a slight generalization \citep{Zank_etal_2020} of the typical expression \citep[see e.g.,][]{Zhou_etal_2004_review} to ensure that  spectral transfer mediated by the Alfv\'en term is possible only when counter-propagating Alfv\'enic fluxes are present or $|\sigma_A| \neq 1$ , i.e., unidirectionally Alfv\'en waves do not interact nonlinearly to cascade and produce zero-frequency non-propagating modes.} which will bounded by the dissipation scale $\tau_{diss}^{-1} \sim V_A /\lambda_A  (1 - \sigma_A^2)^{1/2}$ where $\lambda_A$ is the slab correlation length and assuming fully developed turbulence. The slab correlation length is unknown and expected to be larger than the nonlinear correlation length scale for quasi-2D turbulence, $\lambda_c$ but nonetheless we assume $\lambda_A \sim \lambda_c \sim 1300$ km as measured by \cite{Abramenko_etal_2013}, discussed further below. Consequently, we find that $\tau_A < \sim 32$ s, which is less than a crude estimate of the Alfv\'en propagation or dynamical time $O(\tau_D) \sim 50$ s\footnote{As before, consider 1 Mm as more or less representative of the region in which we are interested for the reasons outlined above. At this height, assuming $\bar{B} = 100 \; \mbox{G} = 10^{-2}$ T for an ephemeral region, $V_A \sim 40.7$ km/s, giving a very crude dynamical timescale for Alfv\'en waves in the chromosphere $\tau_D \sim 50$ s (bear in mind that close to the photosphere, $V_A \sim 0.65$ km/s and at the transition region $V_A \sim 2200$ kms illustrating the crudeness of the dynamical time scale for non-interacting Alfv\'enic fluctuations within the chromosphere). The timescale $\tau_D$ provides an estimate of the time spent in the chromosphere by non-interacting Alfv\'en waves before they either encounter the transition region, to be either reflected or transmitted, or if downward propagating, to be absorbed by the photosphere.}. 
Thus Alfv\'en waves with $k^{-1} < \sim 1300$ km should interact nonlinearly while transiting the chromosphere, either upwards or downwards, and in so doing generate zero-frequency modes that augment the dominant magnetic island/small-scale magnetic flux rope population generated by the reconnection of the mixed polarity magnetic carpet loops.   

\begin{figure}
%%\epsscale{.80}
%\begin{center}
\includegraphics[width= 1.0\textwidth]{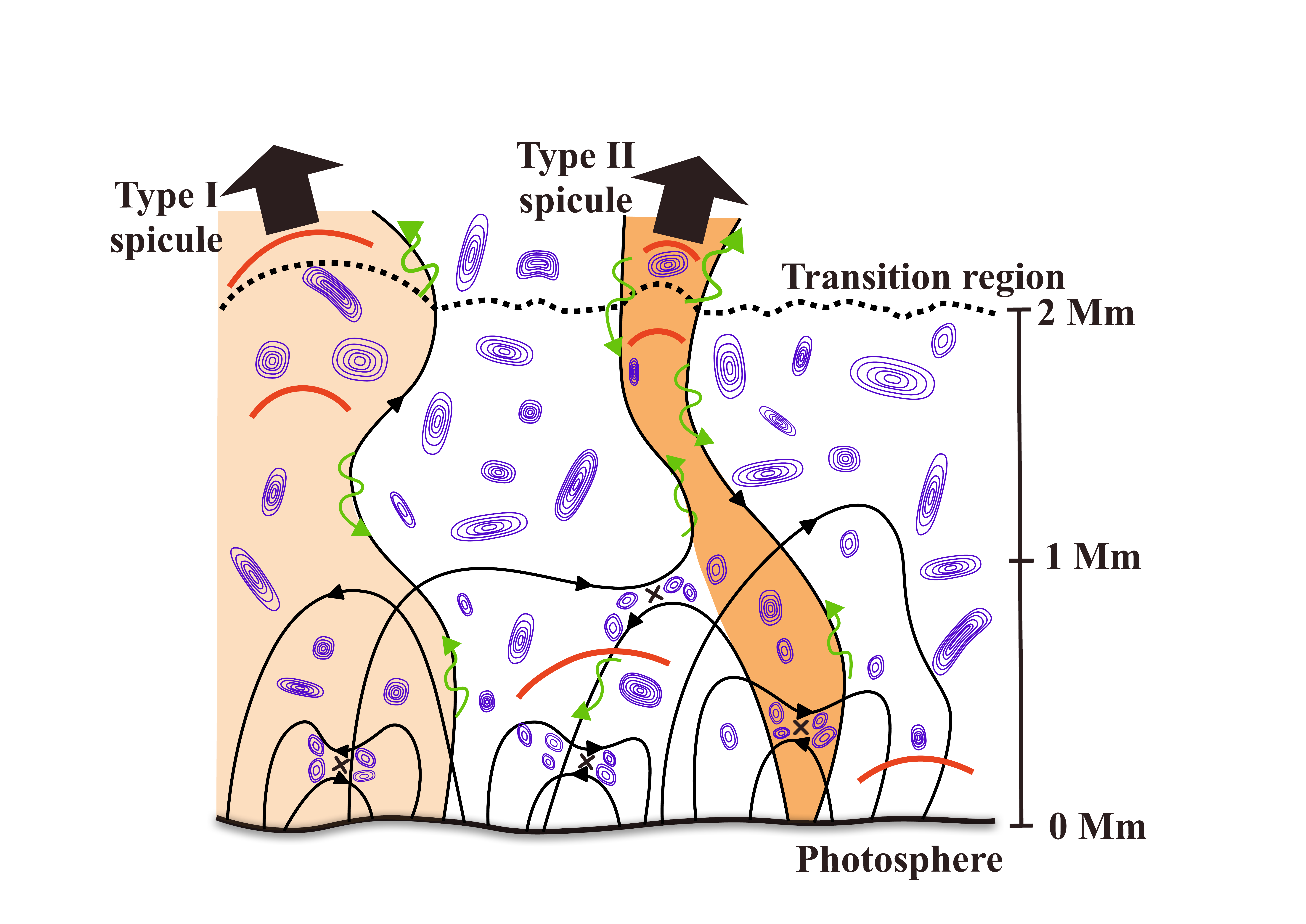}
%\end{center}
\caption{\small A schematic illustrating a section of the highly dynamical chromosphere \citep{Carlsson_etal_2019} showing magnetic field lines (solid black lines with arrows indicating direction), some of which represent closed loops of mixed polarity, including recently reconnected loops (identified by an ``x''), purple-colored randomly oriented small-scale magnetic islands (projections of small-scale magnetic flux ropes) generated by the reconnection of mixed polarity closed magnetic carpet loops, green-colored forward and backward propagating Alfv\'en waves generated by interchange-reconnection of a closed magnetic carpet loop with an open magnetic field, green-colored Alfv\'en waves either reflected or transmitted at the transition region, red-colored shock waves, either acoustic or slow-mode shocks propagating upwards from below the photosphere into the chromosphere \citep[e.g.,][]{Ulmschneider_etal_2005} and thereby entraining chromospheric flows, and high-speed narrow type II and lower speed broader type I spicules \citep{dePontieu_etal_2007} (both possibly driven by shock waves) bounded by open magnetic field extending from the photosphere \citep[e.g.,][]{Sterling_2000}. The spicules and shock waves, and the emergent mixed polarity magnetic carpet loops entrain a randomly distributed set of velocity profiles within the chromosphere, some of which reach the transition region, the fluctuating dotted line located at approximately 2 Mm, creating a randomly fluctuating boundary between the chromosphere and the corona. } \label{fig:cartoon_2}
\end{figure}

The entire process and cycle of turbulence generation is illustrated in the cartoon Figure \ref{fig:cartoon_2}, showing reconnection between mixed polarity magnetic field loops and interchange reconnection between open and closed magnetic field together with outflowing material associated with Type I and II spicules and shock waves that entrain chromospheric material, discussed further below. 

\subsection{Transport of chromospheric turbulence} 
Let us consider the consequences of these results for the transport and dissipation of turbulence and the consequent heating of the chromosphere. We focus primarily on the quasi-2D turbulence that is generated constantly by reconnection interactions of the constantly emerging mixed polarity magnetic carpet. This has a replacement timescale of 1 - 2 hours \citep{Hagenaar_etal_2008}. The timescale on which quasi-2D fluctuations are advected up to the transition region is obtained approximately from $h_{TR}$ (the nominal height of the transition region above the photosphere, taken to be 2 -- 3 Mm) divided by a characteristic ``mean'' flow speed (discussed further below) within the chromosphere. The question of just what a suitable chromospheric mean flow speed is not easily addressed. The chromosphere has been 
 assumed typically to be a static atmosphere, i.e., no upward bulk flow velocity with an exponentially decreasing  density profile with increasing height.  Detailed 1D models used to investigate chromospheric line formation are typically static and semi-empirical \citep{Vernazza_etal_1981, Maltby_etal_1986, Fontenla_etal_1993}. However, steady models of solar acceleration and coronal heating that include the chromosphere \citep{Hollweg_1976, Hammer_1982a, Withbroe_1988} use phenomenological heating models and  require a non-zero small photospheric base upward flow speed. This yields steady and gradually accelerating chromospheric flow speeds that are $< 1$ km s${}^{-1}$. Such models do not admit an exponentially falling density in the chromosphere, which has considerable observational support \citep[e.g.,][]{Vernazza_etal_1981, Maltby_etal_1986, Fontenla_etal_1993}. We take the view here that there does not exist an upward steady bulk chromospheric flow and that the mean density of the chromosphere falls exponentially with height. 

It has long been known but only more recently emphasized \citep{Hansteen_etal_2007, Gudiksen_etal_2011, Carlsson_etal_2019} that the chromosphere is highly temporal and dynamic and that a simple steady-state chromosphere is a poor representation of reality. As already noted, the constant emergence of 
mixed polarity magnetic field loops at granular scales with a cadence of 1 - 2 hours \citep{Hagenaar_etal_2008} was discussed by  \cite{Martinez_Gonzalez_etal_2010}, who estimated that the magnetic carpet loops propagate into the chromosphere at about the sound speed, $\sim 12$ km s${}^{-1}$. This upwardly directed emergence of magnetic carpet will entrain upward flows from the photosphere with scales on the size of the emerging magnetic loop. Based on weak shock theory, the flow entrained by an emerging magnetic carpet loop will have a bulk speed $U \sim 6$ km s${}^{-1}$.  Weak acoustic and magnetoacoustic shocks propagating out of the photosphere appear to be ubiquitous in the chromosphere \citep{Wedemeyer_etal_2004, Mathur_etal_2022} and these too will entrain chromospheric flows, with a similar bulk speed of $U \sim 6$ km s${}^{-1}$ if weak, or larger for stronger shocks. 
And of course, Type I and II spicules are prevalent throughout the chromosphere, exhibiting obvious apparent outflows, and extending to and often well above the transition region. Spicules cover at least 1\% of the Sun's surface at any time \citep[e.g.,][]{Pneuman_Kopp_1978} The upward flow speeds of spicules have historically been measured from the ground, mainly in the H-alpha line, showing them to have outflow speeds of  $10 - 30$ km s${}^{-1}$ \citep[e.g.,][]{Beckers_1968, Beckers_1972} as discussed in \cite{Sterling_2000}.  These spicules can have widths of 300 - 1500 km, lifetimes of 1 - 10 minutes, and a broad distribution of heights with most between 7 - 13 Mm. 
 More recent observations from space have identified much higher upward speeds for chromospheric 
spicules with speeds from 30 - 50 km s${}^{-1}$ to $\sim 100$ km s${}^{-1}$ and sometimes greater  \citep{dePontieu_etal_2007}, well summarized in \cite{Carlsson_etal_2019}, and classified as Type II spicules (here, we use the definitions for type I and II spicules introduced by \cite{dePontieu_etal_2011}).  These values, however, are based on observations in Ca II, and from space rather than from the ground, and so it is not completely clear that these are the ``same''  
chromosphere/spicules as the H-alpha ground-based observations \citep{Sterling_etal_2010, Zhang_etal_2012, Pereira_etal_2012, Pereira_etal_2013}. However, it appears that type I spicules exhibit ``parabolic'' trajectory paths unlike the apparently more vertically propagating type II spicules \citep{Pereira_etal_2012, Hinode_Team_2019}.  Regardless, Type II spicules provide a further dynamical flow input into the chromosphere.  Besides having higher outflow speeds, Type II spicules differ from Type I spicules in being narrower in width ($< 200$ km), shorter lived, and it seems that a fraction of the cold material in the spicule is hot, but not enough to heat the solar corona \citep{Klimchuck_2012, dePontieu_etal_2007, dePontieu_etal_2011}. An important property of spicules is that the upward mass flux of spicules is about 100 times that of the solar wind, with the result 
that not all  the upward-moving spicule material can escape.  As pointed out originally by \cite{Pneuman_Kopp_1978}, most of that material (perhaps $> 99$\%) must fall back to the photosphere. Both blue- and red-shifted UV lines (which correspond to the transition region) have been observed suggestive of outflowing and infalling spicule material \citep{Sterling_2000, Carlsson_etal_2019}. 
 
 For our purposes, the precise details do not matter but the upward emergence of magnetic carpet field, spicules moving upward (and downward), and network regions on the quiet Sun affected by p-mode waves that travel upward from the
photosphere and produce acoustic/slow-mode shocks \citep[e.g.,][]{Carlsson_Stein_1997, Wedemeyer_etal_2004, Ulmschneider_etal_2005}) all act to partially entrain (through both plasma coupling and collisional coupling) and drive large-scale (i.e., energy-containing scale) flows. This complicated, highly temporal and dynamic chromosphere is what appears to be modeled in time-dependent simulations \citep{Freytag_etal_2002, Wedemeyer_etal_2004, Gudiksen_etal_2011, Carlsson_etal_2016}. Besides creating a highly temporal chromosphere populated by stochastic large-scale flows within which turbulence is transported, those upward flows originating from the photosphere will inject photospheric transverse velocity and magnetic field turbulent fluctuations into the chromosphere, through either driving and entrainment by a leading shock or bow wave or by entrainment behind a longer-lived structure such as a spicule. Below, we consider a probability distribution function $f(U)$ describing the distribution of energy-containing scale flows to investigate turbulent heating of the chromosphere and the injection of turbulence across the transition region and into the solar corona. 
 
Based on our simulations and the discussion above, we can consider the transport via advection of turbulence from the photosphere into the chromosphere and up to the transition regions.   The rate at which quasi-2D magnetic turbulent energy is cascaded from large to small scales and the rate at which it is dissipated  is governed by the nonlinear timescale \citep{Zhou_etal_2004_review, Zank_etal_2017a, Zank_etal_2020} for the quasi-2D Els\"asser energy density $\langle {Z^{\infty}}^2 \rangle$. Here, we use a slightly different notation than before to identify incompressible fluctuations in the theoretical model. Specifically, 
 ${{\bf Z}^{\infty}}^{\pm} = \tilde{\bf U}^{\infty} \pm \tilde{\bf B}^{\infty} /\sqrt{\mu_0 \bar{\rho} }$, where the subscript ``$\infty$'' denotes quasi-2D incompressible transverse fluctuations, and $\langle {Z^{\infty}}^2 \rangle = (\langle { {\bf Z}^{\infty}}^+ \cdot { {\bf Z}^{\infty}}^+ \rangle +  \langle { {\bf Z}^{\infty}}^- \cdot { {\bf Z}^{\infty}}^- \rangle)/2 = \langle {\tilde{U}}^2 \rangle + \langle {\tilde{B}}^2 / \mu_0 \bar{\rho} \rangle$ is twice the energy density/volume.  The nonlinear timescale for energy to cascade is given by $\tau_{NL}^{-1} \sim k_{\perp} \langle {Z^{\infty}}^2 \rangle^{1/2}$ s${}^{-1}$, and this is bounded by the dissipation timescale $\tau_{diss}^{-1} \sim \langle {Z^{\infty}}^2 \rangle^{1/2} /\lambda_c$, where $\lambda_c$ is the correlation length. 
For fully developed quasi-2D MHD turbulence in a flowing plasma, we may adopt a von-Karman-Howarth-like phenomenology \citep[e.g.,][and references therein]{zank_etal_1996_evolutionmagfluct, Zank_etal_2017a} in which the flow of energy into the inertial range is balanced by the loss of energy at the dissipation scale, i.e., assuming a Kolmogorov description of fully developed turbulence --  see \cite{Zhou_etal_2004_review} for an excellent discussion. Hence, the 1D transport and dissipation of the advected non-propagating MHD turbulence Els\"asser energy density through the chromosphere is governed by\footnote{One can derive this equation quite easily from the conservation forms of the magnetohydrodynamic (MHD) momentum and induction equations and assuming that the density $\rho$ is independent of time $t$, i.e., $\nabla \cdot \left( \rho U \right) = 0$. On calculating  ${\bf U} \cdot$ of the momentum equation and ${\bf B} \cdot$ of the induction equation, introducing a mean field decomposition for both ${\bf U} = {\bar{\bf U}} + {\bf u}$ and ${\bf B} = {\bf b }$ (assuming no large-scale mean magnetic field) such that  $\langle {\bf u}, {\bf b} \rangle = 0$, summing, and collecting the transport terms on the LHS and the nonlinear terms on the RHS, we obtain equation (\ref{eq:diss}). The third-order nonlinearities are modeled using 2-point correlations and the nonlinear timescale \citep{Zank_etal_2012, Zank_etal_2017a} within an incompressible framework. }
\begin{equation}
\frac{d }{dt} \left( \frac{1}{2}  \langle { \rho Z^{\infty}}^2 \rangle \right)  \simeq \frac{d }{dt} \left( \frac{1}{2} \bar{\rho} \langle { Z^{\infty}}^2 \rangle \right)  = - \alpha \frac{1}{2} \bar{\rho}  \frac{ \langle { Z^{\infty}}^2 \rangle }{\tau_{NL} } = - \alpha \frac{1}{2} \bar{\rho} \frac{ \langle { Z^{\infty}}^2 \rangle^{3/2} }{\lambda_c },  \label{eq:diss}
\end{equation}
where $d/dt$ is the Lagrangian derivative $d/dt = \partial_t + U\partial_h$, $h$ is the height above the photosphere and $\bar{\rho} (h)$ is the mean density profile from the photosphere to the transition region\footnote{We use a simple exponential function to estimate the chromospheric density assuming values of $\rho (h = 0) = 2 \times 10^{-4}$ kg m${}^{-3}$ and $\rho (h = 2 \mbox{Mm}) = 1.6 \times 10^{-11}$ kg m${}^{-3}$, $h$ being the height above the chromosphere and a transition region that begins at a height of $h = 2000$ km. This yields $\bar{\rho} (h) = 2 \times 10^{-4} \exp \left[ h/h_0 \right]$ where the scale height $h_0 = 0.12$ Mm. }.  Unlike the formulation of solar wind turbulence transport models \citep{zhou_matthaeus_1990_swfluctuation, hunana_zank_2010_apj, zank_etal_1996_evolutionmagfluct, Zank_etal_2017a}, we cannot assume a simple steady flow and indeed the exponential variation in the mass density $\bar{\rho} (h)$ is a consequence of assuming a static background atmosphere for the chromosphere.  The choice of $U$ is complicated as discussed above and further below. 

An additional parameter $\alpha$ is present in the transport equation. The application of  the Kolmogorov theory of turbulence requires that the rate of dissipation of energy balance the rate at which energy is injected from the energy-containing range into the inertial range. To maintain the self-similarity of the inertial range as energy is dissipated requires that the correlation length change in way that maintains the balance. A modest generalization allows us to write the von Karman-Dryden equation for the correlation length $\lambda_c$ \citep{karman_howarth_1938_isotropicturb, Dryden_1943, zank_etal_1996_evolutionmagfluct, matthaeus_etal_1996_mhdphenom} as 
\begin{equation}
\frac{d \lambda_c}{dt} = \beta \langle {Z^{\infty}}^2 \rangle^{1/2}, 
\label{eq:1}
\end{equation}
which includes another parameter $\beta$. The presence of the two parameters $\alpha$ and $\beta$ are related, as discussed by \cite{matthaeus_etal_1996_mhdphenom}, to the existence of a single timescale $\tau_c$ governing the turbulence decay law, specifically the relationship $\alpha = 2 - 2\beta$. The particular choice of $\alpha = 2 \beta$ ensures that $\beta = 1/2$ and hence $\alpha = 1$. In many turbulence models, one sometimes retains $\beta = 1/2$ but introduces a ``constant'' $\alpha_K$, for which it is typically assumed that $\alpha_K \neq 1$ and often taken to be smaller ($\alpha_K = 0.1$ or 0.01 are used typically). We will use $\alpha = 2 \beta$ in (\ref{eq:1}) and replace $\alpha$ in ({\ref{eq:diss}) by $\alpha_K$ with possible values 1, 0.1, and 0.01. It is easily seen that (\ref{eq:1}) yields the conservation law 
\begin{equation}
\frac{d}{dt} \left( \left( \bar{\rho} \langle {Z^{\infty}}^2 \rangle \right)^{1/2} \lambda \right) = 0, \label{eq:1a}
\end{equation}
when $\alpha = 2\beta$ and ({\ref{eq:1a}) is an expression of Kolmogorov theory. 

For now, taking $U$ to be a constant unspecified, upward flow speed in the 1D formulation and including $\alpha_K$ together with $\alpha_K = 2 \beta$ allows us to solve the general coupled equations (\ref{eq:diss})  (\ref{eq:1}) in the steady-state ($\partial_t = 0$), i.e., 
 \begin{eqnarray}
 U \frac{d}{dh} \left( \bar{\rho} \langle { Z^{\infty}}^2 \rangle \right)  &=& - \alpha_K \bar{\rho} \frac{ \langle { Z^{\infty}}^2 \rangle^{3/2} }{\lambda_c } \quad \mbox{or} \quad \frac{dy}{dh} = - \frac{\alpha_K}{\lambda_0} \frac{\bar{\rho}^{-1/2} }{y_0^{1/2} U} y^{2}, \label{eq:2} \\
 U \frac{d \lambda_c}{dh} &=& \frac{\alpha_K}{2} \langle  {Z^{\infty}}^2 \rangle^{1/2}, \quad \mbox{or} \quad   \frac{d}{dh} \left( \left( \bar{\rho} \langle { Z^{\infty}}^2 \rangle \right)^{1/2} \lambda_c \right) = 0,       \label{eq:2a} 
 \end{eqnarray}
 where $y \equiv \bar{\rho} \langle { Z^{\infty}}^2 \rangle$ and $\bar{\rho} = \rho_0 \exp [-h/h_0]$. The rate of energy input at the photosphere is a combination of the conversion of the energy in the emerging magnetic carpet to turbulent fluctuations and the turbulent energy associated with the observed photospheric velocity fluctuations observed by \cite{Abramenko_etal_2013} being advected into the chromosphere. The turbulent energy is distributed between kinetic and magnetic energy so that we can assume that the input energy at the lower boundary is given by ${\cal E}(h = 0) \sim {\cal E}_{KE} + {\cal E}_{ME}$ J m${}^{-3}$, which will be entrained and advected by a large-scale flow emerging from the photosphere. However, the injected energy associated with the magnetic carpet is due to the speed with which it emerges from the photosphere ($\sim 12$ km s${}^{-1}$, \cite{Martinez_Gonzalez_etal_2010}) and the strength of the associated magnetic field. As described below, this gives a specific energy injection rate $\dot{S} (h = 0) \sim \dot{S}_{KE} + \dot{S}_{ME}$ J m${}^{-2}$ s${}^{-1}$ from the magnetic carpet. Hence, in terms of the Els\"asser energy density, the magnetic carpet boundary condition for equations (\ref{eq:2}) and (\ref{eq:2a}) at $h = 0$, $y(h = 0) = y_0 = \bar{\rho} (h = 0) 
 \langle { Z^{\infty}}^2 \rangle (h=0)$ J m${}^{-3}$ will be given by $\dot{S}/U$. By contrast, entrainment of the photospheric turbulence field will be given by $y(h=0) = y_0 = {\cal E}_{ph}$ where ${\cal E}_{ph}$ is the photospheric turbulent energy density for a flow such as a spicule, the post-shock flow of an acoustic shock emerging from the photosphere, or even the post-emergent flow behind a magnetic carpet event.  Solving equations (\ref{eq:2}) and (\ref{eq:2a}) with the boundary condition $y_0$ at $h = 0$ yields 
 \begin{eqnarray}
 y(h) = \bar{\rho}(h) \langle { Z^{\infty}}^2 \rangle (h) &=& \frac{  y_0}{1 + 2\alpha_K (h_0/\lambda_0) \left( y_0/\rho_0 U^2 \right)^{1/2} \left( \exp \left[ h/2h_0 \right] - 1 \right)  }, \quad \mbox{[J m${}^{-3}$]}, \label{eq:3} \\
 \lambda (h) &=& \left( y/y_0 \right)^{1/2} \lambda_0, \label{eq:3a}
 \end{eqnarray}
 where $y(h)$ [J m${}^{-3}$] is the turbulent Els\"asser (kinetic plus magnetic) energy per volume over the height $h \in [0,h_{TR}]$ Mm, decaying with increasing $h$, $h_{TR}$ is the height of the transition region, and $\lambda_0$ is the correlation length at the photosphere. The height at which the transition region is located is taken nominally to be $h_{TR} = 2$ Mm. 
 
 An important quantity is  the turbulent heating rate function, 
 \begin{equation} 
 \dot{\cal H} \equiv \alpha_K \frac{\bar{\rho} (h) \langle {Z^{\infty}}^2 \rangle^{3/2} (h)}{2 \lambda_c} = \alpha_K \frac{y^{3/2} }{2 \lambda_c} (h) \bar{\rho}^{-1/2} (h). \quad \mbox{[J m${}^{-3}$ s${}^{-1}$] } \label{eq:4}
 \end{equation}
Note that $U$ is an unspecified 1D flow velocity from the base of a photospheric region that may encompass various emerging events within it. Thus, we need to consider a ``typical'' area and period of the photospheric base, as depicted in Figure \ref{fig:cartoon_2}, in which a mixture of dynamical events are active concurrently. Hence, we need a probabilistic description of the flow speed $U$ within a patchwork region of different kinds of vertical flows and flow speeds over a unit surface area of the Sun in both quiet and open magnetic field regions. Suppose we have a probability distribution function $f(U)$ of flow speeds for the unit surface area for any given time, possibly with some solar cycle variation. Although no ``bulk'' flow speed exists for the patch in the sense of a large-scale, gradually accelerating flow that emerges from the photosphere, we can nevertheless define $\bar{U} = \int U f(U) dU$ and introduce a mean and fluctuating description for the chromospheric flow according to $U = \bar{U} + u$ such that the expectation $E[U] = \bar{U} = \langle U \rangle$ and hence $E[u] = \langle u \rangle = 0$ as usual. Here, $\bar{U}$ is simply the (weighted) mean of the random variable $U$. Similarly, $E[U^2] = \langle U^2 \rangle = \bar{U}^2 + \langle u^2 \rangle$, where $\langle u^2 \rangle = E[U^2]  -  \bar{U}^2$ is the variance. 
To estimate the dissipation, transport, and heating of turbulence in the chromosphere with a distribution of velocities $U$, i.e., to derive $E[y(h)] = E[\bar{\rho}(h) \langle {z^{\infty}}^2 \rangle (h) ]$, we need to calculate the expectation of the nonlinear function (\ref{eq:3}) in a highly inhomogeneous medium. Unfortunately, a general expression as a function of $h$ does not seem to be analytically derivable so we content ourselves with an estimate of the expectation at the transition region, $h = 2$ Mm and compare the result to numerically evaluated expectations. 

In the following, we will consider four families of chromospheric flows: 1) flows associated with magnetic carpet emergence and entrainment in the post-emergent flow; 2) the inducement of flows due to the emergence of shock waves from the photosphere into the chromosphere; and 3) and 4) flows associated with the formation and propagation of type I and II spicules, respectively. We may generally express the boundary condition $y_0 = \dot{S}/U + {\cal E}_{ph}$ for each family of flows. It is useful to rewrite equation (\ref{eq:3}) as 
\begin{equation}
y(h) = \frac{ \left( \dot{S} + {\cal E}_{ph} U \right) U^{1/2} }{U^{3/2} + 2 \alpha_K h_0/\lambda_0 \left( \dot{S} /\rho_0 + {\cal E}_{ph} U /\rho_0 \right)^{1/2} \left( \exp [h/2h_0] - 1 \right) }, 
\label{eq:5}
\end{equation}
thereby allowing us to evaluate $E[ \langle {Z^{\infty}}^2 \rangle ] (h \sim 2 \mbox{Mm}) = \langle \langle {Z^{\infty}}^2 \rangle \rangle (h \sim 2 \mbox{Mm})$. 

Notice that the magnetic carpet energy injection term can be expressed as $\left( \dot{S}/\rho_0 \right)^{1/3} \equiv C_{inj}$ m s${}^{-1}$, and hence $\left( \rho_0 U^3 /\dot{S} \right)^{1/3}= U/C_{inj}$ is the ratio of the turbulence transport speed to the energy injection speed $C_{inj}$. For a typical turbulence injection rate $\dot{S} \sim 2 \times 10^4$ J m${}^{-2}$ s${}^{-1}$ (see below), $C_{inj} \sim 4.6 \times 10^2$ m s${}^{-1}$.

\subsection{Possible sources of chromospheric turbulence}
Evidently, the energy density (\ref{eq:3}) of turbulence depends on the injection energy term $\dot{S}$, the photospheric turbulent energy density ${\cal E}_{ph}$  and the  flow velocity $U$, suitably defined. Consider potential sources of chromospheric turbulence. These can be 1) the emergence of the magnetic carpet and its conversion to turbulent energy; 2) transverse incompressible fluctuating velocity and magnetic fields in the photosphere itself, generated by the constantly moving granules, as observed by \cite{Abramenko_etal_2013}; 3) turbulence generated by or amplified by acoustic/slow mode shocks, whether formed below the photosphere and then propagating into the chromosphere \citep{Wedemeyer_etal_2004} or shocks driven by spicules \citep{Sterling_2000} or even magnetic loops emerging from the photosphere at supersonic speeds \citep{Martinez_Gonzalez_etal_2010}, and 4) the interaction of upward propagating and downward falling spicule material. The possible role of prominences/filaments is not considered here for reasons discussed in \S \ref{sec:prominences}. 
We focus here on the first two possibilities primarily, i.e., the constant  emergence  
of magnetic loops and their conversion to quasi-2D turbulence, and secondly the transport of inter-granular photospheric turbulence into the chromosphere through entrainment by emergent flows such as spicules, shocks, and even the magnetic carpet. 

Consider first the injection of turbulent energy by the magnetic carpet. \cite{Martinez_Gonzalez_etal_2010} investigated the injection of magnetic energy into the chromosphere due to the emerging magnetic carpet. They found that the typical strength of an emerging small-scale loop is $\sim 5.5 \times 10^{-3}$ T (55 G) giving a loop magnetic energy density $E_{mag} = B^2 /2\mu_0 = 12.04$ J m${}^{-3}$. Since the magnetic loop emerges from the photosphere at about an observed speed of $V_L \sim 12$ km s${}^{-1}$ (roughly the sound speed) \citep{Martinez_Gonzalez_etal_2010}, the magnetic energy injection rate per loop is $\sim E_{mag} V_L = 1.44 \times 10^5$ J m${}^{-2}$ s${}^{-1}$. \cite{Martinez_Gonzalez_etal_2010}  assume conservatively that 
only 1\% of the solar surface experiences emergent magnetic carpet events at any given time and correct for the possibility that visible and infrared loops have the same characteristics. In so doing, they estimate the net magnetic energy injection rate at the lower boundary of the chromosphere due to the emergence of the magnetic carpet to be from about 1 - 20\% of $\sim E_{mag} V_L $ or $\dot{S}_{MC} \sim 1.44 \times 10^3$ -- $2.2 \times 10^4$ J m${}^{-2}$ s${}^{-1}$. The \cite{Martinez_Gonzalez_etal_2010}  estimate for magnetic energy input is interesting since \cite{Anderson_Athay_1989a} revised upward the older and widely used radiative loss estimate for the chromosphere \citep{Athay_1966} quoted by \cite{Withbroe_Noyes_1977} ($4 \times 10^3$ J m${}^{-2}$ s${}^{-1}$ or $4 \times 10^6$ erg cm${}^{-2}$ s${}^{-1}$) to $1.4 \times 10^4$ J m${}^{-2}$ s${}^{-1}$. Hence, $\dot{S}_{MC}$ is of the order of or slight larger than the radiative loss rate. 

Much of the magnetic carpet input energy has to be converted to turbulence after which it can then be dissipated in the chromosphere and beyond in the corona, depending critically on the transport characteristics in the chromosphere. One of the important conclusions to emerge from the simulations described above is that reconnection of mixed polarity magnetic field in the presence of both a moderately strong guide magnetic field (open regions)  and a weak guide field (quiet Sun) results in the annihilation of the  input field to be replaced by a fully turbulent plasma comprised primarily of non-propagating small-scale magnetic flux ropes or magnetic islands and transverse incompressible velocity fluctuations. The magnetic energy density in the fluctuating quasi-2D turbulent field far exceeds that in the ``mean field,'' having a harder spectrum ($\propto k_{\perp}^{-3/2}$ versus $k_{\perp}^{-11/3}$) and a greater spectral amplitude. 

Separate from the emergence of the magnetic carpet, transverse velocity fluctuations are driven at the photospheric surface by the displacement of magnetic footpoints by inter-granular flows. \cite{Abramenko_etal_2013} measured these fluctuations to have velocities $\sim 1.21$ km s${}^{-1}$ based on the tracking of motions in the photosphere. Assuming these transverse velocity fluctuations to be incompressible yields the volumetric kinetic energy to be $\sim 1/2 \bar{\rho} \langle u_{\perp}^2 \rangle \sim 1.44 \times 10^2$ J m${}^{-3}$, where $u_{\perp}$ is the fluctuating velocity measured by \cite{Abramenko_etal_2013} and we have taken $\bar{\rho} (h = 0) = 2 \times 10^{-4}$ kg m${}^{-3}$.  Assuming  equipartition of the magnetic and kinetic energy densities (although as found by \cite{Nakanotani_Zank_2025a}, the energy in magnetic fluctuations tends to exceed the kinetic energy in a partially ionized plasma), implies that the turbulent Els\"asser energy density in the photosphere is $E_{ph} \sim 2.88 \times 10^2$ J m${}^{-3}$.  Furthermore, 
 \cite{Abramenko_etal_2013}  use the velocity observations to derive the correlation length for transverse velocity and magnetic field fluctuations as approximately $\lambda_0 \sim 1300$ km just above the photosphere. More recently, \cite{Bailey_etal_2025} find that $\lambda_0 \sim 1500$ km at the photosphere too, rising to perhaps $\sim 5$ Mm at the transition region (their Figure 3). 
 We will assume that the correlation length is $\lambda_0 = 1.3$ Mm at the photosphere and allow $\lambda_c$ evolve according to equation (\ref{eq:2a}) through the chromosphere up to the transition region and the coronal base.  

To estimate the  energy injection rate, we need characteristic upflow speeds $U_{in}$ from the photosphere that advect the transverse fluctuations into the chromosphere. The possible ways in which photospheric turbulence can be advected into the chromosphere, outlined above, can be enumerated.
\begin{enumerate}
\item Entrainment of photospheric turbulence by emerging magnetic carpet loops. Taking as before, an emergent speed of $\sim 12$ km s${}^{-1}$ \citep{Martinez_Gonzalez_etal_2010}, weak shock theory implies an entrained flow speed of $U_{in} \sim 6$ km s${}^{-1}$. 
The entrained flow will inject photospheric turbulence into the chromosphere at a rate $\dot{S}_{MC}^{ph} \sim \alpha \cdot 2.88 \times 10^2 U_{in} = \alpha 1.73 \times 10^6$ J m${}^{-2}$ s${}^{-1}$. On assuming as before \citep{Martinez_Gonzalez_etal_2010} that $\alpha = 1 - 20$\%, we obtain $\dot{S}_{MC}^{ph} \sim  1.73 \times 10^4$ -- $3.46 \times 10^5$ J m${}^{-2}$ s${}^{-1}$. 
\item The ubiquitous presence of upwardly propagating acoustic shocks \citep{Wedemeyer_etal_2004, Mathur_etal_2022} will entrain photospheric turbulence and carry it into the chromosphere. We take $U_{in} = 6$ km s${}^{-1}$ \citep{Mathur_etal_2022}, which is consistent with weak shock theory, although the variance may well be quite large. Hence, $\dot{S}_{SH}^{ph} \sim \alpha {\cal E}_{ph} U_{in} \sim \alpha 1.73 \times 10^6 \sim 1.73 \times 10^4$ J m${}^{-2}$ s${}^{-1}$ assuming a value of $\alpha = 1$\%. Note that we neglect the generation and amplification of pre-existing turbulence by chromospheric shock waves \citep[e.g.,][]{Zank_etal_2021a}.
\item For type I spicules, we take a mid-range value for the flow speed $U_{in} = 20$ km s${}^{-1}$, yielding $\dot{S}_{I}^{ph} \sim \alpha 5.76 \times 10^6$ J m${}^{-2}$ s${}^{-1}$. For $\alpha$, we suppose that at any given time, the Sun is emitting type I spicules 1\% of the time which implies that $\dot{S}_{I}^{ph} \sim  5.76 \times 10^4$ J m${}^{-2}$ s${}^{-1}$. This is related to the estimate that approximately 1\% of the Sun at any given time is covered by type 1 spicules \citep[e.g.,][]{Pneuman_Kopp_1978}. 
\item For higher speed type II spicules, we again choose a low to mid-range value for $U_{in} = 60$ km s${}^{-1}$, which yields a photospheric turbulence energy injection rate of $\dot{S}_{II}^{ph} \sim \alpha 1.72 \times 10^7 \sim 1.72 \times 10^5$ J m${}^{-2}$ s${}^{-1}$ if we assume that only $\alpha = 1$\% of the Sun emits a type II spicule at any given time. It has been suggested that type I spicules are rarer than type II's \citep{Pereira_etal_2012} but here we assume that approximately 1\% of the Sun is covered by type II spicules at an given time.  
%\item Finally, unlike the discrete events associated with the emergence of spicules, shocks, or magnetic carpet, the constant upward buffeting by small-scale vertical fluctuations and waves would likely constantly transport transverse photospheric turbulence into the low chromosphere. Taking an upward ``evaporative'' flow speed $U_{in} = 0.05$ km s${}^{-1}$ (a factor of $\sim 5$ greater than the thermal proton speed at the photosphere) gives $\dot{S}_{E}^{ph} \sim  1.44 \times 10^4$ J m${}^{-2}$ s${}^{-1}$. 
\end{enumerate}

All these energy injection rates exceed  the estimated radiative loss rate observed in the chromosphere \citep{Anderson_Athay_1989a}. 
The boundary conditions needed for the solution (\ref{eq:3}) of the turbulence transport equation (\ref{eq:2}) can be assembled from  combinations of $\dot{S}_{MC}$ and ${\cal E}_{ph}$ together with the distributions of the flow speeds associated with each of the possible chromospheric flows 1) - 4) above, i.e., these provide us with a set of approximate conditions with which to model $y_0$ and hence the transport and dissipation of quasi-2D turbulence in the chromosphere. We emphasize that the spicules and shocks simply advect transverse incompressible turbulent kinetic and magnetic photospheric fluctuations out of the photosphere and into the chromosphere, and are not treated as a specific source of turbulence, unlike the magnetic carpet. 

\subsection{Modeling the dynamical chromosphere statistically} 
We take the view that the chromosphere is highly dynamical \citep{Carlsson_etal_2019}, being comprised of a patchwork of upflows and downflows, hot and cold regions, threaded by multiple acoustic shocks, and a constantly emerging and dynamical mixed polarity magnetic carpet, crudely illustrated in the cartoon Figure \ref{fig:cartoon_2}. This multitude of chromospheric flows will act to both mix turbulent fluctuations throughout the chromosphere as well as carry turbulence up to and through the transition region.  To incorporate flows within a dynamical chromosphere that possesses multiple different origins and characteristics requires a complex (and largely unknown) distribution function. 

In their study of the statistics of type I and II spicules, \cite{Pereira_etal_2012} found that a log-normal distribution was a reasonable description of some of the properties, such as the observed distribution of maximum and transverse speeds, with the distributions typically exhibiting a rapid increase to a maximum followed by a slower decrease. For this reason, and because we focus on $U > 0$ (upwardly propagating flows), we will assume a log-normal distribution for each of the magnetic carpet, shock, and type I and II spicules flow speed $U$ distribution functions, i.e., for each flow $i = 1 \ldots 4$, we will assume an independent log-normal distribution $f_i (U)$, each with their mean $\mu_i$ and variance $\sigma_i^2$. Further observational support for the use of log-normal statistics for plasma flow quantities (e.g., density) and magnetic field observations are presented by \cite{Zhao_etal_2025}. For example, the right panel of their Figure 4 shows the probability distribution of the
proton density fluctuations at different distances derived from Parker Solar Probe and Solar Orbiter {\it in situ} data sets. At each of the three distances, the pdf corresponds to a log-normal distribution \citep[see also][]{Burlaga_Szabo_1999}. 

\subsubsection{Uni-flow statistical model}
Consider first a possibly unreasonable simplification of the problem and consider only magnetic carpet turbulence and the entrainment of photospheric turbulence by the post-emergent flow field. We assume 
a single log-normal flow speed probability distribution function (pdf) $f(U)$, 
\begin{equation}
 f(U) = \frac{1}{U \sigma \sqrt{2 \pi }} e^{-(\ln U - \mu)^2/ 2 \sigma^2}, \label{eq:8}
 \end{equation}
with $U \in [0,U]$. Note that in (\ref{eq:8}), $U$ is normalized to a speed $U_n$ that has no physical significance, $\mu$ is the mean of $\ln U$ and $\sigma$ is the standard deviation of $\ln U$, i.e., $\sigma^2 = E[(\ln U)^2] - \mu^2$. For all $n \in \Re$, we have the expectation $E[U^n] = \exp [n\mu + n^2 \sigma^2/2]$, from which we have 
\begin{equation}
\mu = \ln \left( \frac{E[U]^2}{\sqrt{E[U^2] } } \right)  \quad \mbox{and} \quad \sigma^2 = \ln \left( \frac{E[U^2]}{\sqrt{E[U]^2 } } \right). \label{eq:9}
\end{equation}
On expressing the non-normalized speed $U$ as a mean and fluctuating part, $U = \bar{U} + u$ such that $\langle U \rangle = \bar{U}$, $\langle u \rangle = 0$ and $\langle U^2 \rangle = \bar{U}^2 + \langle u^2 \rangle = E[U]^2 + \langle u^2 \rangle$, it follows that 
\begin{equation}
e^{\mu} = \frac{1}{U_n} \frac{ \bar{U}^2 }{ \sqrt{ \bar{U}^2 + \langle u^2 \rangle } } \quad \mbox{and} \quad e^{\sigma^2} = \frac{ \bar{U}^2 + \langle u^2 \rangle }{ \bar{U}^2 }, \label{eq:9a}
\end{equation}
can be expressed entirely in terms of the flow mean speed and variance. The physically irrelevant normalization $U_n$  cancels out exactly in all expressions for $E[ \langle {Z^{\infty}}^2 \rangle ]$ and $\langle \lambda \rangle$.   

\begin{figure}
%%\epsscale{.80}
\includegraphics[width= 1.0\textwidth]{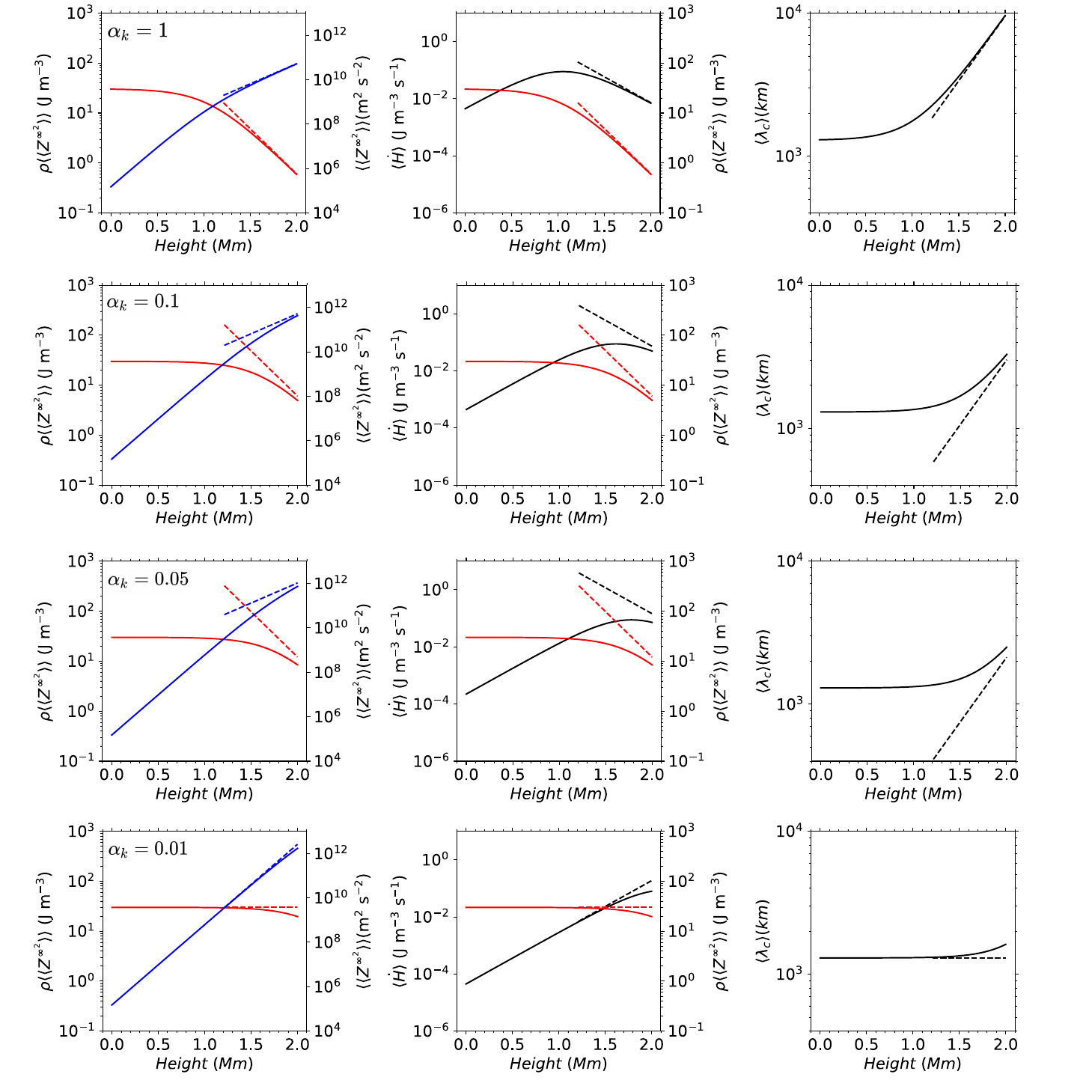}
%$$\includegraphics[width= 1.0\textwidth]{Fig_13_alpha=1.pdf}$$
%$$\includegraphics[width= 1.0\textwidth]{Fig_13_alpha=0.1.pdf}$$
%$$\includegraphics[width= 1.0\textwidth]{Fig_13_alpha=0.05.pdf}$$
%$$\includegraphics[width= 1.0\textwidth]{Fig_13_alpha=0.01.pdf}$$
\caption{\small The figure illustrates the case of turbulence that is generated by the magnetic carpet only and that photospheric turbulence is entrained only by post-emergent flows associated with the emergence of magnetic carpet loops into the chromosphere \citep{Martinez_Gonzalez_etal_2010}.  Plots from $h = 0$ -- 2 Mm showing the expectations of the total energy per unit volume $\langle y \rangle (h)$ J m${}^{-3}$ (red curves, left and middle columns), the Els\"asser specific energy $\langle \langle {Z^{\infty}}^2 \rangle \rangle (h)$ m${}^2$ s${}^{-2}$ (or J kg${}^{-1}$ - energy per unit mass) (blue curves, left column), the heating rate function $\langle \dot{\cal H} \rangle (h)$ J m${}^{-3}$ s${}^{-1}$ (black curves, middle), and the correlation length $\langle \lambda \rangle (h)$ km (black curve, right) as functions of height $h$. Analytic estimates of the expectations near the transition region $\langle \langle {Z^{\infty}}^2 \rangle \rangle$, $\bar{\rho}\langle \langle {Z^{\infty}}^2 \rangle \rangle$ and $\langle \lambda \rangle$, defined by equations (\ref{eq:12a}) and (\ref{eq:12b}) for $\alpha_K = 1$ and 0.1 are overplotted as dashed lines in the corresponding colors. The $\alpha_K = 0.01$ analytic expressions are listed in the text. The rows correspond to decreasing values of the constant $\alpha_K = 1$, 0.1, 0.05, and 0.01. 
The following parameters are adopted: ${\cal E} = 30$ J m${}^{-3}$ and the magnetic carpet post-emergent mean flow speed $\bar{U} = 6$ km s${}^{-1}$ and standard deviation $\langle u^2 \rangle^{1/2} = 2$ km s${}^{-1}$.  } \label{fig:13}
\end{figure}

For this particular problem, we will assume that the energy density of magnetic carpet turbulence generated by reconnection of mixed polarity field is ${\cal E}_{MC} = 12$ J m${}^{-3}$ and that the energy density of photospheric turbulence ${\cal E}_{ph} = 2.88 \times 10^2$ J m${}^{-3}$. We will assume that the post-emergent magnetic carpet eruption flow speed has a mean speed $\bar{U} = 6$ km s${}^{-1}$ and variance 2 km s${}^{-1}$. Hence, $y_0 = \alpha (2 {\cal E}_{MC} + {\cal E}_{ph}) \equiv {\cal E}_{MC}^{tot}$ (the factor of 2 is due to $U_{in} = 2U$ for the injected loop of the magnetic carpet, \cite{Martinez_Gonzalez_etal_2010}). Hence, from equations (\ref{eq:3}) and (\ref{eq:3a})
 \begin{equation}
 y(h) = \frac{  {\cal E}_{MC}^{tot} U}{U + 2\alpha_K (h_0/\lambda_0) \left( {\cal E}_{MC}^{tot} /\rho_0 \right)^{1/2} \left( \exp \left[ h/2h_0 \right] - 1 \right)  }.  \label{eq:9b}  
 \end{equation}
We can use equation (\ref{eq:9b}) and the pdf (\ref{eq:8}) together with equations (\ref{eq:9a}) to obtain numerically the expectations $\langle y \rangle (h)$, $\langle \langle {Z^{\infty}}^2 \rangle \rangle (h)$, $\langle \lambda \rangle (h)$, and $\langle \dot{\cal H} \rangle (h)$. 
It is useful to derive general analytic estimates of the expectations in the vicinity of the transition region so that different parameterizations can be explored, particularly in the context of modeling the boundary conditions at the coronal base to apply to models describing the driving of the solar wind by turbulence. 
For this case, we derive estimated expectations of $\langle \langle {Z^{\infty}}^2 \rangle \rangle$ and $\langle \lambda \rangle$ in the vicinity of the transition region. Consider $h \gg h_0$ and determine when the second term of the denominator $a(h)$ exceeds the first term, $U$.  Taking an intermediate value of $\alpha = 0.1$ \citep{Martinez_Gonzalez_etal_2010} and the values above gives ${\cal E}_{MC}^{tot} \sim 30$ J m${}^{-3}$ from which we find $a(h) \sim \alpha_K 2.87 \times 10^2$ km s${}^{-1}$. Hence, choices of the constant $\alpha_K$ ranging from 1 to $\sim 0.05$ all satisfy $a(h = 2 \mbox{Mm}) \gg 6$ km s${}^{-1}$. Provided $\alpha_K > \sim 0.05$, we have in the vicinity of the transition region 
 \begin{equation}
 \langle {Z^{\infty}}^2 \rangle (h) \simeq \frac{1}{2 \alpha_K} \frac{\lambda_0}{h_0}  \left( {\cal E}_{MC}^{tot} /\rho_0 \right)^{1/2} U e^{h/2h_0}. \label{eq:10}
 \end{equation}
 In similar vein, we find that the correlation length in the vicinity of the transition region is approximately
 \begin{equation}
 \lambda (h) = \sqrt{2 \alpha_K} \left( h_0 \lambda_0 \right)^{1/2} \left( {\cal E}_{MC}^{tot} /\rho_0 \right)^{1/4} U^{-1/2}  e^{h/4h_0}. \label{eq:11}
 \end{equation}
To compute the expected values of $\langle {Z^{\infty}}^2 \rangle$ and $\lambda$, we need to calculate $E[U]$ and $E[U^{-1/2}]$. By virtue of the log-normal statistics expressed through (\ref{eq:9a}), we obtain
\begin{eqnarray}
\langle \langle {Z^{\infty}}^2 \rangle \rangle (h) &\simeq& \frac{1}{2 \alpha_K} \frac{\lambda_0}{h_0}  \left( {\cal E}_{MC}^{tot} /\rho_0 \right)^{1/2} e^{h/2h_0} \bar{U}; \label{eq:12a} \\
\langle \lambda \rangle (h) &\simeq& \sqrt{2 \alpha_K} \left( h_0 \lambda_0 \right)^{1/2} \left( {\cal E}_{MC}^{tot} /\rho_0 \right)^{1/4} e^{h/4h_0} \frac{ \left( \bar{U}^2 + \langle u^2 \rangle \right)^{3/8} }{ \bar{U}^{5/4} }.  \label{eq:12b}
 \end{eqnarray}
On using the parameters listed above, we can compute both the numerical expectations and the analytic estimates that apply near the transition region, both of which are shown in Figure \ref{fig:13}. 
Figure \ref{fig:13} illustrates the numerical expectations for $\langle y \rangle (h)$, $\langle \langle {Z^{\infty}}^2 \rangle \rangle (h)$, $\langle \lambda \rangle (h)$, and $\langle \dot{\cal H} \rangle (h)$ from the turbulence transport solutions (\ref{eq:3}), (\ref{eq:3a}), (\ref{eq:4}), and (\ref{eq:5}), i.e., the total energy per unit volume $\langle y \rangle (h)$ J m${}^{-3}$ (red curves, left and middle columns), the Els\"asser specific energy $\langle \langle {Z^{\infty}}^2 \rangle \rangle (h)$ m${}^2$ s${}^{-2}$ (or J kg${}^{-1}$ - energy per unit mass) (blue curves, left column), the heating rate function $\langle \dot{\cal H} \rangle (h)$ J m${}^{-3}$ s${}^{-1}$ (black curves, middle), and the correlation length $\langle \lambda \rangle (h)$ km (black curve, right) as functions of height $h \in [0,2]$ above the photosphere. Also plotted as dashed lines in the corresponding colors are the analytic estimates of the expected values $\langle \langle {Z^{\infty}}^2 \rangle \rangle$, $\langle y \rangle = \bar{\rho}\langle \langle {Z^{\infty}}^2 \rangle \rangle$ and $\langle \lambda \rangle$, which are valid in the region close to the transition region. The four rows of figures in descending order correspond to four choices of the  constant $\alpha_K = 1$, 0.1. 0.05, and 0.01. The behavior of the expected values mirrors quite well the behavior in the exact solutions of the transport equations when using a specified value of $U$. 
The expectation of the total energy density $\langle y \rangle (h)$ is initially quite flat with increasing height before experiencing strong dissipation higher in the chromosphere, with the height of the turnover evidently dependent on $\alpha_K$. This is not unexpected of course since smaller values of $\alpha_K$ represent weaker dissipation. This behavior is reflected in the expectation of the heating rate profile, which for $\alpha_K \ge 0.05$, peaks closer and closer to $h = 2$ Mm as $\alpha_K$ decreases from 1. The heating rate peak value is nonetheless almost the same for all values of $\alpha_K$. Although we do not compare quantitively, the heating function $\dot{\cal H}$ is qualitatively similar to the heating flux function $\Lambda$ derived from the code VAL-C and shown in Figure 7 of \cite{Anderson_Athay_1989a}, together with the overplotted values of $\Lambda$ given by \cite{Maltby_etal_1986}. Note that the function $\Lambda$ is plotted as a function of column mass $m$, which, although related to height $h$, makes direct comparison difficult but the general behavior between the turbulence heating function and $\Lambda$ is consistent.  
For $\alpha_K = 0.01$, the heating rate peaks at a height greater than 2 Mm. By contrast, the expectation of the Els\"asser specific energy $\langle \langle {Z^{\infty}}^2 \rangle \rangle (h)$ varies relatively little as $\alpha_K$ changes, increasing monotonically from the photosphere to reach values ranging from $\sim 5 \times 10^{10}$ m${}^2$ s${}^{-2}$ to as much as a little more than $10^{12}$ m${}^2$ s${}^{-2}$. The increase is ``geometric'' in that it is due to the exponential change in density with height for a static chromosphere. The third column of Figure \ref{fig:13} illustrates the monotonically increasing change in the expected correlation length with increasing height. Depending on the value of $\alpha_K$, there is a more-or-less extended plateau  that increases in extent with decreasing values of $\alpha_K$, followed by an almost exponential increase to the transition region. It's interesting that the data points of the measured correlation length plotted by \cite{Bailey_etal_2025} from the photosphere to the coronal base/end of the transition region in their Figure 3 resembles the correlation length height profile for the $\alpha_K = 0.1$ case, Their Figure 3 also suggests that our numerical value of  $\langle \lambda \rangle \sim 3$ Mm is very close to their value observed in the vicinity of the transition region. 

From the parameters used above, we use the analytic estimates (also plotted in Figure \ref{fig:13}) to find that the Els\"asser specific energy $\langle \langle {Z^{\infty}}^2 \rangle \rangle \sim \alpha_K^{-1} 5.23 \times 10^{10}$ m${}^2$ s${}^{-2}$ at $h = 2$ Mm, the correlation length $\langle \lambda \rangle \sim \sqrt{\alpha_K}  9.52 \times 10^6$ m, the total energy density $\langle y \rangle = \bar{\rho} \langle \langle {Z^{\infty}}^2 \rangle \rangle \sim \alpha_K^{-1} 6.05 \times 10^{-1}$ J m${}^{-3}$, and the turbulent energy injection rate $\langle \dot{S}_{MC} \rangle \sim \alpha_K^{-1} 3.63 \times 10^3$ J m${}^{-2}$ s${}^{-1}$. 
For $\alpha_K = 1$ and 0.1 respectively, the expected values at the base of the corona are $\langle \langle {Z^{\infty}}^2 \rangle \rangle \sim 5.24 \times 10^{10}, \; 5.23 \times 10^{11}$ m${}^2$ s${}^{-2}$, $\langle \lambda \rangle \sim 9.52, \; 3$ Mm, $\langle y \rangle = \bar{\rho} \langle \langle {Z^{\infty}}^2 \rangle \rangle \sim 0.6, \; 6$, and hence the magnetic carpet turbulent injection rate $\langle \dot{S}_{MC} \rangle \sim 3.63 \times 10^3, \; 3.63 \times 10^4$ J m${}^{-2}$ s${}^{-1}$. 
These values at the transition region provide in principle the coronal base conditions for a solar wind model driven by turbulence created by the magnetic carpet and photospheric turbulence advected by flows associated with the emergence of the magnetic carpet loops from the photosphere. Here, we have assumed that the post-emergent flow speed is approximately half the flow speed dragging the loop into the chromosphere and described by a mean and fluctuating (variance) component. We further assumed that essentially 10\% of the Sun's surface is covered by the magnetic carpet at any given time, which may be an underestimate. The energy injection rates  $\dot{S}_{MC} (h = 2\mbox{Mm})$ of ($\alpha_K = 1$) $3.6 \times 10^3$ or ($\alpha_K = 0.1$) $3.6 \times 10^4$ J m${}^{-2}$ s${}^{-1}$ are  particularly interesting since both values are close to the earlier estimate of $\sim 4 \times 10^3$  J m${}^{-2}$ s${}^{-1}$ \citep{Athay_1966, Withbroe_Noyes_1977} and the more recent estimate of \cite{Anderson_Athay_1989a} of $1.4 \times 10^4$ J m${}^{-2}$ s${}^{-1}$. Furthermore, the $\alpha_K = 0.1$ value for $\langle \lambda \rangle \sim 3$ Mm is encouragingly close to the value found by \cite{Bailey_etal_2025} for the correlation length in the vicinity of the transition region. 

Shown in the bottom row of Figure \ref{fig:13} are the numerical and analytic expectations for the $\alpha_K = 0.01$ case. In this case, the speed term $U$ dominates in the denominator of equation (\ref{eq:9b}), so that $y(h) \simeq {\cal E}_{MC}^{tot}$ and hence $\langle \langle {Z^{\infty}}^2 \rangle \rangle \simeq \left( {\cal E}_{MC}^{tot} /\rho_0 \right) \exp [h/h_0]$ and $\langle \lambda \rangle = \lambda_0$. As can be seen from the bottom row of Figure \ref{fig:13}, there is very little dissipation of the expected total energy $\langle y \rangle (h)$ with height and similarly very little change in $\langle \lambda \rangle (h)$. 
Although $\langle {Z^{\infty}}^2 \rangle (h)$ grows large, $> 10^{12}$, relatively little energy is dissipated in the chromosphere (which is true of the 0.05 case too) since $\langle y \rangle (h)$  decreases from 30 to $\sim 20$ J m${}^{-3}$ only, and hence we can expect comparatively little heating of the chromosphere.  This simply illustrates the necessity of balancing the rate of turbulent dissipation, controlled in part by $\alpha_K$, with the dynamical timescale of the flows that advect the turbulence to ensure chromospheric heating, As we show explicitly in the following subsection, this explains the heating of type II spicules at higher elevations. 

The plotted values of the numerical expectations and the analytic estimates of the expectations agree quite reasonably for all four assumed $\alpha_K$ values.  The analytic expressions can be used to estimate suitable turbulence energy and correlation boundary conditions at the coronal base for solar wind models, provided, in this case, that one considers a single pdf to be sufficient in describing the statistics of the temporal energy-containing range flow speeds and if one considers magnetic carpet turbulence only and the associated entrained photospheric turbulence.

\subsubsection{Multi-flow statistical model}
However, while the above evaluation of the transport and dissipation of chromospheric turbulence is promising, it also demonstrates that the overly simple treatment of the flow speed $U$ pdf needs to be addressed properly. As discussed above, let us now consider four specific temporal  types of flow in the 
chromosphere, viz., post-emergent magnetic carpet flows, shocks injected from the photosphere and distinct from possible shocks associated with spicules \citep{Sterling_2000}, and type I and II spicules ($i = 1 \ldots 4$ respectively) for which each category possesses an independent log-normal pdf $f_i(U)$ with mean $\mu_i$ and variance $\sigma_i^2$. To compute expected values for the Els\"asser specific energy $\langle {Z^{\infty}}^2 \rangle$, we introduce an inner product so that 
\begin{equation}
\langle \langle {Z^{\infty}}^2 \rangle \rangle \equiv \int_0^{\infty}  \langle {Z^{\infty}}^2 \rangle_i f_i(U) dU, \label{eq:13} 
\end{equation}
where $\left( \langle {Z^{\infty}}^2 \rangle_i \right)$ is the Els\"asser specific energy for each flow component $i$ and $\left( f_i (U) \right)$ the corresponding log-normal pdfs. The equations (\ref{eq:8}) - (\ref{eq:9a}) carry through with appropriate subscripts $i$ although we now need to introduce the relative weighting of events. Thus, (\ref{eq:8}) becomes
\begin{equation}
f_i(U) = \frac{n_i/N}{U \sigma_i \sqrt{2\pi} } e^{-(\ln U - \mu_i)^2/2\sigma_i^2}, \label{eq:14}
\end{equation}
and $\sum_{i =1}^4 \int_0^{\infty} f_i(U) dU = \sum_{i =1}^4 n_i/N = 1$ where $n_i/N$ is the relative number of flows $i$ at a given time. 

As with the ``single flow or uni-flow'' case, to estimate $\langle \langle {Z^{\infty}}^2 \rangle \rangle$, we need to establish which of the two terms in the denominator of (\ref{eq:5}) dominates when $h \sim 2$ Mm, i.e., what values of $U$ satisfy 
\[ U^{3/2} < a \left( \dot{S}/\rho_0 + {\cal E}_{ph} U/\rho_0 \right)^{1/2} e^{h/2h_0}, \quad \mbox{or} \quad U^3 - b^2 U < c^2, \]
where $a \equiv 2 \alpha_K (h_0/\lambda_0)$, $b^2 \equiv a^2 ({\cal E}_{ph} /\rho_0) \exp [h/h_0]$, and $c^2 \equiv a^2 (\dot{S}/\rho_0) \exp [h/h_0]$. The bound on $U$ can be estimated as 
\begin{equation}
U \leq \sim 2\alpha_K \frac{h_0}{\lambda_0} \left( {\cal E}_{ph} /\rho_0 \right)^{1/2} e^{h/2h_0} + \frac{ \dot{S} }{2 {\cal E}_{ph} }, \label{eq:5a}
\end{equation}
which ensures that the denominator in (\ref{eq:5}) is dominated by the second term proportional to the  constant $\alpha_K$. If (\ref{eq:5}) is not satisfied by the flow speed $U$, then the flow term dominates in the denominator of equation (\ref{eq:5}). Using the same parameters as before for the single flow case, we obtain $U < \sim \alpha_K 3 \times 10^5 + 690$ m s${}^{-1}$. The choice of $\alpha_K = 1$ ($U \leq 300$ km s${}^{-1}$) ensures that the $a(h)$ term in the denominator is dominant for all possible flows $i = 1 \ldots 4$. By contrast, $\alpha_K = 0.1$ ($U \leq 30$ km s${}^{-1}$)) implies that for flows $i = 1,2,3$, $a(h)$ is dominant but for type II spicules, $U$ dominates. For $\alpha_K = 0.05$, $U$ dominates in the denominator of (\ref{eq:5}) for the type I and II spicules and for $\alpha_K = 0.01$, $U$ dominates for all flows $i = 1 \ldots 4$. 

When the second term in the denominator of (\ref{eq:5}) dominates,
\begin{equation}
\langle {Z^{\infty}}^2 \rangle_i \simeq \frac{1}{2 \alpha_K} \frac{\lambda_0}{h_0} \left( \left( \dot{S} + {\cal E}_{ph} U \right) / \rho_0 \right)^{1/2} U^{1/2} e^{h/2h_0}, \label{eq:15}
\end{equation}
and when the flow speed $U$ dominates,
\begin{equation}
\langle {Z^{\infty}}^2 \rangle_i \simeq \rho_0^{-1} e^{h/h_0} \left(   \dot{S} / U + {\cal E}_{ph} \right). \label{eq:16}
\end{equation}

\begin{figure}
%%\epsscale{.80}
\includegraphics[width= 1.0\textwidth]{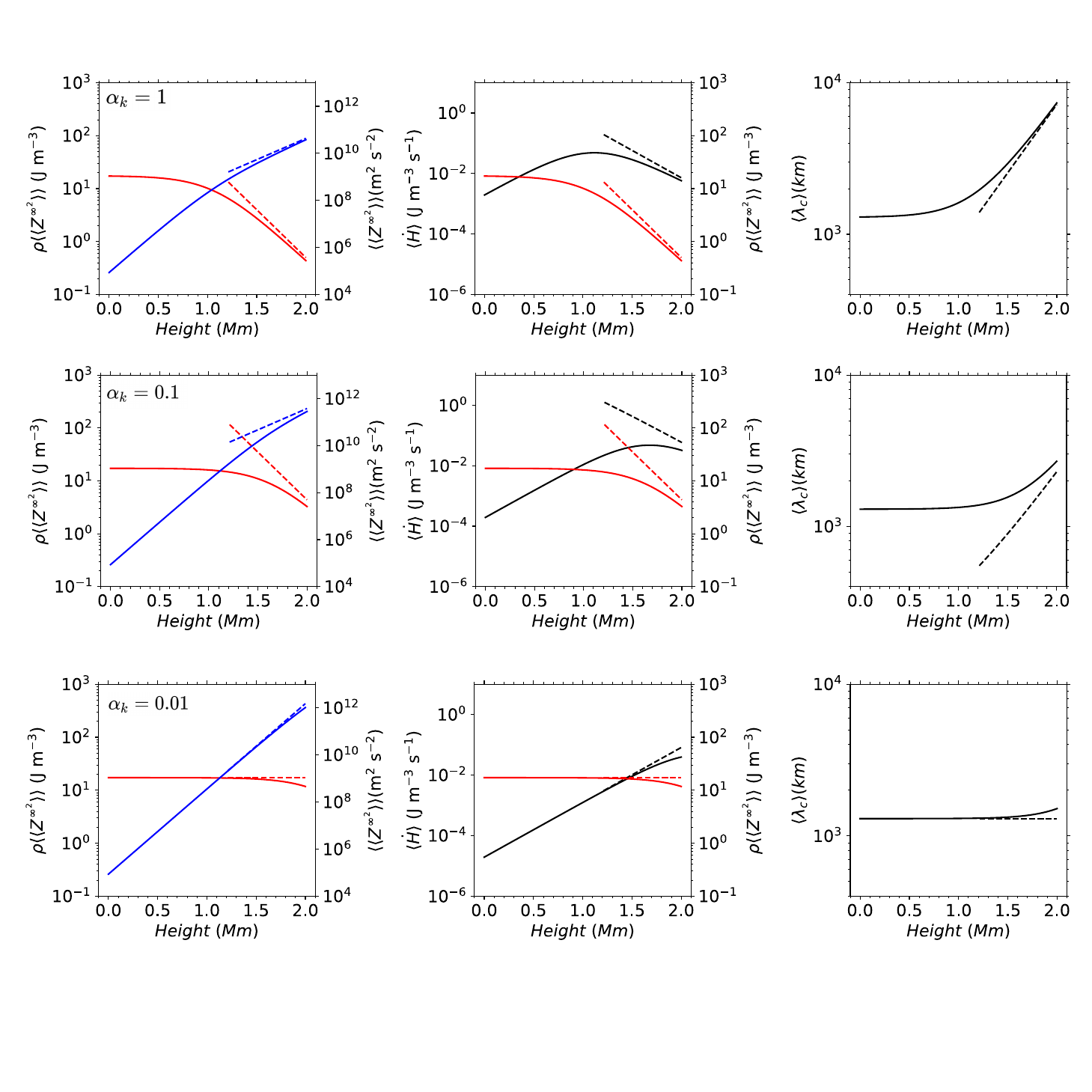}
%$$\includegraphics[width= 0.5\textwidth]{H_and_rhozinf2_for_u=2.0kps.pdf}
%\includegraphics[width= 0.35\textwidth]{H_and_zinf2_for_u=2.0kps.pdf}
%\includegraphics[width= 0.5\textwidth]{zinf2_and_rhozinf2_for_u=2.0kps.pdf}$$
%$$\includegraphics[width= 0.5\textwidth]{H_and_rhozinf2_for_u=5.0kps.pdf}
%\includegraphics[width= 0.35\textwidth]{H_and_zinf2_for_u=5.0kps.pdf}
%\includegraphics[width= 0.5\textwidth]{zinf2_and_rhozinf2_for_u=5.0kps.pdf}$$
\caption{\small The figure illustrates the case of turbulence that is generated by the magnetic carpet only and that entrained photospheric turbulence is not included \citep{Martinez_Gonzalez_etal_2010}.  
The more general form of the distribution $f(U)$ equation (\ref{eq:14}) that distinguishes between possible flows in the chromosphere is used here instead of the simple (and unrealistic) single log-normal distribution used in Figure \ref{fig:13}. 
Plots from $h = 0$ -- 2 Mm showing the numerical (solid lines) and analytic estimates (dashed lines) of the expectations in the same format as Figure \ref{fig:13}. The panels are ordered by decreasing values of $\alpha_K$ from 1 to 0.01. The parameters used here and in Figure \ref{fig:14b} are listed below equation (\ref{eq:20}).   } \label{fig:14a}
\end{figure}

Consider now two cases, one in which we include only magnetic carpet turbulence with no entrained photospheric turbulence, and a second case that includes entrained photospheric turbulence as well. For each, we consider three possible values of the  constant, $\alpha_K = 1$, 0.1, and 0.01. 
As for the magnetic carpet-only turbulence case, we obtain numerically the exact expectations for $\langle y \rangle (h)$, $\langle \langle {Z^{\infty}}^2 \rangle \rangle (h)$, $\langle \lambda \rangle (h)$, and $\langle \dot{\cal H} \rangle (h)$ and we derive the corresponding analytic estimates of the expectations. This requires a few more approximations than the simpler single pdf description. 

{\it Case 1, $\alpha_K = 1$.} In the absence of entrained photospheric turbulence and for $\alpha_K = 1$, equation (\ref{eq:15}) applies to all four of the flows and becomes (we retain $\alpha_K$ in the expressions below for generality since the results apply to values intermediate to 1 and 0.1 as well)
\begin{equation}
\langle {Z^{\infty}}^2 \rangle_i \simeq \frac{1}{2 \alpha_K} \frac{\lambda_0}{h_0} \left( \dot{S} / \rho_0 \right)^{1/2} U^{1/2} e^{h/2h_0}, \qquad i = 1 \ldots 4.  \label{eq:17}
\end{equation}
For the magnetic carpet flow, this becomes as before, 
\begin{equation}
\langle {Z^{\infty}}^2 \rangle_1 \simeq \frac{1}{2 \alpha_K} \frac{\lambda_0}{h_0} \left( {\cal E}_{MC}^{tot} / \rho_0 \right)^{1/2} U e^{h/2h_0},\label{eq:18}
\end{equation}
and $\langle {Z^{\infty}}^2 \rangle_i$, $i = 2$, 3 and 4 are given by (\ref{eq:17}). Since $E_1 [U] = \bar{U}_1$ and $E_i [U^{1/2}] = \bar{U}_i^{3/4} / \left( \bar{U}_i^2 + \langle u_i^2 \rangle \right)^{1/8}$, we obtain, using (\ref{eq:13}), 
\begin{equation}
\langle \langle {Z^{\infty}}^2 \rangle \rangle (h) \simeq \frac{1}{2 \alpha_K} \frac{\lambda_0}{h_0} e^{h/2h_0} \left[ \frac{n_1}{N} \left( 2{\cal E}_{MC} / \rho_0 \right)^{1/2} \bar{U}_1 + \left( \dot{S} / \rho_0 \right)^{1/2} \sum_{i = 2}^4 \frac{n_i}{N} \frac{\bar{U}_i^{3/4} }{ \left( \bar{U}_i^2 + \langle u_i^2 \rangle \right)^{1/8} } \right]. \label{eq:19}
\end{equation}
To evaluate the expected correlation length, we have to evaluate the expectation $\langle \left( y_0/y \right)_i^{1/2} \rangle$ (equation (\ref{eq:3a})), or
\[ \int_0^{\infty}\left( y_0/y \right)_i^{1/2}  \langle {Z^{\infty}}^2 \rangle_i^{-1/2} e^{h/2h_0} f_i(U) dU. \] 
On using $E_1[U^{-1/2}] = \left( \bar{U}_1^2 + \langle u_1^2 \rangle \right)^{3/8} / \bar{U}_1^{5/4}$ and $E_i[U^{-3/4}] = \left( \bar{U}_i^2 + \langle u_i^2 \rangle \right)^{21/32} / \bar{U}_i^{33/16}$, $ i = 2 \ldots 4$, we obtain the expectation of the correlation length as,
\begin{equation} 
\langle \lambda \rangle (h) \simeq \sqrt{2 \alpha_K} \left( h_0 \lambda_0 \right)^{1/2} e^{h/4h_0} \left[ \frac{n_1}{N} \left( 2{\cal E}_{MC} / \rho_0 \right)^{1/4} \frac{ \left( \bar{U}_1^2 + \langle u_1^2 \rangle \right)^{3/8} }{ \bar{U}_1^{5/4} } + \left( \dot{S} / \rho_0 \right)^{1/4} \sum_{i = 2}^4 \frac{n_i}{N} \frac{ \left( \bar{U}_i^2 + \langle u_i^2 \rangle \right)^{21/32} }{ \bar{U}_i^{33/16} } \right]. \label{eq:20}
\end{equation}
Plotted in Figure \ref{fig:14a} are the numerical and analytic estimates of the expectations for Case 1 and from top to bottom, $\alpha_K = 1$, 0.1, and 0.01, following the same format as used in Figure \ref{fig:13}. 
The expected values can be estimated after making the following assumptions, which we will use for all the multi-flow estimates. We will assume that magnetic carpet events are the dominant flows by number in the chromosphere and adopt $n_1/N = 0.7$ and $n_i/N = 0.1$ for $i = 2$, 3, and 4. As before, ${\cal E}_{MC}^{tot} = 30$ J m${}^{-3}$, $\dot{S} = 2 \times 10^4$ J m${}^{-2}$ s${}^{-1}$, $[\bar{U}_i, \langle u_i^2 \rangle^{1/2}]$ are given by $[6,2]$ ($i = 1$), $[6,2]$ ($i = 2$), $[20,5]$ ($i = 3$), and $[60,10]$ ($i = 4$) km s${}^{-1}$ respectively, $h_0 = 0.12$ Mm, and $\lambda_0 = 1.3$ Mm. The expected values are estimated at the transition region $h = 2$ Mm using expressions (\ref{eq:19}) and (\ref{eq:20}). We find the following estimates for the expectations, 
\begin{eqnarray}
\langle \langle {Z^{\infty}}^2 \rangle \rangle (h = 2 \; \mbox{Mm}) &\sim& 4.7 \times 10^{10} \; \mbox{m}^2 \; \mbox{s}^{-2}; \qquad \langle \lambda \rangle \sim 7.19 \; \mbox{Mm}; \nonumber \\
\langle y \rangle &\sim& 4.99 \times 10^{-1} \; \mbox{J} \; \mbox{m}^{-3}; \qquad 
\langle \dot{S}_{TR} \rangle \sim 6.39 \times 10^3 \; \mbox{J} \; \mbox{m}^{-2} \; \mbox{s}^{-1}, \label{eq:21} 
\end{eqnarray}
where in (\ref{eq:21}) we used the expectation of the speed at the transition region $E[U] = \bar{U} \equiv \sum_{i=1}^4 (n_i/N) E[U_i] = \sum_{i=1}^4 (n_i/N) \bar{U}_i = 12.8$ km s${}^{-1}$. The expectation of the energy injection rate $\langle \dot{S}_{TR} \rangle$ in (\ref{eq:21}) at the coronal base is intermediate to the \cite{Athay_1966, Withbroe_Noyes_1977} and \cite{Anderson_Athay_1989a} values, and the correlation length expectation is perhaps a little less than a factor of $\sim 2$ larger than suggested by Figure 3 of \cite{Bailey_etal_2025}. The numerical and analytical estimates agree well above about 1.5 Mm and at the transition region. 

{\it Case 1, $\alpha_K = 0.1$.} For this case, we need to modify the expression for turbulence carried and dissipated by the high-speed type II spicules. In this case, we need to use expression (\ref{eq:16}) since the speed $U$ is dominant in the denominator of (\ref{eq:5}). In the absence of entrained photospheric turbulence, we have $\langle {Z^{\infty}}^2 \rangle_3 = (\dot{S}/\rho_0) U^{-1} \exp [h/h_0]$ from which we obtain
\begin{eqnarray}
\langle \langle {Z^{\infty}}^2 \rangle \rangle (h) &\simeq& \frac{1}{2 \alpha_K} \frac{\lambda_0}{h_0} e^{h/2h_0} \left[ \frac{n_1}{N} \left( 2{\cal E}_{MC} / \rho_0 \right)^{1/2} \bar{U}_1 + \left( \dot{S} / \rho_0 \right)^{1/2} \sum_{i = 2}^3 \frac{n_i}{N} \frac{\bar{U}_i^{3/4} }{ \left( \bar{U}_i^2 + \langle u_i^2 \rangle \right)^{1/8} } \right] \nonumber \\
&\mbox{}& \mbox{} + \frac{n_4}{N} \left( \dot{S}/\rho_0 \right) \frac{ \bar{U}_4^2 + \langle u_4^2 \rangle }{\bar{U}_4^3 } e^{h/h_0}; \label{eq:22} \\
\langle \lambda \rangle (h) &\simeq& \sqrt{2 \alpha_K} \: \left( h_0 \lambda_0 \right)^{1/2} e^{h/4h_0} \left[ \frac{n_1}{N} \left( 2{\cal E}_{MC} / \rho_0 \right)^{1/4} \frac{ \left( \bar{U}_1^2 + \langle u_1^2 \rangle \right)^{3/8} }{ \bar{U}_1^{5/4} } + \left( \dot{S} / \rho_0 \right)^{1/4} \sum_{i = 2}^3 \frac{n_i}{N} \frac{ \left( \bar{U}_i^2 + \langle u_i^2 \rangle \right)^{21/32} }{ \bar{U}_i^{33/16} } \right] \nonumber \\
&\mbox{}& + \frac{n_4}{N} \lambda_0. \label{eq:23}
\end{eqnarray}
The numerical and analytic estimates of the expectations are plotted in the middle row panels of Figure \ref{fig:14a}. 
On assigning values to the various parameters, we find that the $\alpha_K = 0.1$ case for magnetic carpet turbulence only yields 
\begin{eqnarray}
\langle \langle {Z^{\infty}}^2 \rangle \rangle (h = 2 \; \mbox{Mm}) \sim 3.8 \times 10^{11} \; \mbox{m}^2 \; \mbox{s}^{-2}; \quad 
 \langle \lambda \rangle \sim 2.3 \; \mbox{Mm}; \quad 
 \langle y \rangle \sim 4.39 \; \mbox{J} \; \mbox{m}^{-3}; \quad 
\langle \dot{S}_{TR} \rangle \sim 5.6 \times 10^4 \; \mbox{J} \; \mbox{m}^{-2} \; \mbox{s}^{-1}. \label{eq:24} 
\end{eqnarray}
The values for the correlation length at the transition region are in reasonable accord with values from \cite{Bailey_etal_2025} and the turbulence energy injection rate at the coronal base exceeds the \cite{Anderson_Athay_1989a} estimate by a factor of $\sim 5$. 

{\it Case 1, $\alpha_K = 0.01$.}  For this case, illustrated in the third row of Figure \ref{fig:14a}, the denominator is dominated always by the $U$ term and one can neglect the $a(h)$ term in calculating the expectations of $\langle {Z^{\infty}}^2 \rangle$ and $\lambda$. We need to use (\ref{eq:16}) for each term with ${\cal E}_{ph} = 0$, or $\langle \lambda \rangle_i \simeq \left( \dot{S}/\rho_0 \right) U^{-1} \exp [h/h_0]$ for all flows. For $i = 1$, this reduces to $\langle \lambda \rangle_0 \simeq \left( 2{\cal E}_{MC} /\rho_0 \right) \exp [h/h_0]$, as before. We then find 
\begin{equation}
\langle \langle {Z^{\infty}}^2 \rangle \rangle (h) \simeq \frac{n_0}{N}\left( 2{\cal E}_{MC} /\rho_0 \right) e^{h/h_0} + \left( \dot{S} /\rho_0 \right) \sum_{i=2}^3 \frac{n_i}{N} \frac{\bar{U}^2 + \langle u_i^2 \rangle }{\bar{U}_i^{3} } e^{h/h_0}, \quad \mbox{and} \quad 
\langle \lambda \rangle \simeq \lambda_0, \label{eq:25}
\end{equation}
from which the following estimates are derived, 
\begin{eqnarray}
\langle \langle {Z^{\infty}}^2 \rangle \rangle (h = 2 \; \mbox{Mm}) &\sim& 1.5 \times 10^{12} \; \mbox{m}^2 \; \mbox{s}^{-2}; \quad 
\langle \lambda \rangle \sim 1.3 \; \mbox{Mm};  \quad 
\langle y \rangle \sim 17.31 \; \mbox{J} \; \mbox{m}^{-3}; \quad 
\langle \dot{S}_{TR} \rangle \sim 2.22 \times 10^5 \; \mbox{J} \; \mbox{m}^{-2} \; \mbox{s}^{-1}. \nonumber \\
\mbox{} \label{eq:26} 
\end{eqnarray}
From the numerical and estimated expectations above, it appears that the choice of $\alpha_K = 0.01$ renders the dissipation rate too small, implying an unreasonably large turbulent energy flux enters the lower corona, and that the chromosphere experiences insufficient heating. This is evident from the almost negligible changers in the change of the total energy/volume $\langle y \rangle (h)$ and correlation length $\langle \lambda \rangle (h)$ from the photosphere to $h = 2$ Mm. In short, the dissipation rate is too small compared to the dynamical timescale of the flows when $\alpha_K = 0.01$. 

\begin{figure}
%%\epsscale{.80}
\includegraphics[width= 1.0\textwidth]{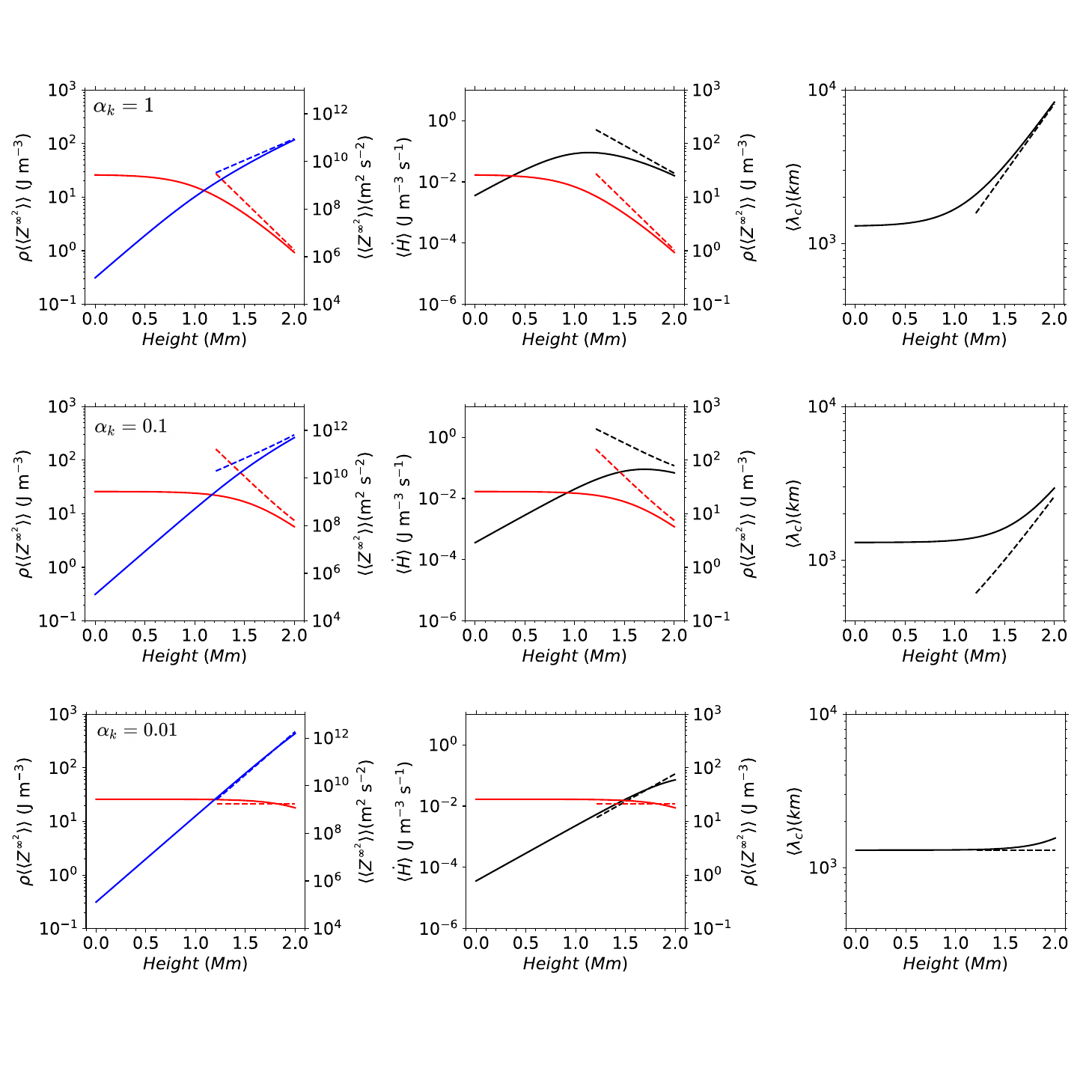}
%$$\includegraphics[width= 0.5\textwidth]{H_and_rhozinf2_for_u=2.0kps.pdf}
%\includegraphics[width= 0.35\textwidth]{H_and_zinf2_for_u=2.0kps.pdf}
%\includegraphics[width= 0.5\textwidth]{zinf2_and_rhozinf2_for_u=2.0kps.pdf}$$
%$$\includegraphics[width= 0.5\textwidth]{H_and_rhozinf2_for_u=5.0kps.pdf}
%\includegraphics[width= 0.35\textwidth]{H_and_zinf2_for_u=5.0kps.pdf}
%\includegraphics[width= 0.5\textwidth]{zinf2_and_rhozinf2_for_u=5.0kps.pdf}$$
\caption{\small  The figure illustrates the case of turbulence that is generated by the magnetic carpet  and photospheric turbulence entrained by all four of the flows under consideration is included.  
We use the more general form of the distribution $f(U)$ equation (\ref{eq:14}) that distinguishes between four possible flows in the chromosphere. 
Plots from $h = 0$ -- 2 Mm showing the numerical (solid lines) and analytic estimates (dashed lines) of the expectations in the same format as Figure \ref{fig:13}. The panels are ordered by decreasing values of $\alpha_K$ from 1 to 0.01. The parameters used here and in Figure \ref{fig:14a} are listed below equation (\ref{eq:20}).  } \label{fig:14b}
\end{figure}

Let us consider now the second case in which both magnetic carpet turbulence and entrained photospheric turbulence is included in a multi-flow model. As before, we consider three cases for the  constant $\alpha_K = 1$, 0.1, and 0.01. The presence of both magnetic carpet turbulence $\dot{S}/U$ and the photospheric energy density ${\cal E}_{ph}$ renders the problem more complicated analytically, although the numerically computed expectations remain straightforward, and we need to make some modest but reasonable simplifying assumptions. This case is illustrated in Figure \ref{fig:14b} in exactly the same format as Figure \ref{fig:14a}. 

{\it Case 2, $\alpha_K = 1$.} Formally, for this case we have 
\begin{equation}
\langle {Z^{\infty}}^2 \rangle_i \simeq \frac{1}{2 \alpha_K} \frac{\lambda_0}{h_0} e^{h/2h_0} \rho_0^{-1/2} \left( \dot{S}/U + {\cal E}_{ph} \right)^{1/2} U = a(h) \rho_0^{-1/2} \left( \dot{S}/U + {\cal E}_{ph} \right)^{1/2} U, \quad i = 1, \ldots 4, \label{eq:27}
\end{equation}
together with $y_0 = \dot{S}/U + {\cal E}_{ph}$. Hence,
\begin{equation}
\left( y_0/y \right)_i = a^{-1} (h) \rho_0^{-1/2} e^{h/h_0} \left( \dot{S}/U + {\cal E}_{ph} \right)^{1/2} U^{-1}, \quad i = 1, \ldots 4. \label{eq:28}
\end{equation}
For the $i = 1$ flow (post-emergent magnetic carpet flow), we have $\dot{S}/U = 2 {\cal E}_{MC}$ which implies that $y_0 = \dot{S}/U + {\cal E}_{ph} = 2 {\cal E}_{MC} + {\cal E}_{ph} = {\cal E}_{MC}^{tot}$ as before.  This gives 
\begin{equation}
\langle {Z^{\infty}}^2 \rangle_1 \simeq \frac{1}{2 \alpha_K} \frac{\lambda_0}{h_0} e^{h/2h_0} \left( {\cal E}_{MC}^{tot} /\rho_0 \right)^{1/2} U; \quad 
\left( y_0/y \right)_1^{1/2} = \sqrt{2 \alpha_K} \left( \frac{h_0}{\lambda_0} \right)^{1/2} e^{h/4h_0} \left( {\cal E}_{MC}^{tot} /\rho_0 \right)^{1/4} U^{-1/2}. \label{eq:29}
\end{equation}
For $i = 2$, shock waves emerging from the photosphere, we can assume that ${\cal E}_{ph} \sim \dot{S}/U$, implying that $y_0 = \dot{S}/U + {\cal E}_{ph} = 2 \dot{S}/U$. In similar vein, since $U \gg U_{in}$ for type I and II spicules, i.e., flows $i = 3$ and 4 respectively, ${\cal E}_{ph} \gg \dot{S}/U$ and hence $y_0^{1/2} = \left( {\cal E}_{ph} + \dot{S}/U \right)^{1/2} = {\cal E}_{ph}^{1/2} \left( 1 + \dot{S}/{\cal E}_{ph} U^{-1} \right)^{1/2} \simeq {\cal E}_{ph}^{1/2} \left( 1 + (1/2) \dot{S}/{\cal E}_{ph} U^{-1} \right)$. It follows then that 
\begin{eqnarray}
\langle {Z^{\infty}}^2 \rangle_2 &\simeq& \frac{1}{2 \alpha_K} \frac{\lambda_0}{h_0} e^{h/2h_0} \left( 2 \dot{S} /\rho_0 \right)^{1/2} U^{1/2}; \quad \left( y_0/y \right)_2^{1/2} = \sqrt{2 \alpha_K} \left( \frac{h_0}{\lambda_0} \right)^{1/2} e^{h/4h_0} \left( 2 \dot{S} /\rho_0 \right)^{1/4} U^{-3/4}. \label{eq:30} \\ 
\langle {Z^{\infty}}^2 \rangle_{3,4} &\simeq& \frac{1}{2 \alpha_K} \frac{\lambda_0}{h_0} e^{h/2h_0} \left( {\cal E}_{ph} /\rho_0 \right)^{1/2} \left( U + \frac{\dot{S} }{2 {\cal E}_{ph} } \right) ; \quad \left( y_0/y \right)_{3,4}^{1/2} = \sqrt{2 \alpha_K} \left( \frac{h_0}{\lambda_0} \right)^{1/2} e^{h/4h_0} \left( {\cal E}_{ph} /\rho_0 \right)^{1/4} \nonumber \\
&\mbox{}& \qquad \qquad \qquad \qquad \qquad \qquad \qquad \qquad \qquad \qquad \qquad \qquad \times \left( U^{-1/2} + \frac{\dot{S} }{4 {\cal E}_{ph} } U^{-3/4} \right). \label{eq:31} \end{eqnarray}
 Expressions (\ref{eq:29}) - (\ref{eq:31}) allow us to assemble the $\alpha_K = O(1)$ expectations for $\langle {Z^{\infty}}^2 \rangle$ and $\lambda$ in the vicinity of the transition region for a dynamical chromosphere populated by multiple temporal flows as
\begin{eqnarray}
\langle \langle {Z^{\infty}}^2 \rangle \rangle (h) &\simeq& \frac{1}{2 \alpha_K} \frac{\lambda_0}{h_0} e^{h/2h_0} \left[ \frac{n_1}{N} \left( {\cal E}_{MC}^{tot} /\rho_0 \right)^{1/2} \bar{U}_1 + \frac{n_2}{N} \left( 2 \dot{S} /\rho_0 \right)^{1/2} \frac{\bar{U}_2^{3/4} }{\left( \bar{U}_2^2 + \langle u_2^2 \rangle \right)^{1/8} } \right. \nonumber \\
&\mbox{}& \qquad \qquad \mbox{} + \left. \left( {\cal E}_{ph} /\rho_0 \right)^{1/2} \sum_{i=3}^4 \frac{n_i}{N} \left( \frac{\dot{S} }{2 {\cal E}_{ph} }  + \bar{U}_i \right) \right]; \label{eq:32} \\
\langle \lambda \rangle (h) &\simeq& \sqrt{2 \alpha_K} \left( h_0 \lambda_0 \right)^{1/2} e^{h/4h_0} \left[ \frac{n_1}{N} \left( {\cal E}_{MC}^{tot} /\rho_0 \right)^{1/4} \frac{ \left( \bar{U}_1^2 + \langle u_1^2 \rangle \right)^{3/8} }{\bar{U}_1^{5/4} } + \frac{n_2}{N} \left( 2 \dot{S} /\rho_0 \right)^{1/4} \frac{ \left( \bar{U}_2^2 + \langle u_2^2 \rangle \right)^{21/32} }{ \bar{U}_2^{33/16} } \right. \nonumber \\ 
&\mbox{}& \qquad \qquad \mbox{} + \left. \left( {\cal E}_{ph} /\rho_0 \right)^{1/4} \sum_{i=3}^4 \frac{n_i}{N} \left( \frac{\dot{S} }{4 {\cal E}_{ph} } \frac{ \left( \bar{U}_i^2 + \langle u_i^2 \rangle \right)^{15/8} }{\bar{U}_i^{21/4} }  + \frac{ \left( \bar{U}_i^2 + \langle u_i^2 \rangle \right)^{3/8} }{\bar{U}_i^{5/4} } \right) \right]. \label{eq:33}
\end{eqnarray}
As for the examples of Case 1, we provide estimates for expectations of the turbulent Els\"asser specific energy, correlation length, energy density, and turbulent injection energy at the base of the corona for the multi-flow $\alpha_K = 1$ example. On assuming the same parameters as used in Case 1 with the addition of a nonzero ${\cal E}_{ph} \sim 15$ J m${}^{-3}$, we obtain 
\begin{eqnarray}
\langle \langle {Z^{\infty}}^2 \rangle \rangle (h = 2 \; \mbox{Mm}) \sim 8.93  \times 10^{10}  \; \mbox{m}^2 \; \mbox{s}^{-2}; \quad 
\langle \lambda \rangle \sim 8.1 \; \mbox{Mm}; 
\langle y \rangle \sim 1.03 \; \mbox{J} \; \mbox{m}^{-3}; \quad 
\langle \dot{S}_{TR} \rangle \sim 1.32 \times 10^4 \; \mbox{J} \; \mbox{m}^{-2} \; \mbox{s}^{-1}. \label{eq:34} 
\end{eqnarray}
Although the estimates ((\ref{eq:34}) are reasonable, like the previous $\alpha_K = 1$ estimates, the correlation length is roughly twice that found by \cite{Bailey_etal_2025} although a value of $\alpha_K = 0.25$ would reduce $\langle \lambda \rangle \sim 4$ Mm and increase $\langle \langle {Z^{\infty}}^2 \rangle \rangle \sim 3.56 \times 10^{11}$ m${}^2$ s${}^{-1}$. The choice of $\alpha_K = 0.25$ is suitable for the constraints under which the expectations (\ref{eq:32}) and (\ref{eq:33}) were derived, and the turbulent energy injection rate $\langle \dot{S}_{TR} \rangle$  (\ref{eq:34}) meets the \cite{Anderson_Athay_1989a} threshold. 

{\it Case 2, $\alpha_K = 0.1$.} The results from the $\alpha_K = 1$ case above carry through unchanged for $i = 1 \dots 3$ flows. For $i = 4$ (type II spicules),  $\langle {Z^{\infty}}^2 \rangle_4 (h)$ is given by equation (\ref{eq:16}) and $y_0 = \dot{S}/U + {\cal E}_{ph}$, from which we have $\left( y_0/y \right)_4 = 1$. Hence, 
\begin{eqnarray}
\langle \langle {Z^{\infty}}^2 \rangle \rangle (h) &\simeq& \frac{1}{2 \alpha_K} \frac{\lambda_0}{h_0} e^{h/2h_0} \left[ \frac{n_1}{N} \left( {\cal E}_{MC}^{tot} /\rho_0 \right)^{1/2} \bar{U}_1 + \frac{n_2}{N} \left( 2 \dot{S} /\rho_0 \right)^{1/2} \frac{\bar{U}_2^{3/4} }{\left( \bar{U}_2^2 + \langle u_2^2 \rangle \right)^{1/8} } + \frac{n_3}{N} \left( {\cal E}_{ph} /\rho_0 \right)^{1/2} \left( \frac{\dot{S} }{2 {\cal E}_{ph} }  + \bar{U}_3 \right) \right] \nonumber \\
&\mbox{}& \qquad \qquad \mbox{} + \frac{n_4}{N} e^{h/h_0} \left( {\cal E}_{ph} /\rho_0 \right) \left( 1 + \left( \dot{S}/{\cal E}_{ph} \right) \frac{ \bar{U}_4^2 + \langle u_4^2 \rangle }{\bar{U}_4^3 } \right); \label{eq:35} \\
\langle \lambda \rangle (h) &\simeq& \sqrt{2 \alpha_K} \left( h_0 \lambda_0 \right)^{1/2} e^{h/4h_0} \left[ \frac{n_1}{N} \left( {\cal E}_{MC}^{tot} /\rho_0 \right)^{1/4} \frac{ \left( \bar{U}_1^2 + \langle u_1^2 \rangle \right)^{3/8} }{\bar{U}_1^{5/4} } + \frac{n_2}{N} \left( 2 \dot{S} /\rho_0 \right)^{1/4} \frac{ \left( \bar{U}_2^2 + \langle u_2^2 \rangle \right)^{21/32} }{ \bar{U}_2^{33/16} } \right. \nonumber \\ 
&\mbox{}& \qquad \qquad \mbox{} + \left. \frac{n_3}{N} \left( {\cal E}_{ph} /\rho_0 \right)^{1/4} \left( \frac{\dot{S} }{4 {\cal E}_{ph} } \frac{ \left( \bar{U}_3^2 + \langle u_3^2 \rangle \right)^{15/8} }{\bar{U}_3^{21/4} }  + \frac{ \left( \bar{U}_3^2 + \langle u_3^2 \rangle \right)^{3/8} }{\bar{U}_3^{5/4} } \right) \right] + \frac{n_4}{N} \lambda_0. \label{eq:36}
\end{eqnarray}
For $\alpha_K = 0.1$, we estimate the expected values at the transition region to be 
\begin{eqnarray}
\langle \langle {Z^{\infty}}^2 \rangle \rangle &\sim& 6.5  \times 10^{11}  \; \mbox{m}^2 \; \mbox{s}^{-2};  \qquad \langle y \rangle \sim 7.53 \; \mbox{J} \; \mbox{m}^{-3};  \quad 
\langle \lambda \rangle \sim  2.6 \; \mbox{Mm}; \quad 
\langle \dot{S}_{TR} \rangle \sim 9.63 \times 10^4 \; \mbox{J} \; \mbox{m}^{-2} \mbox{s}^{-1}.  \label{eq:37} 
\end{eqnarray}
The equations illustrate the obvious point that little of the turbulence entrained in high-speed type II spicules has had time to cascade and dissipate in the chromosphere and therefore carries much of its entrained turbulence energy into the lower corona by which time the spicule has had sufficient time to experience heating by turbulent dissipation. This provides a natural explanation for why type II spicules are observed to become hot at heights at and above the transition region \citep{dePontieu_etal_2009, dePontieu_etal_2011, Klimchuk_2012}, and is discussed further in Section \ref{sec:spicules}. By contrast, these results show that 1) the slower flows are critical in heating the chromosphere via the dissipation of turbulence, and 2) the strength of the dissipation, as controlled by the variable correlation length $\lambda$ and the size of t$\alpha_K$ are equally important in determining how effective are the various temporal chromospheric flows in allowing for the dissipation of turbulence in the chromosphere and coronal base. 

{\it Case 2, $\alpha_K = 0.01$.} This case is illustrated in the bottom panel of Figure \ref{fig:14b} and is the same, apart from the slightly different parameters, as that illustrated in Figure \ref{fig:14a}. The same analytic expressions as derived for Case 1, $\alpha_k = 0.01$ apply and the same conclusions hold. 

\subsection{Possible heating of spicules} \label{sec:spicules}

In closing this discussion, following the discovery of type II spicules by \cite{dePontieu_etal_2007}, a further result was confirmation that some fraction of the cold entrained spicule material was heated \citep{Sterling_Hollweg_1984, dePontieu_etal_2009}. This observation led \cite{dePontieu_etal_2009, dePontieu_etal_2011} to suggest that the hot plasma originating from type II spicules might be responsible for the high coronal temperatures needed to explain the acceleration of the solar wind.  This suggestion was examined closely by \cite{Klimchuk_2012} who found 
``that only a small fraction of the hot plasma can be supplied by spicules ($< 2$\% in active regions, $< 5$\% in the quiet Sun, and $< 8$\% in coronal holes.'' As emphasized already, we do not advocate for the heating of the corona via the deposition of heated chromospheric material in the corona but instead argue that randomly distributed chromospheric flows, which includes type I and II spicules, transport turbulence throughout the chromosphere where some portion is dissipated and the remainder can be injected above the transition region to contribute to coronal heating. We comment that it is not universally accepted that type I and II spicules form distinct categories \citep{Zhang_etal_2012, Sterling_etal_2010} but \cite{Pereira_etal_2012} pointed out that the classical type 1 spicules appear to be rarer than type II events and exhibit parabolic trajectories. The discussion here will be restricted to essentially upward or vertically propagating spicules with examples having speeds from 10 km s${}^{-1}$ to 100 km s${}^{-1}$. 

Despite \cite{Klimchuk_2012} concluding that the deposition of hot plasma in the corona by type II spicules cannot be responsible for the heating of the corona, the question of how spicules might be heated has not been well explained. The turbulence transport theory developed here lends itself well to  this problem. Klimchuk's Figure 1 provides a very nice graphical conceptualization of the apparently gradual/distributed heating of a type II spicule and its evolution with height. 

Consider a high speed spicule expanding into a flux tube with a variable cross-sectional factor $A(h) = g(h) A(0)$ where $g(h = 0) = A(0)$. Following \cite{Hollweg_1982}, we assume that the expansion factor at the transition region is $\sim 200$ times that at the photosphere, i.e., $g(h = 2 \: \mbox{Mm}) \simeq 200 A(0)$. A simple power law expansion for the flux tube is 
\begin{equation}
g(h) = \left( h/h_0 + 1\right)^{\alpha} = A(h)/A(0),  \label{eq:s12}
\end{equation}
where $h_0 = 1$ Mm and $\alpha = 4.8$ satisfies the constraint at $h_{TR} = 2$ Mm. We further assume that the spicule flow speed $U$ is constant and steady. Hence, the density profile is simply 
\begin{equation}
\bar{\rho} (h) = \rho_0 \frac{A(0)}{A(h)} = \rho_0 \left( h/h_0 + 1\right)^{-\alpha}, \label{eq:s13}
\end{equation}
and $\rho_0 = 2 \times 10^{-4}$ kg m${}^{-3}$ as before. The steady turbulence transport equation (\ref{eq:2}) can be solved using (\ref{eq:s13}) to obtain
\begin{equation}
y(h) =  \frac{ y_0 }{1 + \frac{2 \alpha_K}{\alpha + 2} (h_0/\lambda_0 )\left(  y_0 /(\rho_0 U^2) \right)^{1/2} \left( \left( h/h_0 + 1 \right)^{(\alpha + 2)/2} - 1 \right)  }; \qquad \lambda_c = \left( y_0 /y(h) \right)^{1/2} \lambda_0, 
\label{eq:s14}
\end{equation}
where $y_0 = y(h = 0)$ is the Els\"asser turbulent injection energy at the photosphere. As before, we restrict our attention to turbulence generated by the magnetic carpet due to the transformation of  magnetic carpet loops to turbulence, i.e., $\dot{S}_{MC}$. This injection rate will be mediated by the speed of propagation $U$ of the spicule i.e., $\dot{S}_{MC}/U$. As in the case of randomly distributed flows discussed above, the injection of photospheric Els\"asser turbulence energy ${\cal E}_{ph}$ J m${}^{-3}$ will be entrained by the spicule flow that has a speed $U$. Hence, we use the same expression for $y_0$ as used for spicules in the randomly distributed flows case, i.e., $y_0 = \dot{S}/U + {\cal E}_{ph}$ J m${}^{-3}$ with the same parameters, $\dot{S} = 2 \times 10^4$ J m${}^{-2}$ s${}^{-1}$ and ${\cal E}_{ph} = 15$ J m${}^{-3}$. 

\begin{figure}
%%\epsscale{.80}
\includegraphics[width= 1.0\textwidth]{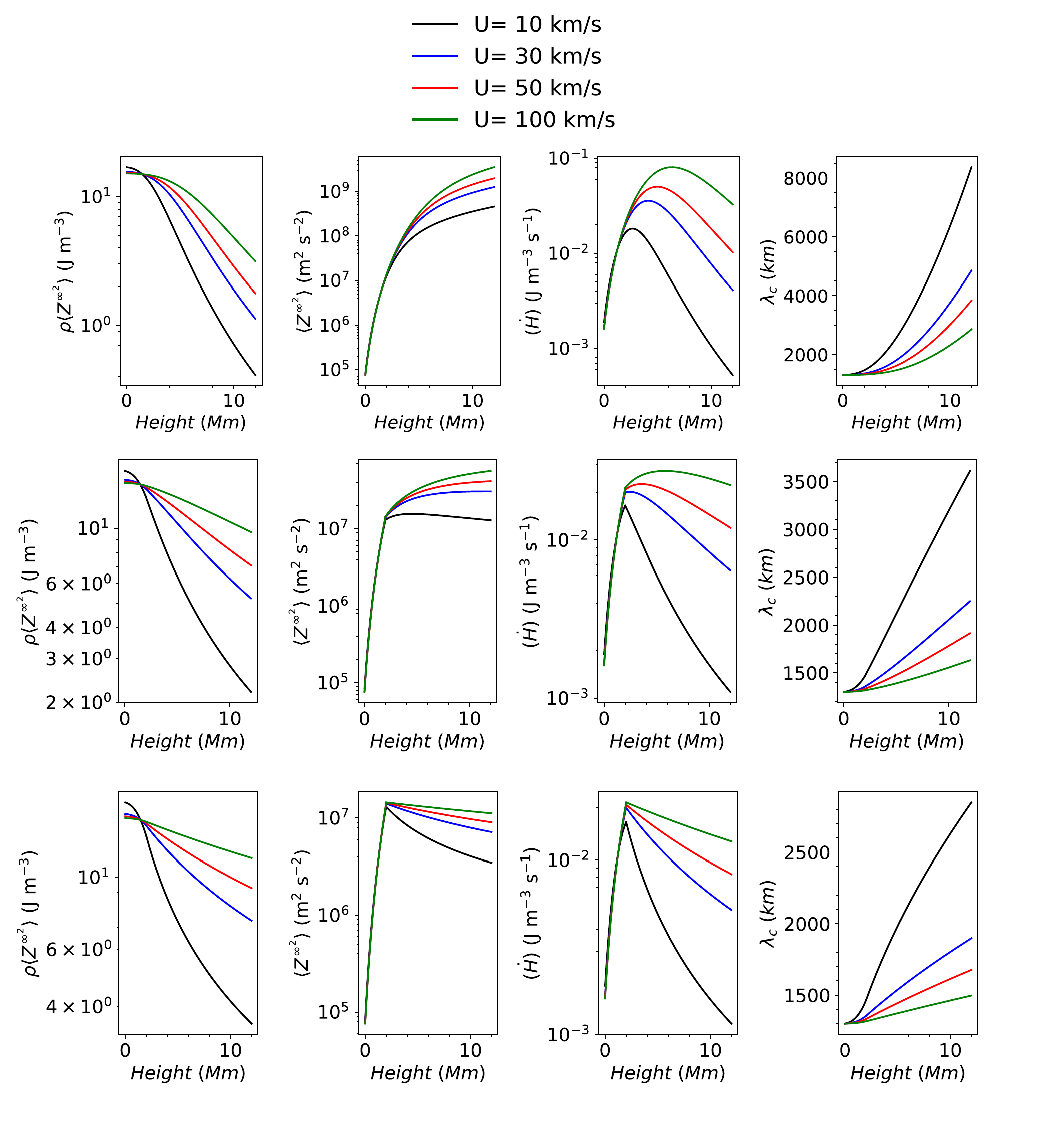}
\caption{\small  Representative plots with the Kolmogorov factor $\alpha_K = 1$ showing the 
 total energy density $\bar{\rho}(h) \langle {Z^{\infty}}^2 \rangle (h)$ J m${}^{-3}$ (left column), the Els\"asser specific energy $ \langle {Z^{\infty}}^2 \rangle (h)$ m${}^2$ s${}^{-2}$ (or J kg${}^{-1}$, i.e.,  energy per unit mass) (left middle column), the heating rate $\dot{\cal H}$ J m${}^{-3}$ s${}^{-1}$ (right middle column), and the correlation length $\lambda_c (h)$ km for four example type I and II spicules, differentiated by speed, $U = 10$ (black curve), 30 (blue), 50 (red), and 100 (green) km s${}^{-1}$. The transition region is assumed to be located at 2 Mm and the various quantities are plotted up to a height of 12 Mm. The top panel shows solutions for a strongly expanding flux tube with expansion factor $A(h)/A(0)$ given by expression (\ref{eq:s12}) up to 12 Mm. The middle panel shows results for a flux tube  expanding weakly above the transition region with the expansion factor given by (\ref{eq:s15}) for $\beta = 1$, and finally the bottom panel corresponds to a flux tube that is not expanding above the transition region, i.e.,  with $\beta = 0$ in (\ref{eq:s15}). 
Here we use the following parameters: $\dot{S} = 2 \times 10^4$ J m${}^{-2}$ s${}^{-1}$, ${\cal E}_{ph} = 15$ J m${}^{-3}$ J m${}^{-1}$; $\rho_0 = 2 \times 10^{-4}$ kg m${}^{-3}$; $h_0 = 1$ Mm; $\lambda_c = 1.3$ Mm; $h_{TR} = 2$ Mm; $\alpha = 4$, and $\beta = 1$ (middle panel) and 0 (bottom panel).    } \label{fig:15}
\end{figure}

Illustrated in the four panels across the top row of Figure \ref{fig:15} are four examples of spicules with different speeds, $U = 10$, 30, 50, and 100 km s${}^{-1}$. The parameters are listed in the figure caption and the Kolmogorov factor $\alpha_K = 1$. The top panel of curves is more illustrative than realistic since we have assumed that the expansion factor (\ref{eq:s12}) and hence the density profile (\ref{eq:s13}) applies in both the chromosphere and in the corona until about 12 Mm. We panels in the two rows below use a more reasonable form of $A(h)$ that resembles that used by \cite{Hollweg_1982} and illustrated by e.g., the flux tubes shown in Figure 2 of \cite{Sterling_2000}. The left column shows the 
 total energy density $\bar{\rho}(h) \langle {Z^{\infty}}^2 \rangle (h)$ J m${}^{-3}$, the left middle column the the Els\"asser specific energy $ \langle {Z^{\infty}}^2 \rangle (h)$ m${}^2$ s${}^{-2}$, the right middle column the heating rate $\dot{\cal H}$ J m${}^{-3}$ s${}^{-1}$, and the right column shows the correlation length $\lambda_c (h)$ km.  

At the photosphere, the Els\"asser energy density $\bar{\rho} \langle {Z^{\infty}}^2 \rangle$ [J m${}^{-3}$]  is modestly different for each case, ordered inversely with the assumed spicule speed $U$ because of the fixed injection rate for the magnetic carpet turbulence. Each of the energy density curves decay at a different rate however, with the faster spicules advecting the turbulence further during a dissipative time scale and hence carrying more turbulent energy per cubic meter to greater heights than slower spicules. This is reflected too in the middle right panel that shows the heating rate $\dot{\cal H}$ [J m${}^{-3}$ s${}^{-1}$].  The heating rate for the slowest spicule ($U = 10$ km s${}^{-1}$) peaks at the lowest height, just above the transition region ($\sim 3$ Mm). The heating peak location is ordered by spicule speed -- the 100 km s${}^{-1}$ spicule heating rate peaks at about 6 Mm. The left middle panel shows the Els\"asser specific energy as a function of height. Because of the assumed continued expansion of the flux tube, the specific energy continues to increase with height with the value at 12 Mm ordered by spicule speed. 
This behavior of the Els\"asser specific energy is not realistic, as illustrated in the panels below, since one needs a flux tube that is either constant or expands more slowly above the transition region than given by the expressions (\ref{eq:s12}) and (\ref{eq:s13}). Finally, the correlation length too is ordered by $U$ with the slowest speed spicule having the largest correlation length $\lambda_c$ by 12 Mm and the fastest spicule having the smallest correlation length. This of course is simply due to the rate of dissipation that is controlled by the size of $\lambda_c$ in the Kolmogorov theory.  

\cite{Hollweg_1982} used an expansion factor $A(h)/A(0)$ that becomes constant above the transition region, this to reflect the structure of magnetic funnels \citep[e.g.,][]{Sterling_2000}. We  introduce a slight generalization of Hollweg's flux tube model by defining 
\begin{equation}
A(h) = \left\{ \begin{array}{ll}
                     A(0) \left( h/h_0 + 1 \right)^{\alpha} & h \leq h_{TR} \\
                     A(h_{TR}) \left( h/h_{TR} \right)^{\beta} & h > h_{TR}
                     \end{array} \right.,                                                           \label{eq:s15}
\end{equation}
where $\beta < \alpha$ (and $\beta = 0$ would approximate the \cite{Hollweg_1982} form with a vertical  upper section of the funnel), $\bar{\rho} (h) = \rho_{TR} \left( h_{TR} / h \right)^{\beta}$, $h > h_{TR}$, and $\rho_{TR}$ is evaluated from equation (\ref{eq:13}) with  $\rho_{TR} = \bar{\rho} (h = h_{TR})$. This then yields the solution of $y(h)$ for $h > h_{TR}$ as 
\begin{equation}
 y(h) =  \frac{ y_{TR}^0 }{1 + \frac{2 \alpha_K}{\beta + 2} (h_{TR}/\lambda_0 )\left(  y_{TR}^0 /(\rho_0 U^2) \right)^{1/2} \left( \left( h/h_{TR} \right)^{(\beta + 2)/2} - 1 \right)  }, \qquad  \lambda_c = \left( y_0 /y(h) \right)^{1/2} \lambda_0. 
\label{eq:s16}
\end{equation}
Here we connect the solution below the transition region (\ref{eq:s14}) to the lower boundary of the new solution (\ref{eq:s16}), i.e., $y(h = h_{TR}) \equiv y_{TR}^0$. The total solution with a more complex flux tube model is therefore given by equations (\ref{eq:s14}) and (\ref{eq:s16}) together with the flux tube functions (\ref{eq:s12}) and (\ref{eq:s15}) and the corresponding density profiles. 

Illustrated in the second and third rows of panels of Figure \ref{fig:15} are solutions for funnel-shaped flux tubes. The middle row solutions use a slow expansion factor $\beta = 1$ and the bottom row of panels use non-expanding upper sections with $\beta = 0$. The curves obviously do not change smoothly in passing from pre- to post-transition regions because of the abrupt change in the solutions (\ref{eq:s14}) and (\ref{eq:s16}) induced by the expansion factor (\ref{eq:s15}). Nonetheless, the basic character of the solutions is captured. It is evident that there is a distinct difference in the evolution with height of the Els\"asser energy density $\bar{\rho} \langle {Z^{\infty}}^2 \rangle$, specific energy $\langle {Z^{\infty}}^2 \rangle$ and heating $\dot{\cal H}$ behavior. The rate at which the total energy decays, while still ordered by spicule speed $U$ as before, decays more slowly as the flux tube opening decreases ($\beta$ becomes smaller). In the middle row, the ``geometric'' increase in the specific energy is also slowed for the three higher speed spicules, and even reversed for $U =10$ km s${}^{-1}$ example. When the flux tube is no longer expanding, bottom row, the specific energy for all the example spicules decays immediately on crossing the transition region. The middle row shows too that the heating function for $U = 10$ km s${}^{-1}$ is strongest at the transition region and the higher speed spicules peak a little above the transition region, at most at about 4 - 5 Mm. For the non-expanding flux tube, bottom row, the heating maximizes at the transition region for values of $U$. The correlation lengths are quite significantly different when the flux tube above the the transition region expands slowly or not at all. From the top row to the middle row, the correlation lengths for each $U$ decrease by about half, and decrease again from the middle to the bottom row. Much of the behavior illustrated in the plots of Figure \ref{fig:15} can be understood from the absence or near-absence of adiabatic expansion above the transition region. Evidently, the speed of the spicules is important in advecting greater levels of turbulent energy up to much greater heights above $h_{TR}$ to ensure heating in coronal regions that are no longer dominated by collisional radiative losses. 

\begin{figure}
%%\epsscale{.80}
\includegraphics[width= 1.0\textwidth]{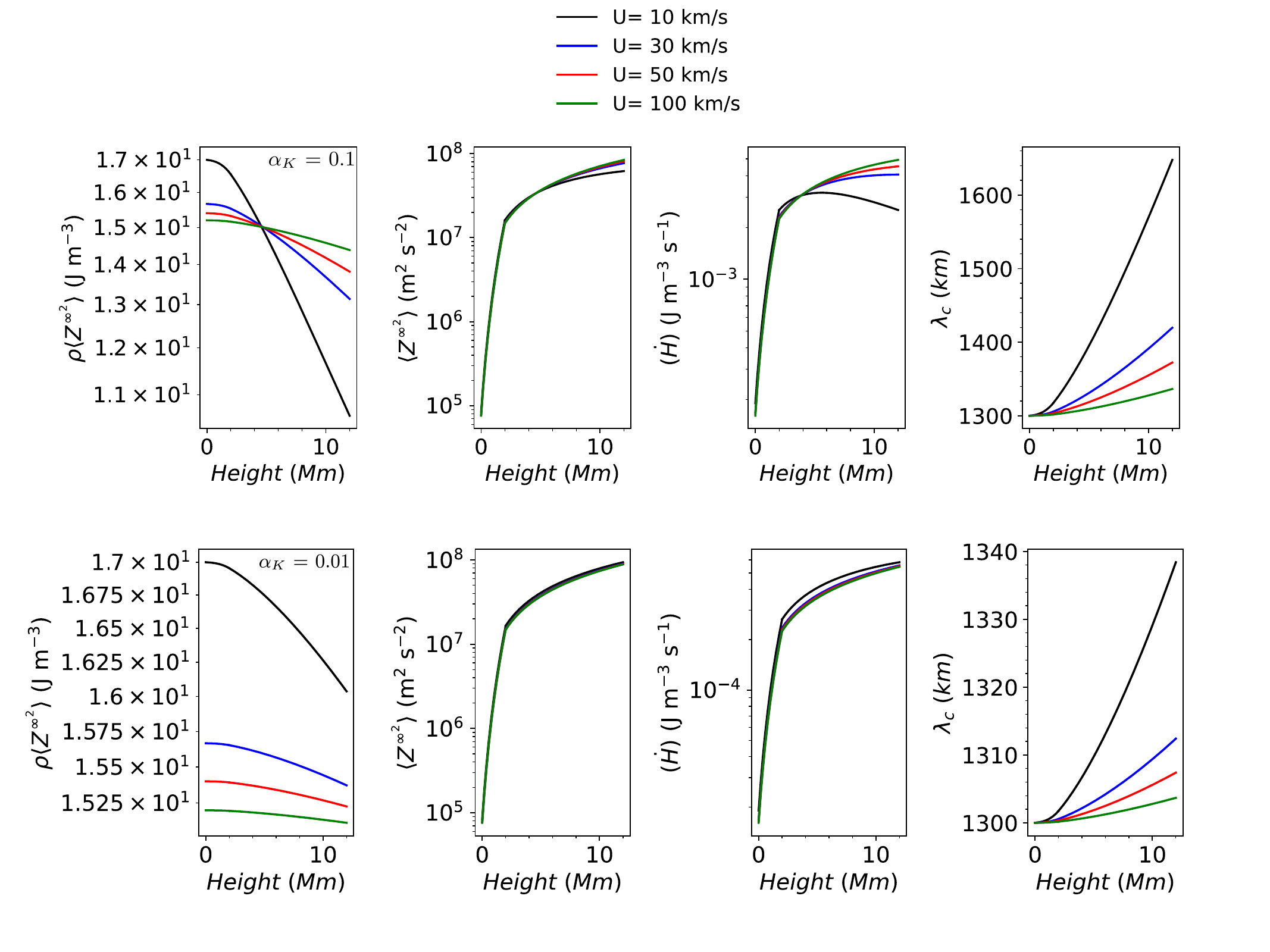}
\caption{\small  Plots that correspond to the middle panel of Figure \ref{fig:15} (a weakly expanding flux tube above the transition region with $\beta = 1$) showing (top panels) the case for a  constant $\alpha_K = 0.1$ and (bottom panels) $\alpha_K = 0.01$. Otherwise, the format and parameters are identical to those used in  Figure \ref{fig:15}.   } \label{fig:16}
\end{figure}

The final Figure \ref{fig:16} illustrates the role of $\alpha_K$ with the top row adopting a value $\alpha_K = 0.1$ and the bottom row 0.01. Here, the figures correspond to the middle row of Figure \ref{fig:15}, i.e., a weakly expanding flux tube above the transition region with $\beta = 1$. The first column shows quite strikingly different behavior of the total energy $\bar{\rho} \langle {Z^{\infty}}^2 \rangle$ in the rate at which it decays (note that the boundary values at $h = 0$ are identical for all cases of $\alpha_K$, the rate at which the energy decays is markedly slower as $\alpha_K$ decreases, and the total energy decreases very little by 12 Mm. The slow energy decay is reflected in the correlation length $\lambda_c$ scarcely changing with increasing $h$. The specific energy continues to increase for all speeds $U$ above the transition region and, with one exception, the heating rate continues to increase too above the transition region. In summary, high speed flows such as spicules are sensitive to changes in the assumed values of $\alpha_K$. 

\subsection{Possible role of prominences} \label{sec:prominences} 

Although spicules share properties with solar prominences, or filaments, in that both features consist of ``chromospheric'' material, spicules and (eruptive/active region and quiescent) prominences/filaments represent quite different classes, in properties, possible generation mechanisms, and the extent of photospheric coverage \citep[e.g.,][]{Engvold_Ch_2_2015}. 
Unlike the magnetic carpet, and to some extent spicules, the percentage of the solar surface area covered by prominences is small. Filaments/prominences tend to be found above filament channels, and these channels, rooted in the photosphere, typically extend into the lower corona. The magnetic environment of the low coronal extended filament channels shields cooler filamentary material thermally from the surrounding hot corona as well as providing magnetic support against gravity. Filament channels are oriented along the ``polarity reversal boundary (PRB)’’ that divides opposite polarities in the line-of-sight magnetic fields measured in the photosphere \citep[e.g.,][]{Martin_2015, Mackay_2015}. Quiescent and active region prominences exhibit a large-scale spine oriented along the PRB and smaller-scale barbs extend from the spine into the chromosphere \citep[e.g.,][]{ Martin_etal_2008}. Filament barbs appear to be connected or rooted in enhanced concentrations of magnetic flux located at photospheric supergranulation cell boundaries \citep{Plocieniak_Rompolt_1973, Martin_Echols_1994, Lin_etal_2005}. High-resolution H$\alpha$ images identify thin threads within spines and barbs and are thought to comprise the fundamental structures of all solar filaments \citep{Lin_etal_2008}. Indeed, \cite{Engvold_Ch_2_2015} suggests that a barb consists of numerous thin threads, with neighboring threads within barbs apparently rooted in separate but closely spaced locations in the chromosphere. At the barb base, the volume density of the threads is so high that individual threads cannot be resolved. 

The global dimensions of quiescent prominences are typically $< 5$ Mm wide, 20 - 50 Mm high by 200 Mm long, but this can vary \citep{Engvold_Ch_2_2015}, and they reach both higher and lower solar latitudes (greater than and less than $\sim 40^{\circ}$) with an 
involved solar-cycle dependence. Filaments, and certainly the quiescent class particularly, are typically very stable and long-lived globally yet are highly dynamical on fine spatial scales and short-time scales. \cite{Berger_etal_2010} describe observations of small-scale turbulent upflows seen in several prominences, beginning from between the bottom of the prominence and the top of the chromospheric spicule layer, rising to $\sim 10$ - 15 Mm. Maximum initial upflow speeds range from  20 - 30 km s${}^{-1}$ and last for some 5 – 15 minutes. \cite{Ryutova_etal_2010} use high-resolution and high-cadence Hinode satellite data to study observationally various instabilities associated with quiescent prominences. Eclipse images \citep{Druckmuller_etal_2014, Habbal_etal_2014, Habbal_etal_2026} offer an extraordinary glimpse into the dynamical nature of filaments, including ``smoke-rings,'' larger-scale Kelvin-Helmholtz and possibly Rayleigh-Taylor instabilities, and more.  

In concluding this brief discussion about prominences, we note that a class of abundant, very small-scale filaments, ``mini-filaments,’’ has been identified \citep{Wang_etal_2000}, apparently existing in much larger numbers than the classical filaments/prominences described above \citep[see e.g., Figure 2 of][although strictly speaking, that plot applies to erupting filaments and mini-filaments, and additional categories of non-eruptive versions may exist]{Sterling_etal_2024}. However, we should be cautious in interpreting mini-filaments as a form of quiescent filaments since the data is not yet available for extensive comparison between typical filaments and mini-filaments. Based on their characteristic size and properties, and if indeed mini-filaments are highly prevalent and widely distributed throughout the chromosphere, they may well be a source of chromospheric turbulence while enhancing its transport in much the same way as spicules do.  

Related to the mechanism for chromospheric (and coronal) heating presented in this paper, the question is whether prominences/filaments play an important role in 1) generating turbulence in the chromosphere, and 2) in transporting turbulence and mixing it throughout the chromosphere for it to dissipate the heat the chromospheric plasma. The prominence morphology suggests 1) that filaments are rooted in the chromosphere through their barb structure, and 2) that the connection to the chromosphere lies primarily along supergranulation cell boundaries. Starting with the second point, since the feet of prominence barbs appear to be localized to supergranulation cell boundaries, very little of the chromospheric plasma above the photosphere will be mediated by filaments or their dynamics. By contrast, the magnetic carpet, which covers almost the entire surface of the Sun, and similarly spicules, has access to the entire volume of the chromospheric plasma. Nonetheless, the first point, including the possibility of turbulent upflows originating from the the feet of prominence barbs, suggests that turbulence transported and generated by the barbs of filamentary structures may play a very localized role in both generating and transporting turbulence in the vicinity of supergranular boundaries. 

Perhaps a more interesting question is if the turbulence generation, entrainment, and transport mechanisms described in this work are relevant to the initiation and dynamics of prominences. The formation mechanism for prominences is not settled, and one popular approach is described as ``chromospheric-evaporation condensation,’’ in which localized heating occurs at the coronal loop footpoints. The heating drives plasma evaporation. However, the characteristics of the heating mechanism itself strongly influences the occurrence of condensation. In a very interesting study, \cite{ Yoshihisa_etal_2025} considered a single heating event along a single field line, treating the single field line as an elemental unit of a coronal loop. Specifically, they used a 1.5D MHD model that included the effects of 
radiative cooling, thermal conduction, gravity, and energy dissipation by shock waves and Alfv\'en wave turbulence. \cite{Yoshihisa_etal_2025} found that incorporating additional energy dissipation through a 
phenomenological Alfv\'en wave turbulent cascade \citep{Shoda_etal_2018} resulted in vertical velocities consistent with observations. \cite{Yoshihisa_etal_2025} further identified parametrically the heating rate characteristics that led to condensation of the filament flows. Although a more complex and fundamentally multi-dimensional problem than that of spicule heating, it is possible that the role of turbulence as described in this paper may provide an explicit turbulence formulation for the ``chromospheric-evaporation condensation’’ model advanced by \cite{Yoshihisa_etal_2025}. 

\section{Conclusions}

Our understanding of low-frequency turbulence in the solar chromosphere remains poorly understood. This work addresses two of the most important questions about chromospheric turbulence: 1) what is the source(s) of low-frequency turbulence that potentially contributes to the heating of the chromosphere, and 2) once generated, how is the turbulence transported and dissipated throughout the chromosphere and possibly up to the lower boundary of the solar corona? In the first part of the paper, we focus on part 1) of the question, using a particle-in-cell (PIC) code to examine reconnection-driven turbulence in a mixed polarity magnetic field that corresponds physically to the constant emergence of the magnetic carpet in both open coronal hole and quiet Sun environments, as advocated by e.g., \cite{Martinez_Gonzalez_etal_2010, Cranmer_vanBallegooijen_2010, Cranmer_etal_2013, Zank_etal_2018, Zank_etal_2021}. In the second part of the paper, we utilize the results from the first part to develop a simple transport model for turbulence generated by the magnetic carpet and entrained photospheric turbulence based on a random distribution of energy-containing range temporal flows within the chromosphere.

Magnetic loops of mixed polarity emerge constantly from the photosphere, replenished on a very short timescale of $\sim 1$ -- 2 hours \citep{Hagenaar_etal_2008}, and cover the entire surface of the quiet and active Sun. A little more than about 50\% of the emergent loops have heights below about 1 Mm \citep{Cranmer_vanBallegooijen_2010}. We use PIC simulations to model the evolution of mixed polarity magnetic field for two cases, defined by the initial imbalance of the polarity. In one case, we include some open field, assuming that the initial guide to reconnection plane magnetic field ratio $B_g/B_0 = 1$ (i.e., imbalanced), and in the other we assume predominantly mixed polarity field ($B_g/B_0 = 0.2$ or balanced). These choices can be regarded as surrogates for open coronal holes (or large-scale magnetic loops) and the quiet Sun, respectively. In both cases, the initial mixed polarity magnetic field transitions rapidly to a highly turbulent state dominated by small-scale nonlinear structures. The reconnection-driven turbulence is initiated by the destabilization of the current sheet by the tearing mode instability, and the current sheet begins to break up into multiple magnetic flux ropes of different scales.  The multi-scale magnetic flux ropes interact and merge,  producing secondary flux ropes or magnetic islands continuously in the 3D reconnection layer. The simulation box for both imbalanced and balanced states forms a highly turbulent reconnection layer that is very inhomogeneous. 

In the case of open coronal holes or at the base of large-scale loop structures, since a mean field  can be identified approximately when $B_g /B_0 = 1$, the simulations show that the low-frequency turbulence is anisotropic, i.e., the turbulence characteristics are different in directions perpendicular and parallel to the mean magnetic field directions. The fluctuations in the simulation are  dominated by transverse magnetic fluctuations with wavenumbers perpendicular to the mean magnetic field. These fluctuations correspond to structures and can be thought of as quasi-2D advected fluctuations and form the majority population.  The PSD for advected fluctuations $\tilde{B}_{\perp} (k_{\perp})$, $\langle \tilde{B}_{\perp}^2 \rangle (k_{\perp}) = E_{\tilde{B}_{\perp}} (k_{\perp})$ has a fairly flat spectrum $\propto k_{\perp}^{-3/2}$ and is of larger amplitude than the minority slab spectrum population, which has $\langle \tilde{B}_{\perp}^2 \rangle (k_{\parallel}) \propto k_{\parallel}^{-2}$.  Besides the spectral anisotropy, the low-frequency turbulence exhibits a variance anisotropy with a quasi-2D to slab ratio of $\sim 1.25$. 

Regions of the chromosphere contained within coronal holes or very large loops will possess magnetic turbulence that is generated by 1) repeated reconnection between the predominant mixed magnetic carpet loops, and 2)  interchange reconnection between some magnetic carpet loops with open magnetic field. 
For these regions, although a large-scale mean magnetic field can be identified, the fluctuating magnetic field and kinetic energy are dominant since the the initial mixed polarity magnetic field of the magnetic carpet has been largely annihilated to be replaced by fully developed turbulence.
The  reconnection process between mixed polarity magnetic carpet loops generates quasi-2D turbulence whereas loop-open magnetic field interchange reconnection can initiate downward and upward propagating Alfv\'enic fluctuations that form the slab turbulence component. We therefore expect approximately balanced slab turbulence, i.e., the  normalized slab cross helicity $\sigma_A \simeq 0$. However, Alfv\'en waves will experience strong reflection at the transition region (perhaps $> 95$\% of the incident flux).  
The counter-propagating reflected Alfv\'en flux will interact with the upward Alfv\'en flux to initiate a nonlinear cascade that generates 2D zero-frequency non-propagating modes \citep{Shebalin_etal_1983} in the chromosphere. 
The efficient reflection of Alfv\'en waves at the transition region presents an additional further possibility for the origin of 2D turbulence \citep{Zank_etal_2021}. We therefore expect that most of the interchange reconnection generated Alfv\'enic fluctuations will be trapped in the chromosphere, unable to easily transmit through the transition region and thus enhance the levels of advected turbulent structures. 

Simulations for the $B_g/B_0 = 0.2$ or quiet Sun case show that the simulation becomes dominated rapidly by magnetic and velocity fluctuations and the mean fields are largely annihilated. Since the simulation began with almost balanced mixed polarity magnetic field, and hence no clearly defined guide magnetic field, the reconnection-driven turbulent field is essentially isotropic. The fluctuating magnetic field is composed primarily of non-propagating structures that are randomly oriented. The  quiet Sun and open Sun regions are therefore distinguished by the geometry of the structures, the former being distributed isotropically and the latter being quasi-2D. The central result from the $B_g/B_0 = 0.2$ case is that the low-frequency turbulence is dominated by randomly oriented magnetic structures such as small-scale magnetic flux ropes. 

Having addressed the question of the origin of turbulence in the chromosphere via the reconnection-driven annihilation of magnetic carpet loops, we address the equally critical question of how turbulence is transported throughout the chromosphere. Since the turbulence is found to be dominated by small-scale structures, it is only through advection that turbulence can be distributed and dissipated throughout the chromosphere, and possibly transported into the lower corona. A large-scale mean flow does not exist in the chromosphere and the solar atmosphere is more appropriately described as an atmosphere with an exponentially decreasing density profile with increasing height. The assumption of a static atmosphere is well supported by observations and detailed 1D models that investigate chromospheric line formation \citep[e.g.,][]{Vernazza_etal_1981, Maltby_etal_1986, Fontenla_etal_1993}. However, it is now recognized that the chromosphere is highly dynamical \citep[e.g.,][]{Sterling_2000, Freytag_etal_2002, Wedemeyer_etal_2004, Hagenaar_etal_2008, Martinez_Gonzalez_etal_2010, Gudiksen_etal_2011, Carlsson_etal_2016, Carlsson_etal_2019}, filled with flows on scales that can be associated with an energy-containing range, such as post-emergent flows associated with the magnetic carpet, shock waves and post-shock flows, and type I and II spicules emerging from the photosphere, for example. We introduce a turbulence transport model that is based on Kolmogorov theory and related to a von Karman-Howarth-Dryden description \citep{zank_etal_1996_evolutionmagfluct, matthaeus_etal_1996_mhdphenom, Zank_etal_2017a}. We introduce a statistical description for the random flow field in the chromosphere. We assume log-normal statistics for the pdf $f(U)$ of the flow speeds and use the solutions to the turbulence transport equations to obtain expectations for the evolution of the total energy per unit volume $\langle y \rangle (h)$ J m${}^{-3}$, the 
Els\"asser specific energy $\langle \langle {Z^{\infty}}^2 \rangle \rangle (h)$ m${}^2$ s${}^{-2}$ (or J kg${}^{-1}$, i.e., the energy per unit mass), the heating rate function $\langle \dot{\cal H} \rangle (h)$ J m${}^{-3}$ s${}^{-1}$, and the correlation length $\langle \lambda \rangle (h)$ km  
as functions of height $h$. The numerically evaluated expectations were complemented with analytic estimates of the expectations near the transition region, $h = 2$ Mm, defined by equations (\ref{eq:12a}) and (\ref{eq:12b}), and we investigated the dependence on the Kolmogorov parameter $\alpha_K = 1$, 0.1, and 0.01. 

In summary, the results illustrated in Figures \ref{fig:13} - \ref{fig:14b} show that the injection of turbulence via the annihilation of the magnetic carpet along with the entrainment of photospheric turbulence by the post-emergent loop flow and its transport via advection by random chromospheric flows (flows associated with type I and II spicules, and post-shock and post-emergent magnetic carpet loop flows)  both actively heat the chromosphere and inject the remaining turbulent energy into the base of the corona at levels that exceed the \cite{Withbroe_Noyes_1977} - \cite{Anderson_Athay_1989a} constraint. 
The injection of turbulent energy into the low corona is not uniformly distributed but occurs rather as a  random ``patchwork'' of injection sites across the surface of the transition region. We considered several models describing the transport and dissipation of turbulence by random flow fields, ranging from a simple uni-flow pdf to a multi-flow case with four independent pdfs that considered either only magnetic carpet-generated turbulence or magnetic carpet-generated plus entrained photospheric turbulence. For each of these models, we computed, via both numerical and analytic approaches, the expected energy injection rates $\langle \dot{S} \rangle$ J m${}^{-2}$ s${}^{-1}$ for the chromosphere and at the base of the corona, together with the expected total energies, specific Els\"asser energies, and correlation lengths. 
 We find that the expected injection rates exceed the estimated energy requirement of \cite{Withbroe_Noyes_1977} and \cite{Anderson_Athay_1989a} at the lower coronal boundary, for example ($\sim 4 \times 10^3$  J m${}^{-2}$ s${}^{-1}$ \citep{Athay_1966, Withbroe_Noyes_1977} or the more recent estimate of \cite{Anderson_Athay_1989a} of total heat flux $1.4 \times 10^4$ J m${}^{-2}$ s${}^{-1}$). 
 We emphasize that it is the generation, transport, and dissipation of MHD turbulence that results in the heating of the chromosphere (and eventually the corona) and not the direct heating of plasma via  reconnection processes associated with magnetic carpet annihilation and then subsequent flows of hot plasma into the corona, the latter resembling the ``furnace'' model of \cite{Axford_McKenzie_1992, McKenzie_etal_1997} or the nanoflare heating model \citep{Parker_1988, Klimchuk_2015}. The  heating is due to the energy cascade associated with small scale quasi-2D magnetic flux ropes or magnetic islands to very small dissipation scales, and the nonlinear structures themselves are created by the reconnection of mixed polarity magnetic loops present in the continually emerging magnetic carpet. 
The approach described here appears to meet the constraints required for low-frequency turbulence to heat both the chromosphere and corona. The next step will be to integrate the turbulence heating model presented here into an appropriate temperature description of the chromosphere and corona.

As a final remark, we suggest that the dissipation of entrained turbulence, both magnetic carpet-generated and photospheric, by type I and II spicules will lead to the gradual heating of those flows with increasing height. This may provide an explanation for the observed heating of spicules that becomes more evident with increasing height \citep{dePontieu_etal_2009, Klimchuk_2012}. As illustrated in Figures \ref{fig:15} and \ref{fig:16}, the effectiveness of the dissipation and thus heating is  differentiated by the speed of the spicules since the dissipation rate is independent of the spicule flow speed, implying dissipation in high speed flows will be evident only at greater heights than for slower speed spicules. 

%% Please use the acknowledgment and contribution environments. This will 
%% be anonomyized when the "anonymous" style option is used. 
\begin{acknowledgments}
G.P.Z., MN, L.L.Z., and L.A. acknowledge the partial support of NASA Parker Solar Probe contract SV4-84017, NSF EPSCoR RII-Track-1 cooperative agreement OIA-1655280, a NASA IMAP subaward under NASA contract 80GSFC19C0027, and a NASA award 80NSSC20K1783. MY acknowledges the support of an  NSF grant AGS-2230633. X.L. and F.G. acknowledge support from NASA HTMS grant NNH240B72A and LWS grant 80NSSC21K1313. This work used resources of the National Energy Research Scientific Computing Center (NERSC), a U.S. Department of Energy Office of Science User Facility, under award FES-ERCAP0033004, and of the Oak Ridge Leadership Computing Facility at Oak Ridge National Laboratory, supported by the U.S. Department of Energy Office of Science under Contract No. DE-AC05-00OR22725, through an allocation from the ASCR Leadership Computing Challenge (ALCC) program.
\end{acknowledgments}

\end{document}